\documentclass[prx,reprint,amsmath,amssymb,bm,floatfix]{revtex4-2}

\usepackage{graphicx}
\usepackage{multirow}
\usepackage{amsfonts,amsthm}
\usepackage{mathrsfs}
\usepackage{xcolor}
\usepackage{booktabs}
\usepackage{ifthen}
\usepackage{bm}
\usepackage{tikz}
\usepackage{pgfplots}
\usepackage{hyperref}
\usepackage{subcaption}
\usepackage{adjustbox}
\usetikzlibrary{calc}
\usetikzlibrary{math,arrows.meta}

\newcommand{\mvariable}[3]{
  \ifthenelse{\equal{#3}{}}{{#1}(\mbox{#2})}{}
  \ifthenelse{\equal{#3}{leftright}}{{#1}\left(\mbox{#2}\right)}{}
  \ifthenelse{\equal{#3}{big}}{{#1}\big(\mbox{#2}\big)}{}
  \ifthenelse{\equal{#3}{Big}}{{#1}\Big(\mbox{#2}\Big)}{} }

\newcommand{\xvar}[2]{\mvariable{x}{#1}{#2}}
\newcommand{\xjvar}[2]{\mvariable{x^{(j)}}{#1}{#2}}
\newcommand{\xop}[2]{\mvariable{\hat{x}}{#1}{#2}}
\newcommand{\zop}[2]{\mvariable{\hat{z}}{#1}{#2}}
\newcommand{\zopd}[2]{\mvariable{\hat{z}^{\dagger}}{#1}{#2}}
\newcommand{\statesingle}[2]{\ifthenelse{\equal{#2}{}}{{\{\!\!\{{#1}\}\!\!\}}}{{\{\!\!\{{#1}\}\!\!\}}_{{#2}}}}
\newcommand{\stateset}[2]{\ifthenelse{\equal{#2}{}}{{\{{#1}\}}}{{\{{#1}\}_{{#2}}}}}
\newcommand{\occupation}[2]{\ifthenelse{\equal{#2}{}}{{#1}}{} }
\newcommand{\statevectorL}[2]{\ifthenelse{\equal{#2}{}}{{\langle{#1}|}}{{{}_{{#2}}\langle{#1}|}}}
\newcommand{\statevectorR}[2]{\ifthenelse{\equal{#2}{}}{{|{#1}\rangle}}{{|{#1}\rangle_{{#2}}}}}
\newcommand{\statevectorLR}[3]{\ifthenelse{\equal{#3}{}}{{\langle{#1}|{#2}\rangle}}{{}_{{#3}}{\langle{#1}|{#2}\rangle}_{{#3}}}}

\begin{document}
\title{Hidden correlations in quantum jumps: A theory of individual trials}
\author{Hiroshi Ishikawa}
\email[]{Contact author: hishihvt@outlook.jp}
\affiliation{College of Systems Engineering and Science, Shibaura Institute of Technology, Saitama 330-8570, Japan}
\date{\today}

\begin{abstract}
Quantum trajectory theories have not fully reconciled discrete quantum jumps with continuous unitary evolution.
We address this challenge by developing a hidden variable formulation that reveals hidden correlations in individual trials.
We represent jump sequences as logical propositions on energy eigenstate occupations, where logical variables describe their truth values and chain operators describe their probabilities.
To circumvent established no-go theorems, we introduce a proposition selection rule that requires well-defined joint probabilities for a given density operator.
This rule unifies compatibility criteria for quantum and hidden variable theories, eliminating incompatible propositions that lead to value assignment paradoxes.
Unlike conventional criteria mandating operator commutativity, it permits propositions that become deterministic for specific initial conditions.
This framework systematically captures deterministic relations in individual trials, including the exclusivity of state occupation implied by the observed jumps.
This exclusivity, distinct from the Pauli exclusion principle, enables the formalization of a quantum version of Boolean logic as laws for quantum propositions.
These assumptions circumvent Bell-type inequalities by prohibiting arithmetic operations between incompatible logical variables, while enabling classical descriptions within the complementarity.
The resulting theory reconciles discrete jumps with continuous evolution, providing a direct description of individual trials while preserving standard probability predictions.
\end{abstract}
\maketitle

\section{Introduction}
\label{sIntro}
Quantum jumps have been observed in diverse systems, including photon counters~\cite{Hamamatsu1,Bachor}, trapped ions~\cite{QuantumJump2,QuantumJump1}, and radiation fields in high-Q cavities~\cite{Haroche1,Haroche3}.
These discrete transitions have provoked considerable theoretical debate~\cite{Gisin1992,Zeh1993} because they apparently contradict continuous unitary evolution.
This contradiction is particularly evident in photon counting experiments, where individual measurements yield discrete integers $n(t)\,{=}\,0,1,\ldots$, while quantum theory predicts continuous expectation values $\langle n(t)\rangle\,{\in}\,\mathbb{R}$.
Such integer functions are non-differentiable at jump points and thus cannot be derived from any standard differential equations.
This observation reveals an intrinsic duality of quantum theory: individual outcomes do not emerge as solutions to its equations of motion~\cite{Einstein1905}.
Consequently, integer functions are found experimentally, even though they themselves cannot be derived from any differential laws.
\par
When quantum jumps were first observed for single quantum systems, this fundamental conflict between experiments and theories prompted two distinct interpretations.
The early interpretation~\cite{Cook1985,QuantumJump2} treated jumps as transitions between energy eigenstates, following Bohr's atomic model~\cite{Bohr1913} and Einstein's rate equations~\cite{Einstein1917}.
This view, suggesting that binary logical variables can represent state occupation, directly conflicts with established no-go theorems for hidden variable (HV) theories~\cite{Peres}.
In contrast, modern theoretical frameworks~\cite{Haroche,Daley2014} avoid this conflict by reinterpreting jumps as stochastic interruptions of state evolution.
These approaches, however, face inherent limitations in their definitions of quantum trajectories and do not resolve the coexistence of multiple physical mechanisms for the discontinuities.
\par
Specifically, the open systems and quantum jump approaches~\cite{Carmichael,Plenio1998} define quantum trajectories using quantum master equations, numerically demonstrating state collapse due to environmental decoherence~\cite{Haroche}.
These approaches, however, fail to predict the collapse of coherent states and cannot describe the absorption of coherent light.
While discrete photodetector responses are conventionally attributed to photons, both the photon concept~\cite{Antiphoton,Muthukrishnan} and the mechanism of ``true'' light absorption~\cite{Hopfield1958} remain controversial in quantum optics.
The consistent histories approach~\cite{Griffiths2001,Omnes1992} takes a different approach, defining quantum trajectories through time-ordered products of unitary projectors and focusing on a consistent statistical description.
Yet, this approach also faces challenges in restoring classical responses of apparatus within unitary dynamics.
\par
These limitations now warrant reconsideration of the early view: observed jumps may indicate true discontinuities, despite theoretical prohibitions.
Indeed, detector counts $n(t)$ in photon counting experiments behave as binary logical variables when $\langle n(t)\rangle\,{\ll}\,1$.
This suggests that logical variables emerge in experimental contexts and may require theoretical analyses.
In particular, these logical variables should describe the photoelectric effect at each detection site---single-shot events that initiate counting sequences.
Current experimental techniques for manipulating single quantum systems should enable more quantitative verification of such logical variables.
Meaningful tests, however, require a theoretical framework that at least admits their existence.
We therefore reexamine the premises of major no-go theorems and develop a HV formulation that describes quantum jumps using logical variables.
\par
This paper is organized as follows.
Section \ref{sChallenges} identifies three fundamental challenges for introducing HVs to quantum trajectory theories.
Section \ref{sMethods} describes our strategy: restricting to energy eigenstates, introducing a proposition selection rule, and incorporating Boolean logic.
Section \ref{sFormulation} formalizes these concepts within non-relativistic quantum electrodynamics (QED), presenting our eightfold assumptions.
Section \ref{sDiscussion} demonstrates how our framework describes individual trials.
We derive deterministic relations in entangled systems and provide three illustrative examples: the EPR correlations, classical trajectories, and the light quantum hypothesis.
We also present a numerical simulation of the double-slit experiment using coherent light.
Section \ref{sConclusion} presents overall conclusions.
Mathematical derivations are provided in Supplemental Material~\cite{Supplemental}, and applications to contemporary problems are discussed in a separate paper.

\section{Challenges}
\label{sChallenges}
We must address three interrelated challenges simultaneously.
First, any self-consistent theory must circumvent established no-go theorems against HV theories.
These theorems concern: (i) value assignments to operators (Kochen-Specker~\cite{Kochen1967}, CHSH~\cite{CHSH}, GHZ~\cite{GHZ}); (ii) continuity of density operators (Gleason~\cite{Gleason1957,Peres}); and (iii) Bell's model of local HVs~\cite{Bell}.
Value assignment constitutes the primary obstacle, while other paradoxes present unique challenges that require careful treatment.
Second, we must identify a proper classical limit to reconcile the discontinuity of logical variables with the continuity of wavefunctions.
While this issue arises in both classical motion~\cite{EPR,EPR-Bohr} and macroscopic two-level systems~\cite{Leggett}, we focus on classical motion, since we do not assume non-invasive measurements.
Third, we must harmonize Boolean logic with quantum theory.
This challenge, partially addressed in the consistent histories approach, is connected to the first through incompatibility problems specific to the theories of state occupations.

\section{Methods}
\label{sMethods}
We address these challenges through three interdependent modifications to the consistent histories approach: (A) restricting projectors to energy eigenstates, (B) revising compatibility criteria, and (C) establishing Boolean logic as laws for individual trials.
These modifications work together to provide a self-consistent framework.

\subsection{Restricting projectors to energy eigenstates}
\label{sMethods1}
The first modification restricts projectors to those specified by the spectral decomposition of a single observable.
This restriction eliminates the need to assign values to all commuting observables.
The same principle~\cite{Held2022} previously justified Bohm's HV theory~\cite{Bohm1952}, which employs position operators to establish an alternative interpretation.
We instead use the energy operator to describe the observed jumps between energy eigenstates~\cite{QuantumJump2,QuantumJump1}.
Specifically, these states are defined as the orthonormal eigenstates of the non-interacting Hamiltonian in non-relativistic QED~\cite{Messiah}.
Our formulation thus concerns the occupation of Bohr's stationary states, where these stationary states are now generalized to represent both atomic systems and radiation fields.
\par
For non-degenerate systems with unique stationary states, assigning 1 to the occupied state and 0 to others yields stochastic logical variables that describe individual trials.
Quantum theory can calculate the occupation probability of any such state.
For degenerate systems, however, paradoxes persist due to the non-uniqueness of stationary states.
For example, a spin-1 system in zero magnetic field has three degenerate stationary states $|a;{\mathrm{x}}\rangle$, where the parameter ${\mathrm{x}}$ represents a combination of three orthogonal measurement axes (``setting''), and $a\,{\in}\,\{(1,0,0),(0,1,0),(0,0,1)\}$ denotes quantum numbers for the squared spin components along these axes (``outcome'').
The degeneracy permits arbitrary choice of setting ${\mathrm{x}}$ and the corresponding projectors $|a;{\mathrm{x}}\rangle\langle a;{\mathrm{x}}|$.
The Kochen-Specker theorem~\cite{Kochen1967} shows that assigning values to all mathematically allowed projectors results in contradiction when the number of projectors exceeds 117.

\subsection{Revising compatibility criteria}
\label{sMethods2}
\subsubsection{Quantum trajectories as logical propositions}
The second modification revises compatibility criteria to reduce the number of projectors requiring value assignment.
To this end, we represent stationary state occupations as logical propositions, describing their probabilities and truth values using projectors and logical variables, respectively.
Here, projectors yield probabilities through the Born rule, while binary logical variables directly encode truth values.
As described above, this dual representation becomes paradoxical when binary values are assigned to projectors for incompatible settings.
We resolve this paradox by imposing a selection rule on propositions, which disallows incompatible settings.
Such a rule unifies the compatibility criteria for projectors and logical variables, thereby excluding settings that quantum theory never considers simultaneously.
\par
To describe causal relationships between propositions, we represent time sequences of stationary state occupations as compound logical propositions.
While projectors describe state occupations at each time, chain operators describe whole sequences specified by compound propositions.
We show that this chain operator, specifically defined for energy eigenstates, generalizes the chain operator of a history in the consistent histories approach.
As a new compatibility criterion, we require these compound propositions to be constructed from single-time propositions that have well-defined joint probabilities for a given density operator.
To see why such a simple rule resolves the paradoxes, we must examine conventional compatibility criteria in HV and quantum theories.

\subsubsection{Problems in conventional criteria}
For HV theories, mathematical no-go theorems have employed minimal assumptions, typically including the operator commutativity and the sum and product rules~\cite{Kochen1967}.
These assumptions yield a theory in which the sum and product exist whenever observables have definite values, permitting unphysical observable combinations that ignore measurement contextuality~\cite{Peres}.
In Kochen-Specker's proof for spin-1 systems, the 117+ projectors include incompatible settings never realized simultaneously in any single measurement.
For each setting, each measurement requires value assignment to only three commuting projectors.
Thus, the paradox arises from an incomplete commutativity requirement that does not exclude arbitrary parameter choices.
\par
Quantum theory treats operator commutativity differently from HV theories.
The key requirement is that joint probabilities must be consistently defined for \emph{all} density operators~\cite{Breuer}.
This criterion effectively requires operator commutativity without parameter arbitrariness, as typically seen in the von Neumann-L\"{u}ders theory~\cite{Lueders1951} and quantum information theory~\cite{Heinosaari2010}.
These criteria, however, are too restrictive by uniformly disallowing value assignment to non-commuting operators.
The complementarity principle~\cite{EPR-Bohr}, in contrast, dictates that quantum theory should admit classical descriptions when uncertainty is large.
Thus, conventional compatibility criteria are \emph{too permissive} for HV theories yet \emph{too restrictive} for quantum theory.

\subsubsection{Advantage of the revised criterion}
Our intermediate criterion requires well-defined joint probabilities for \emph{a given} density operator.
This criterion admits two types of propositions: (I) those constructed from single-time propositions with commuting projectors, and (II) those that become deterministic for a specific density operator.
The commutativity in this context eliminates parameter arbitrariness.
For single two-level systems, this rule effectively reproduces the conventional criterion in quantum theory, since all deterministic propositions become trivial.
For the spin-1 systems, criterion (I) eliminates incompatible settings, while criterion (II) permits some trivially deterministic propositions.
Hence, our criterion resolves the Kochen-Specker paradox.
For many-level systems, criterion (II) allows propositions with non-commuting projectors and sufficiently large uncertainty---precisely the type of propositions needed to justify classical descriptions.
\par
Still, to justify classical descriptions, we must bridge the gap between energy eigenstates and observable eigenstates.
We consider measurement apparatus for this purpose.
For example, practical particle detectors determine classical trajectories via localized energy eigenstates, as exemplified by cloud chambers that detect specific thermodynamic phases of probe molecules.
The observed trajectories then define empirical maps from energy eigenstates to position/momentum eigenstates.
Here, the former describes a realistic system including the apparatus, while the latter defines an idealized system excluding the apparatus.
If we assume the equivalence of probabilities between the two descriptions, the maps preserve our selection rule.
Then, position/momentum projectors must have well-defined joint probabilities for a given density operator.
This requirement is fulfilled only under criterion (II)---that is, when the projectors cover the entire wavefunction spread ($\Delta{q}\Delta{p}\,{\gg}\,\hbar$).
Consequently, our framework admits classical descriptions within the complementarity principle.
This discussion also shows that our logical variables can be both stochastic and deterministic, depending on uncertainty levels.
\par
We now briefly discuss how this framework circumvents other no-go theorems.
First, our formulation avoids inconsistencies surrounding Gleason's theorem by not applying the Born rule to individual trials.
We thus \emph{define} logical variables as hidden variables inexpressible by density operators.
Second, our logical variables are distinct from Bell's HVs since they do not satisfy Bell's locality assumption.
In our framework, the locality implies the statistical independence of propositions---an impossible assumption even for non-interacting subsystems with local operators.
Third, when logical variables define instantaneous values of observables, the selection rule prohibits arithmetic operations between incompatible logical variables.
This eliminates contradictions raised by Bell-type inequalities (see Section \ref{sDiscussion}).
To define arithmetic operations between logical variables, however, we need to unravel hidden correlations in individual trials.

\subsection{Incorporating Boolean logic}
\label{sMethods3}
The third modification formalizes Boolean logic as fundamental laws for quantum propositions.
As discussed in Section \ref{sIntro}, logical variables are indifferentiable and do not follow the equations of motion.
Applying the Born rule to binary variables also leads to inconsistencies, as indicated by Gleason's theorem.
Thus, our framework requires laws for individual trials beyond the equations of motion and the Born rule.
Although such laws remain unexplored due to the long-standing prohibition of HVs, our dual theory allows us to infer the laws for logical variables from the algebraic rules for chain operators.
Crucially, the consistent histories approach has established that chain operators follow Boolean logic~\cite{Griffiths2001}; however, its utility was limited by the rejection of HVs and the need for additional consistency conditions.
\par
To establish Boolean logic in the framework of HV theories, we postulate the exclusivity of state occupations---a direct consequence of the observed quantum jumps.
This exclusivity, requiring that only one stationary state be occupied at each time, is indispensable to guarantee that the addition of logical variables again yields a logical variable.
Moreover, it differs from the Pauli exclusion principle, as the jumps are observed in both fermionic and bosonic systems~\cite{QuantumJump1,Haroche1}.
Given this exclusivity and the new proposition selection rule, Boolean logic (negation $\lnot$, conjunction $\wedge$, and disjunction $\vee$) holds strictly as transformation rules for chain operators and logical variables.
These rules have the same mathematical forms as in digital computing, with the sole exception that a conjunction for chain operators includes a time-ordering symbol ${\mathcal{T}}$.
This symbol is necessary to restore the standard formulas for quantum coherence functions~\cite{Glauber1963}.
\par
Previously, the consistent histories approach introduced the consistency condition to guarantee the additivity of chain operators for mutually exclusive histories.
In our dual framework, however, this additivity follows naturally from Boolean logic and the exclusivity, together with the additivity of logical variables.
If we start with Boolean logic and the exclusivity, our proposition selection rule turns out to be a consistency requirement between Boolean logic and the Born rule.
In this way, HVs, exclusivity, Boolean logic, and our proposition selection rule form a tightly constrained theoretical framework, each supporting the others.
This self-consistent logical structure circumvents established no-go theorems, enabling a consistent HV formulation that captures deterministic relations in individual trials.

\subsection{Scope and physical background}
Based on the three modifications described above, we construct a theory of logical propositions that calculates probabilities using chain operators, while representing truth values using logical variables.
Since this theory appears more fundamental than initially anticipated, we explicitly define our scope.
This paper aims to provide a direct description of individual trials for quantum jumps observed in quantum optics experiments.
We formulate our theory within non-relativistic QED, particularly based on the traditional approaches, as presented in Mandel~\cite{Mandel} and Milonni~\cite{Milonni2004,Milonni1993}.
We address optical coherence, entanglement, and wave-particle duality, with straightforward extensions to include relativistic effects in a fixed frame.
On the other hand, we do not consider covariant formulation, mass renormalization, and gauge ghosts.
These limitations imply that our theory remains phenomenological, awaiting investigations from more fundamental perspectives.
\par
From a practical viewpoint, our analysis builds on the Heisenberg-picture description of macroscopic quantum optical channels.
In these systems, light interference and detection are more naturally described using field operators than density operators~\cite{Loudon}.
In rigorous modeling of complete transmitter-receiver systems~\cite{Milonni2004}, two-level spins, or projectors to atomic states, play even more fundamental roles.
For two-level atoms, this approach successfully derives Einstein causality~\cite{Milonni2004} and spontaneous emission~\cite{Milonni1993} without external reservoirs~\footnote{In the normal ordering, retarded self-fields account for spontaneous emission, whereas non-relativistic approximation yields inaccurate expressions for Lamb shift~\cite{Milonni1993}.}.
Our HV formulation emerges when we extend this framework to encompass \emph{light absorption} by multi-level atoms.
Conventionally, the quantum trajectory methods have focused on quantum jumps in light emission processes.
In contrast, our theory directly describes light absorption by considering the occupation of atomic scattering states.
In this context, the origin of ``true absorption'' has been debated in the theory of dielectric functions~\cite{Hopfield1958}.
For photoionization and photochemical reactions, however, the spatial separation of subsystem wavefunctions~\cite{Zeweil} accounts for the irreversibility of the absorption process.
\par
It should be noted that this theory depends on conventional frameworks in describing statistical mixtures.
When describing individual trials in a noisy environment, conditional probabilities must be conditioned on both inter-state transitions and external decoherence.
The former, representing the quantum jumps in our theory, describes jumps associated with unitary evolution by known Hamiltonians.
The latter, representing the quantum jumps in modern frameworks, is necessary to describe environmental effects and information leakage.
Although we do not provide such a unified theory in this paper, our formulation, explicitly modeling apparatus, should complement the general information theoretic approach~\cite{Haroche,IntroQIS}.
In this respect, we only object to the unrestricted prohibition of HVs, which stems from the unphysical premises in conventional HV theories.

\section{Formulation}
\label{sFormulation}
Our theory consists of the following eight assumptions: (1) Born rule, (2) Unitarity, (3) Non-relativistic QED, (4) Duality, (5) Boolean logic, (6) Stationary states, (7) Proposition, and (8) Compatibility.
While conventional formulations use only (1)--(3) and prohibit (4), our formulation justifies (4) and (5) through constraints imposed by (6)--(8).
We present these assumptions in an order that most naturally introduces our basic concepts.
\par
To implement these assumptions specifically, we introduce a compact notation for representing time sequences of state occupations.
We use the symbols `$\occupation{\stateset{\mathrm{C}}{}}{},t$' to represent the proposition `One of the stationary states in a set $\stateset{\mathrm{C}}{}$ is occupied at time $t$', while `$\occupation{\statesingle{\phi}{}}{},t$' represents a singleton case for a single state $|\phi\rangle$.
The bold-face symbol `${\bf C},{\bm t}$' denotes a jump sequence specified by multiple sets ${\bf C}\,{\equiv}\,\{{\mathrm{C}}_{1},\ldots,{\mathrm{C}}_{N}\}$ and multiple times ${\bm t}\,{\equiv}\,\{t_{1},\ldots,t_{N}\}$.
More specific definitions are presented below.
\par
We now start with two assumptions specifying propositions whose probabilities are determined by quantum theory.
\paragraph*{Duality:}
A binary logical variable represents the truth value of the proposition `${\bf C},{\bm t}$' by
\begin{align}
\xjvar{`${\bf C},{\bm t}$'}{} = \begin{cases}
  1, & \mbox{proposition `${\bf C},{\bm t}$' is True}\\
  0, & \mbox{proposition `${\bf C},{\bm t}$' is False},
\end{cases}
\label{4-11}
\end{align}
where $j\,{=}\,1,2,\ldots$ indexes individual trials.
The expectation value is the statistical average:
\begin{align}
\langle \xvar{`${\bf C},{\bm t}$'}{} \rangle 
\equiv \lim_{{\mathfrak{N}}\to\infty} \frac{1}{{\mathfrak{N}}} \sum_{j=1}^{\mathfrak{N}} \xjvar{`${\bf C},{\bm t}$'}{}
\in [0,1],
\label{4-12}
\end{align}
where the sum is taken over $\mathfrak{N}$ typical samples drawn from a given ensemble.
\paragraph*{Born rule:}
The expectation value can be predicted using the Born rule:
\begin{align}
\langle \xvar{`${\bf C},{\bm t}$'}{} \rangle 
& = {\mathrm{tr}}\big[ \,\xop{`${\bf C},{\bm t}$'}{}\,\hat{\rho}\,\big]
\label{4-21}\\
\xop{`${\bf C},{\bm t}$'}{}
& \equiv \zopd{`${\bf C},{\bm t}$'}{} \zop{`${\bf C},{\bm t}$'}{},
\label{4-22}
\end{align}
where the Heisenberg-picture density operator $\hat{\rho}$ is a positive Hermitian operator that has a unit trace (${\mathrm{tr}}[\hat{\rho}]\,{=}\,1$, $\hat{\rho}\,{>}\,0$, $\hat{\rho}^{\dagger}\,{=}\,\hat{\rho}$).
The observable $\xop{`${\bf C},{\bm t}$'}{}$ is also positive Hermitian, whereas the chain operator $\zop{`${\bf C},{\bm t}$'}{}$ is not necessarily Hermitian.
For brevity, we suppress the $\hat{\rho}$-dependence of the expectation value.
\par
The Duality assumption ensures that the expectation value represents the probability that the proposition `${\bf C},{\bm t}$' is true in the frequentist sense:
\begin{align}
{\mathrm{Pr}}\left(\mbox{`${\bf C},{\bm t}$'}\right)
\equiv \lim_{{\mathfrak{N}}\to\infty} \frac{\sum_{j}^{\mbox{\scriptsize `${\bf C},{\bm t}$' is True}} 1}{\sum_{j} 1}
= \big\langle \xvar{`${\bf C},{\bm t}$'}{} \big\rangle.
\label{4-24}
\end{align}
The Born rule connects this probability to the corresponding chain operator and density operator through Eqs.\eqref{4-21}--\eqref{4-22}.
Note that we do not apply the Born rule to logical variables themselves; our density operators always describe statistical ensembles, where the ensemble may be characterized by a specific truth value of `${\bf C},{\bm t}$'.
Accordingly, our logical variables are inherently stochastic, while becoming deterministic if and only if the proposition `${\bf C},{\bm t}$' becomes invariably true or false for the given density operator.
\par
Next, we specify ensemble evolution through two assumptions shared with conventional formulations.
\paragraph*{Unitarity:}
For each elementary single-time proposition `$\occupation{\statesingle{\phi}{}}{},t$', the chain operator is a projector
\begin{align}
\zop{`$\occupation{\statesingle{\phi}{}}{}$'}{} 
\equiv |\phi\rangle\langle\phi|,
\label{4-32}
\end{align}
which evolves as a Heisenberg operator according to
\begin{align}
& \zop{`$\occupation{\statesingle{\phi}{}}{},t$'}{} 
= \hat{U}^{\dagger}(t)\,\zop{`$\occupation{\statesingle{\phi}{}}{}$'}{}\,\hat{U}(t),
\label{4-33}
\end{align}
where $\hat{U}(t)\,{\equiv}\,\exp(-{\mathrm{i}}\hat{H}t/\hbar)$ denotes the unitary time evolution operator.
\paragraph*{Non-relativistic QED:}
The total Hamiltonian $\hat{H}$ is the minimal-coupling Hamiltonian~\cite{Mandel,Messiah,Cohen}, defined by
\begin{align}
\hat{H} = 
& \sum_{k}^{K}\frac{1}{2m_{k}}\left\|{\hat{\bm p}}_{k} - e_{k}\hat{\bm A}_{\mathrm{T}}(\hat{\bm r}_{k})\right\|^{2}
+ \sum_{k}^{K}\frac{g_{k}e_{k}}{2m_{k}}{\hat{\bm S}}_{k}\cdot\hat{\bm B}(\hat{\bm r}_{k})
\nonumber\\
& + \frac{1}{2}\sum_{k}^{K}\sum_{k'}^{k'{\neq}k}\frac{e_{k}e_{k'}}{4\pi\varepsilon_{0}}\frac{1}{\left\|\hat{\bm r}_{k}-\hat{\bm r}_{k'}\right\|}
\nonumber\\
& + \iiint\frac{\varepsilon_{0}}{2}\left[\left\|\hat{\bm E}_{\mathrm{T}}({\bm x})\right\|^{2}+{c^{2}}\left\|\hat{\bm B}({\bm x})\right\|^{2}\right] d{\bm x},
\label{4-41}
\end{align}
where ${\hat{\bm r}}_{k}$ and ${\hat{\bm p}}_{k}$ denote the position and momentum operators for the $k$-th particle ($1\,{\le}\,k\,{\le}\,K$), respectively.
$\hat{\bm A}_{\mathrm{T}}({\bm x})$ and $\hat{\bm E}_{\mathrm{T}}({\bm x})$ represent the vector potential operator and electric field operator satisfying $0\,{=}\,{\bm \nabla}\cdot\hat{\bm A}_{\mathrm{T}}({\bm x})\,{=}\,{\bm \nabla}\cdot\hat{\bm E}_{\mathrm{T}}({\bm x})$.
Other symbols follow standard SI notation.
The commutation relations are given by 
\begin{align}
[\hat{r}_{k\xi},\,\hat{p}_{k'\xi'}]_{-} 
& = {\mathrm{i}}\hbar\,\delta_{k k'}\delta_{\xi \xi'}
\label{4-42}\\
[{\hat{E}_{{\mathrm{T}}\xi}}({\bm x}),\,{\hat{A}_{{\mathrm{T}}\xi'}}({\bm x}')]_{-} 
& = \frac{{\mathrm{i}}\hbar}{\varepsilon_{0}}\,\delta^{{\mathrm{T}}}_{\xi \xi'}({\bm x}-{\bm x}'),
\label{4-43}
\end{align}
where $\xi,\xi'\,{\in}\,\{x,y,z\}$ are Cartesian indices, and $\delta^{{\mathrm{T}}}_{\xi \xi'}({\bm x}-{\bm x}')$ is the transverse delta function~\footnote{One can evaluate the formal expression $\hat{\bm A}_{\mathrm{T}}(\hat{\bm r}_{k})$ and $\hat{\bm B}(\hat{\bm r}_{k})$ in the position representation through the replacement $\hat{\bm r}_{k}\,{\to}\,{\bm r}_{k}$ and $\hat{\bm p}_{k}\,{\to}\,-{\mathrm{i}}\hbar{\bm \nabla}_{k}$~\cite{Mandel}. To ensure the unitarity of $\hat{U}(t)$ and the Hermiticity of $\hat{H}$, we impose periodic boundary conditions on the surface of a very large volume $V$.}.
\par
These assumptions describe the statistical behavior of the system, ensuring the conservation of probability and total energy.
To enable the direct description of individual trials, our formulation additionally defines stationary states as bare energy eigenstates.
\paragraph*{Stationary States:}
Stationary states are orthonormal eigenstates of the non-interacting Hamiltonian:
\begin{align}
\hat{H}_{0} \equiv 
& \sum_{k}^{K}\frac{1}{2m_{k}}\|{\hat{\bm p}}_{k}\|^{2}
+ \sum_{k}^{K}\frac{g_{k}e_{k}}{2m_{k}}{\hat{\bm S}}_{k}\cdot\hat{\bm B}(\hat{\bm r}_{k})
\nonumber\\
& + \frac{1}{2}\sum_{k}^{K}\sum_{k'}^{k'{\neq}k}\frac{e_{k}e_{k'}}{4\pi\varepsilon_{0}}\frac{1}{\left\|\hat{\bm r}_{k}-\hat{\bm r}_{k'}\right\|}
\nonumber\\
& + \iiint\frac{\varepsilon_{0}}{2}\left[\|\hat{\bm E}_{\mathrm{T}}({\bm x})\|^{2}+{c^{2}}\|\hat{\bm B}({\bm x})\|^{2}\right]d{\bm x}.
\label{4-51}
\end{align}
For degenerate systems, the choice of the eigenstates may vary with time but must remain unique within each compound proposition.
If the compound proposition involves multiple propositions `$\occupation{\stateset{\mathrm{C}_{n,i}}{}}{},t_{n}$' ($i=1,\ldots,I$) describing time $t_{n}$, the corresponding set must have the form:
\begin{align}
\stateset{\mathrm{C}_{n,i}}{}
\equiv \{\,|\phi\rangle\,|\,|\phi\rangle\,{\in}\,\stateset{\mathfrak{E}(t_{n})}{}
,~\mbox{$|\phi\rangle$ satisfies $\mathrm{C}_{n,i}$}\,\},
\label{4-62}
\end{align}
where
\begin{align}
\stateset{\mathfrak{E}(t_{n})}{}
\equiv \{\,|\phi\rangle\,|\,\hat{H}_{0}|\phi\rangle = E(\phi)|\phi\rangle, \langle \phi | \phi' \rangle = \delta_{\phi,\phi'}\,\}
\label{4-61}
\end{align}
is the complete set of stationary states uniquely assigned to $t_{n}$ for each compound proposition `${\bf C},{\bm t}$'.
Set operations (orthogonal complement space $\perp$, intersection $\cap$, and union $\cup$) on stationary states are defined by
\begin{align}
& \stateset{\mathrm{C}_{n}}{}^{\perp}
\equiv \{\,|\phi\rangle\,|\,|\phi\rangle\,{\in}\,\stateset{\mathfrak{E}(t_{n})}{}~\mathrm{and}~|\phi\rangle\,{\notin}\,\stateset{\mathrm{C}_{n}}{} \}
\label{4-63}\\
& \stateset{\mathrm{C}_{n}}{}\,{\cap}\,\stateset{\mathrm{C}'_{n}}{}
\equiv \{\,|\phi\rangle\,|\,|\phi\rangle\,{\in}\,\stateset{\mathrm{C}_{n}}{} ~\mathrm{and}~ |\phi\rangle\,{\in}\,\stateset{\mathrm{C}'_{n}}{}\,\}
\label{4-64}\\
& \stateset{\mathrm{C}_{n}}{}\,{\cup}\,\stateset{\mathrm{C}'_{n}}{}
\equiv \{\,|\phi\rangle\,|\,|\phi\rangle\,{\in}\,\stateset{\mathrm{C}_{n}}{} ~\mathrm{or}~ |\phi\rangle\,{\in}\,\stateset{\mathrm{C}'_{n}}{}\,\}.
\label{4-65}
\end{align}
\par
These assumptions define stationary states based on experiments that observed quantum jumps in both atomic systems and radiation fields~\cite{QuantumJump1,Haroche1}.
We include the Zeeman interaction to describe atomic systems with spin-orbit splittings, though this choice is debatable and requires future refinement~\footnote{For example, quantum jumps have been observed between $\{6^{2}{\mathrm{P}}_{3/2}, 6^{2}{\mathrm{S}}_{1/2}, 6^{2}{\mathrm{P}}_{1/2}, 5^{2}{\mathrm{D}}_{3/2}\}$ states and a $5^{2}{\mathrm{D}}_{5/2}$ state in $\mathrm{Ba}^{+}$ ions, where split terms $5^{2}{\mathrm{D}}_{5/2}$ and $5^{2}{\mathrm{D}}_{3/2}$ are involved without direct transitions. Theoretically, the Zeeman interaction may be unnecessary if we define $\hat{H}_{0}$ from $\hat{H}$ by a replacement ${\hat{\bm p}}_{k}\,{-}\,e_{k}\hat{\bm A}_{\mathrm{T}}(\hat{\bm r}_{k})\,{\to}\,{\hat{\bm p}}_{k}$. If we maintain the Zeeman interaction, a static part of the transverse electric field may also be necessary to include all time-independent parts of $\tilde{H}_{I}(t)$ into $\hat{H}_{0}$.}.
For consistent set operations, we define sets of stationary states so that propositions with the same time argument always have commuting projectors.
In contrast, we allow propositions defined at different times to have non-commuting projectors---a possibility that conventional formulations have excluded.
\par
We now postulate Boolean logic and provide a formal definition of proposition `${\bf C},{\bm t}$' based on this foundation.
\begin{widetext}
\paragraph*{Boolean Logic:}
Boolean operations (negation $\lnot$, conjunction $\wedge$, disjunction $\vee$) transform logical variables as
\begin{align}
\xjvar{$\lnot$`$\occupation{\stateset{{\mathrm{C}}}{}}{},t$'}{}
& = 1 - \xjvar{`$\occupation{\stateset{{\mathrm{C}}}{}}{},t$'}{}
\label{4-71}\\
\xjvar{`$\occupation{\stateset{{\mathrm{C}}}{}}{},t$'$\wedge$`$\occupation{\stateset{{\mathrm{C}}'}{}}{},t'$'}{}
& = \xjvar{`$\occupation{\stateset{{\mathrm{C}}}{}}{},t$'}{}\,\xjvar{`$\occupation{\stateset{{\mathrm{C}}'}{}}{},t'$'}{}
\label{4-72}\\
\xjvar{`$\occupation{\stateset{{\mathrm{C}}}{}}{},t$'$\vee$`$\occupation{\stateset{{\mathrm{C}}'}{}}{},t'$'}{}
& = \xjvar{`$\occupation{\stateset{{\mathrm{C}}}{}}{},t$'}{}
+ \xjvar{`$\occupation{\stateset{{\mathrm{C}}'}{}}{},t'$'}{}
- \xjvar{`$\occupation{\stateset{{\mathrm{C}}}{}}{},t$'$\wedge$`$\occupation{\stateset{{\mathrm{C}}'}{}}{},t'$'}{}.
\label{4-73}
\end{align}
Transformation rules for projectors have the form:
\begin{align}
\zop{$\lnot$`$\occupation{\stateset{{\mathrm{C}}}{}}{},t$'}{}
& = \hat{1} - \zop{`$\occupation{\stateset{{\mathrm{C}}}{}}{},t$'}{}
\label{4-74}\\
\zop{`$\occupation{\stateset{{\mathrm{C}}}{}}{},t$'$\wedge$`$\occupation{\stateset{{\mathrm{C}}'}{}}{},t'$'}{}
& = \mathcal{T}\zop{`$\occupation{\stateset{{\mathrm{C}}}{}}{},t$'}{}\,\zop{`$\occupation{\stateset{{\mathrm{C}}'}{}}{},t'$'}{}
\label{4-75}\\
\zop{`$\occupation{\stateset{{\mathrm{C}}}{}}{},t$'$\vee$`$\occupation{\stateset{{\mathrm{C}}'}{}}{},t'$'}{} 
& = \zop{`$\occupation{\stateset{{\mathrm{C}}}{}}{},t$'}{}
+ \zop{`$\occupation{\stateset{{\mathrm{C}}'}{}}{},t'$'}{}
- \zop{`$\occupation{\stateset{{\mathrm{C}}}{}}{},t$'$\wedge$`$\occupation{\stateset{{\mathrm{C}}'}{}}{},t'$'}{}.
\label{4-76}
\end{align}
\end{widetext}
Here, the time-ordering symbol ${\mathcal{T}}$ in Eq.\eqref{4-75} arranges projectors from right to left in an ascending order of time.
For $t\,{=}\,t'$ where Boolean logic operations induce set operations, the following relations hold in addition:
\begin{align}
\mbox{$\lnot$`$\occupation{\stateset{{\mathrm{C}}}{}}{},t$'}
& = \mbox{`$\occupation{\stateset{\mathrm{C}}{}^{\perp}}{},t$'}
\label{4-77}\\
\mbox{`$\occupation{\stateset{{\mathrm{C}}}{}}{},t$'$\wedge$`$\occupation{\stateset{{\mathrm{C}}'}{}}{},t$'}
& = \mbox{`$\occupation{\stateset{{\mathrm{C}}}{}\,{\cap}\,\stateset{{\mathrm{C}}'}{}}{},t$'}
\label{4-78}\\
\mbox{`$\occupation{\stateset{{\mathrm{C}}}{}}{},t$'$\vee$`$\occupation{\stateset{{\mathrm{C}}'}{}}{},t$'}
& = \mbox{`$\occupation{\stateset{{\mathrm{C}}}{}\,{\cup}\,\stateset{{\mathrm{C}}'}{}}{},t$'}.
\label{4-79}
\end{align}
\paragraph*{Proposition:}
The compound proposition `${\bf C},{\bm t}$' can always be decomposed as
\begin{align}
\mbox{`${\bf C},{\bm t}$'}
& = \bigwedge_{n=1}^{N} \mbox{`$\occupation{\stateset{\mathrm{C}_{n}}{}}{},t_{n}$'}
\label{4-81}\\
\mbox{`$\occupation{\stateset{\mathrm{C}_{n}}{}}{},t_{n}$'} 
& = \bigvee_{|\phi\rangle}^{\stateset{\mathrm{C}_{n}}{}} \mbox{`$\occupation{\statesingle{\phi}{}}{},t_{n}$'},
\label{4-82}
\end{align}
where $\bigwedge$ and $\bigvee$ represent conjunction and disjunction over all relevant propositions, respectively.
\par
Boolean logic is a strong assumption that holds only for state occupations.
Eqs.\eqref{4-71}--\eqref{4-73} and Eqs.\eqref{4-74}--\eqref{4-76} describe individual trials and statistical ensembles in parallel, while Eqs.\eqref{4-77} and Eqs.\eqref{4-78}--\eqref{4-79} represent the completeness of the Hilbert space and the exclusivity of state occupations, respectively.
If we focus on a singleton case with $\stateset{{\mathrm{C}}}{}\,{=}\,\{|\phi\rangle\}$ and $\stateset{{\mathrm{C}'}}{}\,{=}\,\{|\phi'\rangle\}$, from Eq.\eqref{4-61}, Eq.\eqref{4-64}, Eq.\eqref{4-72}, and Eq.\eqref{4-78}, we obtain an explicit expression for the exclusivity of state occupations:
\begin{align}
\xjvar{`$\occupation{\statesingle{\phi}{}}{},t$'$\wedge$`$\occupation{\statesingle{\phi'}{}}{},t$'}{} = \delta_{\phi,\phi'}\xjvar{`$\occupation{\statesingle{\phi}{}}{},t$'}{}.
\label{4-83}
\end{align}
\par
The Proposition assumption then specifies how to construct compound propositions using Boolean logic.
For logical variables, these assumptions yield
\begin{align}
\xjvar{`${\bf C},{\bm t}$'}{}
& = \prod_{n=1}^{N} \xjvar{`$\occupation{\stateset{{\mathrm{C}}_{n}}{}}{},t_{n}$'}{}
\label{4-84}\\
\xjvar{`$\occupation{\stateset{{\mathrm{C}}_{n}}{}}{},t_{n}$'}{}
& = \sum_{|\phi\rangle}^{\stateset{{\mathrm{C}}_{n}}{}} \xjvar{`$\occupation{\statesingle{\phi}{}}{},t_{n}$'}{}.
\label{4-85}
\end{align}
Note that all logical variables on the right-hand side of Eq.\eqref{4-85} must be zero except for a specific $\phi$, showing how the exclusivity establishes correlations between propositions.
Accordingly, the truth values of propositions within each compound proposition must be treated as a single package for each trial.
\par
Boolean logic and the Proposition assumption also yield the following expressions for the chain operators:
\begin{align}
\zop{`${\bf C},{\bm t}$'}{}
& = \mathcal{T} \prod_{n=1}^{N} \zop{`$\occupation{\stateset{{\mathrm{C}}_{n}}{}}{},t_{n}$'}{}
\label{4-86}\\
\zop{`$\occupation{\stateset{{\mathrm{C}}_{n}}{}}{},t_{n}$'}{}
& = \sum_{|\phi\rangle}^{\stateset{{\mathrm{C}}_{n}}{}} \zop{`$\occupation{\statesingle{\phi}{}}{},t_{n}$'}{}.
\label{4-87}
\end{align}
Combining Eq.\eqref{4-86} with other assumptions, we find that the probability of `${\bf C},{\bm t}$' is the joint probability of constituent propositions and has the following form:
\begin{align}
& {\mathrm{Pr}}\left(\mbox{`${\bf C},{\bm t}$'}\right)
\nonumber\\
& = {\mathrm{Pr}}\left(\mbox{$\bigwedge_{n=1}^{N}$`$\occupation{\stateset{\mathrm{C}_{n}}{}}{},t_{n}$'}\right)
\label{4-90}\\
& = {\mathrm{tr}}\left[ \left({\mathcal{T}}\prod_{n=1}^{N} \zop{`$\stateset{{\mathrm{C}}_{n}}{},t_{n}$'}{}\right) \hat{\rho}
\left({\mathcal{T}}\prod_{n=1}^{N} \zop{`$\stateset{{\mathrm{C}}_{n}}{},t_{n}$'}{}\right)^{\dagger} \right].
\label{4-91}
\end{align}
To ensure the consistency of probability calculations, this joint probability must have a single well-defined value for any combination of $t_{1},\ldots,t_{N}$.
However, the right-hand side of Eq.\eqref{4-91} may converge to different values at $t_{n}\,{\to}\,t$, depending on the time order of $\zop{`$\stateset{{\mathrm{C}}_{n}}{},t_{n}$'}{}$.
We avoid this inconsistency through the compatibility condition.
\paragraph*{Compatibility:}
For any compound proposition `${\bf C},{\bm t}$' consisting of single-time propositions `$\stateset{{\mathrm{C}}_{n}}{},t_{n}$' ($n = 1,\ldots,N$), the joint probability of any combination ${\mathcal{C}}$ of the single-time propositions
\begin{align}
& {\mathrm{Pr}}\left(\mbox{$\bigwedge_{n}^{\mathcal{C}}$`$\occupation{\stateset{\mathrm{C}_{n}}{}}{},t_{n}$'}\right)
\label{4-92}\\
& = {\mathrm{tr}}\left[ \left({\mathcal{T}}\prod_{n}^{\mathcal{C}} \zop{`$\stateset{{\mathrm{C}}_{n}}{},t_{n}$'}{}\right) \hat{\rho}
\left({\mathcal{T}}\prod_{n}^{\mathcal{C}} \zop{`$\stateset{{\mathrm{C}}_{n}}{},t_{n}$'}{}\right)^{\dagger} \right],
\nonumber
\end{align}
must converge to a single well-defined value at $t_{n}\,{\to}\,t$ ($n\,{\in}\,{\mathcal{C}}$) for each ${\mathcal{C}}$.
Here, ${\mathcal{C}}$ represents any combination of indices taken from $\{1,\ldots,N\}$.
\par
If we define a projector product $\hat{K}_{\mathcal{C}}$ by
\begin{align}
\hat{K}_{\mathcal{C}}
\equiv \prod_{n}^{{\mathcal{C}}} \zop{`$\occupation{\stateset{\mathrm{C}_{n}}{}}{}$'}{}
\label{4-93}
\end{align}
and denote any rearrangement of $\hat{K}_{\mathcal{C}}$ by ${\mathcal{P}}\hat{K}_{\mathcal{C}}$, the Compatibility assumption leads to
\begin{align}
{\mathrm{tr}}\left[\,\hat{K}^{\dagger}_{\mathcal{C}}\,\hat{K}_{\mathcal{C}}\,\hat{\rho}(t)\,\right]
= {\mathrm{tr}}\left[ \left({\mathcal{P}}\hat{K}_{\mathcal{C}} \right)^{\dagger} \left({\mathcal{P}}\hat{K}_{\mathcal{C}}\right) \hat{\rho}(t)\,\right],
\label{4-94}
\end{align}
where $\hat{\rho}(t)\,{\equiv}\,\hat{U}^{\dagger}(t)\,\hat{\rho}\,\hat{U}(t)$ is the density operator in the Schr\"{o}dinger picture, and the order of projectors in Eq.\eqref{4-93} is tentatively chosen as the ascending order of $n\,{\in}\,{\mathcal{C}}$ from right to left.
The implication of this new compatibility condition is discussed below.

\section{Discussion}
\label{sDiscussion}
\subsection{Roles of the Compatibility assumption}
\label{sDiscussionA}
To see the roles played by the Compatibility assumption, first consider the simplest case with $N\,{=}\,2$ and ${\mathcal{C}}\,{=}\,\{1,2\}$.
In this case, the right-hand side of Eq.\eqref{4-92} becomes 
\begin{align}
& {\mathrm{tr}}\left[\,\zop{`$\stateset{{\mathrm{C}}_{1}}{},t_{1}$'}{}\zop{`$\stateset{{\mathrm{C}}_{2}}{},t_{2}$'}{} \hat{\rho} \zopd{`$\stateset{{\mathrm{C}}_{2}}{},t_{2}$'}{}\zopd{`$\stateset{{\mathrm{C}}_{1}}{},t_{1}$'}{}\,\right]
\nonumber
\end{align}
for $t_{1}\,{>}\,t_{2}$ and 
\begin{align}
& {\mathrm{tr}}\left[\,\zop{`$\stateset{{\mathrm{C}}_{2}}{},t_{2}$'}{}\zop{`$\stateset{{\mathrm{C}}_{1}}{},t_{1}$'}{} \hat{\rho} \zopd{`$\stateset{{\mathrm{C}}_{1}}{},t_{1}$'}{}\zopd{`$\stateset{{\mathrm{C}}_{2}}{},t_{2}$'}{}\,\right]
\nonumber
\end{align}
for $t_{1}\,{<}\,t_{2}$.
Requiring that these expressions coincide at $|t_{1}\,{-}\,t_{2}|\,{\to}\,0$, we obtain the following form of Eq.\eqref{4-94}:
\begin{align}
& {\mathrm{tr}}\Big[\,\Big(\big(\zop{`$\occupation{\stateset{\mathrm{C}_{1}}{}}{}$'}{} \zop{`$\occupation{\stateset{\mathrm{C}_{2}}{}}{}$'}{}\big)^{\dagger} \zop{`$\occupation{\stateset{\mathrm{C}_{1}}{}}{}$'}{} \zop{`$\occupation{\stateset{\mathrm{C}_{2}}{}}{}$'}{} - 
\nonumber\\
& \big(\zop{`$\occupation{\stateset{\mathrm{C}_{2}}{}}{}$'}{} \zop{`$\occupation{\stateset{\mathrm{C}_{1}}{}}{}$'}{}\big)^{\dagger} \zop{`$\occupation{\stateset{\mathrm{C}_{2}}{}}{}$'}{} \zop{`$\occupation{\stateset{\mathrm{C}_{1}}{}}{}$'}{}\Big)\,\hat{\rho}(t)\,\Big]
= 0.
\label{5-1}
\end{align}
From Eq.\eqref{5-1}, we immediately see that our compatibility condition holds in two specific cases: (I) when the projectors commute ($\zop{`$\occupation{\stateset{\mathrm{C}_{1}}{}}{}$'}{} \zop{`$\occupation{\stateset{\mathrm{C}_{2}}{}}{}$'}{} = \zop{`$\occupation{\stateset{\mathrm{C}_{2}}{}}{}$'}{} \zop{`$\occupation{\stateset{\mathrm{C}_{1}}{}}{}$'}{}$), and (II) when all propositions are deterministic ($\forall n$, ${\mathrm{tr}}[\,\zop{`$\occupation{\stateset{\mathrm{C}_{n}}{}}{}$'}{} \hat{\rho}(t) \zop{`$\occupation{\stateset{\mathrm{C}_{n}}{}}{}$'}{}\,] \in \{0,1\}$).
The former is the conventional commutativity requirement, while the latter provides a state-dependent condition where the projectors effectively act as $\hat{0}$ or $\hat{1}$ on the given density operator.
Note that the deterministic case always applies when the given density operator has support confined to a subspace $\stateset{\mathrm{C}_{n}}{}$.
\par
If we substitute Eq.\eqref{4-87} at $t\,{=}\,0$ into Eq.\eqref{5-1}, Eq.\eqref{5-1} reduces to a detailed balance condition for ${\mathcal{C}}\,{=}\,\{1,2\}$:
\begin{align}
\sum^{\mbox{\scriptsize `$\occupation{\stateset{\mathrm{C}_{1}}{}}{}$'}}_{|\phi_{1}\rangle} \sum^{\scriptsize \mbox{`$\occupation{\stateset{\mathrm{C}_{2}}{}}{}$'}}_{|\phi_{2}\rangle}
\left|\langle\phi_{1}|\phi_{2}\rangle\right|^{2} 
\Big( \langle\phi_{1}| \hat{\rho}(t) |\phi_{1}\rangle - \langle\phi_{2}| \hat{\rho}(t) |\phi_{2}\rangle \Big)
= 0.
\label{5-2}
\end{align}
When we consider all possible index combinations, e.g., ${\mathcal{C}} = \{1,1,\ldots\}$, $\{2,2,\ldots\}$, $\{1,2,1,2,\ldots\}$, $\ldots$, we obtain infinitely many equations for $\langle\phi_{n}| \hat{\rho}(t) |\phi_{n}\rangle$ ($n\,{=}\,1,2$) containing the same transition probabilities and population factors.
For finite-dimensional systems, these heavily overdetermined equations will not have a solution unless each factor independently vanishes.
Thus, the two cases discussed above should be the only cases where the Compatibility condition is satisfied for any index combination.
We therefore assume, without rigorous mathematical justification, that the only meaningful cases for the Compatibility assumption are the commutative case (I) and the deterministic case (II).
Importantly, this conclusion generalizes to arbitrary $N$ due to the identical mathematical structure of Eq.\eqref{4-94}.

\subsection{Correspondence with conventional approaches}
\label{sDiscussionB}
\subsubsection{Consistent histories approach}
We now relate our formulation to established quantum trajectory theories and demonstrate how it overcomes known problems.
Using the assumptions described in Section \ref{sFormulation}, we can express the probability of proposition `${\bf C},{\bm t}$' as
\begin{align}
& {\mathrm{Pr}}\left(\mbox{`${\bf C},{\bm t}$'}\right)
= {\mathrm{tr}}\left[\,\zop{`${\bf C},{\bm t}$'}{}\,\hat{\rho}\,\zopd{`${\bf C},{\bm t}$'}{}\,\right],
\label{5-9}
\end{align}
where the chain operator is given by
\begin{align}
\zop{`${\bf C},{\bm t}$'}{}
& = \hat{U}(t_{N}) \zop{`$\occupation{\stateset{{\mathrm{C}}_{N}}{}}{}$'}{} \hat{U}(t_{N}-t_{N-1})
\nonumber\\
& ~~~~ \cdots \hat{U}(t_{2}-t_{1})\zop{`$\occupation{\stateset{{\mathrm{C}}_{1}}{}}{}$'}{} \hat{U}(t_{1}).
\label{5-10}
\end{align}
We assume $t_{1}\,{<}\,t_{2}\,{<}\,\cdots\,\,{<}\,t_{N}$ without loss of generality.
Eq.\eqref{5-10} shows that the compound proposition `${\bf C},{\bm t}$' defines a quantum trajectory consisting of stationary state projectors and unitary evolution operators.
From Eq.\eqref{4-33}, Eq.\eqref{4-86}, and Eq.\eqref{4-87}, we also obtain
\begin{align}
\zop{`${\bf C},{\bm t}$'}{}
& = \sum_{\mathrm{histories}}^{\scriptsize \mbox{`${\bf C},{\bm t}$'}} \mathcal{T} \prod_{n=1}^{N} \zop{`$\occupation{\statesingle{\phi_{n}}{}}{},t_{n}$'}{}
\label{5-11}\\
& = \sum_{\mathrm{histories}}^{\scriptsize \mbox{`${\bf C},{\bm t}$'}} 
\hat{U}(t_{N}) \zop{`$\occupation{\statesingle{\phi_{N}}{}}{}$'}{} \hat{U}(t_{N}-t_{N-1})
\nonumber\\
& ~~~~~~~~~ \cdots \hat{U}(t_{2}-t_{1})\zop{`$\occupation{\statesingle{\phi_{1}}{}}{}$'}{} \hat{U}(t_{1}),
\label{5-12}
\end{align}
where each term in the sum represents the chain operator of a history~\cite{Griffiths2001}.
The sum runs over all possible sequences of stationary states in `${\bf C},{\bm t}$'.
Substituting Eq.\eqref{5-11} into Eq.\eqref{5-9} and taking advantage of the exclusivity of state occupations, we obtain the additivity of probabilities for mutually exclusive histories---an essential requirement that previously led to the consistency condition.

\subsubsection{Description of individual trials}
In parallel with Eq.\eqref{5-11}, our framework establishes the equality for logical variables:
\begin{align}
\xjvar{`${\bf C},{\bm t}$'}{}
= \sum_{\mathrm{histories}}^{\scriptsize \mbox{`${\bf C},{\bm t}$'}} \prod_{n=1}^{N} \xjvar{`$\occupation{\statesingle{\phi_{n}}{}}{},t_{n}$'}{}
,\quad \forall j.
\label{5-13}
\end{align}
This equality is justified by the exclusivity, which correlates logical variables within the scope defined by $j$ and `${\bf C},{\bm t}$'.
The correlated logical variables must satisfy the Compatibility assumption, while uncorrelated logical variables from different trials must be distinguished by different indices (e.g., $j$ and $j'$) to avoid introducing erroneous correlations.
\par
This correlation naturally leads us to distinguish arithmetic operations for correlated and uncorrelated logical variables.
For correlated logical variables sharing the same trial index $j$, Boolean logic gives
\begin{align}
& \xjvar{`$\stateset{{\mathrm{C}}}{},t$'}{} + \xjvar{`$\stateset{{\mathrm{C}}'}{},t'$'}{} 
\label{5-14}\\
& ~~ = \xjvar{`$\stateset{{\mathrm{C}}}{},t$'$\vee$`$\stateset{{\mathrm{C}}'}{},t'$'}{} 
+ \xjvar{`$\stateset{{\mathrm{C}}}{},t$'$\wedge$`$\stateset{{\mathrm{C}}'}{},t'$'}{}
\nonumber\\
& \xjvar{`$\stateset{{\mathrm{C}}}{},t$'}{} \, \xjvar{`$\stateset{{\mathrm{C}}'}{},t'$'}{}
= \xjvar{`$\stateset{{\mathrm{C}}}{},t$'$\wedge$`$\stateset{{\mathrm{C}}'}{},t'$'}{},
\label{5-15}
\end{align}
where the Compatibility assumption must hold for the propositions `$\stateset{{\mathrm{C}}}{},t$' and `$\stateset{{\mathrm{C}}'}{},t'$'.
Note that Eq.\eqref{5-14} reduces to the exclusivity when $\stateset{{\mathrm{C}}}{}\,{=}\,\{|\phi\rangle\}$, $\stateset{{\mathrm{C}'}}{}\,{=}\,\{|\phi'\rangle\}$, and $t\,{=}\,t'$.
In contrast, Eqs.\eqref{5-14}--\eqref{5-15} do not apply to uncorrelated logical variables with different trial indices.
In this way, trial indices play a crucial role in our framework, determining both correlation scopes and the applicability of Boolean logic and the Born rule.

\subsubsection{Open systems and quantum jump approaches}
The open systems and quantum jump approaches define quantum trajectories based on the non-unitary evolution of environment-coupled systems.
These approaches numerically demonstrate state collapse due to environmental decoherence, although such collapses are believed not to represent real systems owing to the arbitrariness of definitions~\cite{Haroche}.
To illustrate how these trajectories connect to ours, consider an optical cavity surrounded by ideal photodetectors~\cite{Carmichael}.
We take the target system as a single-mode radiation field described by an annihilation operator $\hat{a}$.
In this system, quantum trajectories are constructed from two superoperators ${\mathcal{L}}\,{-}\,{\mathcal{S}}$ and ${\mathcal{S}}$, where $({\mathcal{L}}\,{-}\,{\mathcal{S}})\breve{\rho}_{c}\,{\equiv}\,(\hat{H}_{\mathrm{eff}}\breve{\rho}_{c}\,{-}\,\breve{\rho}_{c}\hat{H}_{\mathrm{eff}}) / ({\mathrm{i}}\hbar)$ represents continuous evolution governed by a non-Hermitian Hamiltonian $\hat{H}_{\mathrm{eff}}$, while ${\mathcal{S}}\breve{\rho}_{c}\,{\equiv}\,\hat{a}\breve{\rho}_{c}\hat{a}^{\dagger}$ represents discrete state collapse caused by the jump operator $\hat{a}$.
$\breve{\rho}_{c}$ represents the conditional density operator of a specific trajectory.
\par
An essential feature of these quantum trajectories is the absence of state collapse for coherent states---the eigenstates of the jump operator $\hat{a}$~\cite{Haroche}.
Specifically, the superoperator ${\mathcal{S}}$ shown above induces jumps from the Fock state $\breve{\rho}_{c}\,{=}\,\statevectorR{1}{\mathrm{R}}\statevectorL{1}{R}$ to the vacuum state ${\mathcal{S}}\breve{\rho}_{c}\,{=}\,\statevectorR{0}{\mathrm{R}}\statevectorL{0}{\mathrm{R}}$.
In contrast, coherent states never collapse because $\breve{\rho}_{c}\,{=}\,\statevectorR{\tilde{\alpha}}{\mathrm{R}}\statevectorL{\tilde{\alpha}}{R}$ yields ${\mathcal{S}}\breve{\rho}_{c}\,{\propto}\,\statevectorR{\tilde{\alpha}}{\mathrm{R}}\statevectorL{\tilde{\alpha}}{\mathrm{R}}$.
This lack of quantum jumps can be justified for light emission processes.
This is because the system is monitored only indirectly through the environment (or photodetectors), and the non-unitary evolution ${\mathcal{L}}\,{-}\,{\mathcal{S}}$ accounts for energy loss through a continuous decrease of $\tilde{\alpha}$.
For light absorption processes, however, photon counting experiments have observed discrete jumps for coherent light~\cite{Gisin2002}.
Thus, describing light absorption necessitates a superoperator ${\mathcal{S}}'$ that predicts jumps for both Fock states and coherent states. 

\subsubsection{Measurement operators for light absorption processes}
Our framework, by contrast, considers unitary quantum trajectories for the entire system including both the source and photodetector (i.e., apparatus).
In ordinary optical measurements, light emitters and photodetectors can be viewed as non-interacting subsystems, allowing us to focus on quantum jumps in photodetectors.
Note that quantum jumps in radiation fields have been observed only through quantum non-demolition measurements (QND) using atomic interferometers~\cite{Haroche1}.
Then, quantum jumps in the photodetector subsystem can be modeled as transitions from the bound to scattering states of the detector material, expressed as
\begin{align}
\mbox{`ionized,\,$t$'}
= \mbox{`$\occupation{\stateset{E>0}{\mathrm{A}}}{},t$'$\wedge$`$\occupation{\stateset{E<0}{\mathrm{A}}}{},0$'}
,\quad t \ge 0,
\label{5-23}
\end{align}
where $\stateset{E<0}{\mathrm{A}}$ and $\stateset{E>0}{\mathrm{A}}$ respectively denote all bound and scattering states of the atomic system Hamiltonian $\hat{H}_{\mathrm{A}}$ of the photodetector subsystem.
\par
We can readily calculate the probability of `ionized,\,$t$' using the Born rule (Eq.\eqref{5-9}) and standard approximations~\footnote{The approximations include the first-order perturbation approximation, long-wavelength approximation, electric-dipole approximation, quasi-monochromatic approximation, long-time approximation, and rotating-wave approximation. We also take advantage of the source-field equation~\cite{Loudon,Carmichael}---the formal solutions of the Heisenberg-Maxwell equation for point-like sources.}, where our chain operator
\begin{align}
\zop{`ionized,$t$'}{} 
= \zop{`$\occupation{\stateset{E>0}{\mathrm{A}}}{},t$'}{} \zop{`$\occupation{\stateset{E<0}{\mathrm{A}}}{},0$'}{}
\label{5-24}
\end{align}
effectively serves as a measurement operator through Eq.\eqref{5-9}.
This calculation yields Mandel's formula~\cite{Mandel} that predicts the correct photoionization probabilities for both Fock states and coherent states.
For the cavity problem, $\zop{`$\occupation{\stateset{E>0}{\mathrm{A}}}{},t$'}{}$ in Eq.\eqref{5-24} is proportional to the Heisenberg-picture annihilation operator $\hat{a}(t')$, while $\zop{`$\occupation{\stateset{E<0}{\mathrm{A}}}{},0$'}{}$ reduces to unity when the detector is initially in bound states.
Thus, the compound proposition in Eq.\eqref{5-23} describes quantum jumps in photodetectors, while keeping a reasonable connection with the conventional jump operator for light emission processes.

\subsection{Deterministic relations in individual trials}
\label{sDiscussionC}
\subsubsection{General properties}
Our HV formulation directly represents the deterministic relations in individual trials.
For the simplest case with two propositions `$\occupation{\stateset{\mathrm{C}_{1}}{}}{},t_{1}$' and `$\occupation{\stateset{\mathrm{C}_{2}}{}}{},t_{2}$', deterministic relations in individual trials can be expressed as
\begin{align}
\xjvar{`$\occupation{\stateset{\mathrm{C}_{1}}{}}{},t_{1}$'}{} = \xjvar{`$\occupation{\stateset{\mathrm{C}_{2}}{}}{},t_{2}$'}{}
,\quad \forall{j}.
\label{5-31}
\end{align}
This equation states that the truth values of `$\occupation{\stateset{\mathrm{C}_{1}}{}}{},t_{1}$' and `$\occupation{\stateset{\mathrm{C}_{2}}{}}{},t_{2}$' are invariably equal in any trial.
A purely statistical theory represents the same relation through conditional probabilities in the form:
\begin{align}
1 & = {\mathrm{Pr}}(\mbox{`$\occupation{\stateset{\mathrm{C}_{1}}{}}{},t_{1}$'}|\mbox{`$\occupation{\stateset{\mathrm{C}_{2}}{}}{},t_{2}$'}) 
\nonumber\\
& = {\mathrm{Pr}}(\mbox{`$\occupation{\stateset{\mathrm{C}_{2}}{}}{},t_{2}$'}|\mbox{`$\occupation{\stateset{\mathrm{C}_{1}}{}}{},t_{1}$'}),
\label{5-32}
\end{align}
where the conditional probability of `$\occupation{\stateset{\mathrm{C}_{2}}{}}{},t_{2}$' given `$\occupation{\stateset{\mathrm{C}_{1}}{}}{},t_{1}$' is defined by
\begin{align}
{\mathrm{Pr}}(\mbox{`$\occupation{\stateset{\mathrm{C}_{2}}{}}{},t_{2}$'}|\mbox{`$\occupation{\stateset{\mathrm{C}_{1}}{}}{},t_{1}$'})
& \equiv \frac{\langle\xvar{`$\occupation{\stateset{\mathrm{C}_{1}}{}}{},t_{1}$'$\wedge$`$\occupation{\stateset{\mathrm{C}_{2}}{}}{},t_{2}$'}{}\rangle}{\langle\xvar{`$\occupation{\stateset{\mathrm{C}_{1}}{}}{},t_{1}$'}{}\rangle}.
\label{5-33}
\end{align}
From Eq.\eqref{5-32}--\eqref{5-33}, we obtain
\begin{align}
\langle \xvar{`$\occupation{\stateset{\mathrm{C}_{1}}{}}{},t_{1}$'$\wedge$`$\occupation{\stateset{\mathrm{C}_{2}}{}}{},t_{2}$'}{} \rangle
& = \langle \xvar{`$\occupation{\stateset{\mathrm{C}_{1}}{}}{},t_{1}$'}{} \rangle
\nonumber\\
& = \langle \xvar{`$\occupation{\stateset{\mathrm{C}_{2}}{}}{},t_{2}$'}{} \rangle,
\label{5-34}
\end{align}
which shows that representing deterministic relations requires well-defined joint probabilities.
Consequently, conventional formulations do not address deterministic relations between propositions with non-commuting projectors.
Furthermore, deterministic relations involving multiple arithmetic operations, such as Eq.\eqref{4-85} and Eq.\eqref{5-13}, are difficult to express in a probabilistic form.
In these cases, a HV formulation is necessary to represent experimental observations.
\par
If we define the conditional density operator given condition `$\occupation{\stateset{\mathrm{C}_{1}}{}}{},t_{1}$' by
\begin{align}
\hat{\rho}(\mbox{`$\occupation{\stateset{\mathrm{C}_{1}}{}}{},t_{1}$'})
\equiv \frac{\zop{`$\occupation{\stateset{\mathrm{C}_{1}}{}}{},t_{1}$'}{}\,\hat{\rho}\,\zopd{`$\occupation{\stateset{\mathrm{C}_{1}}{}}{},t_{1}$'}{}}{{\mathrm{tr}}\left[\,\zop{`$\occupation{\stateset{\mathrm{C}_{1}}{}}{},t_{1}$'}{}\,\hat{\rho}\,\zopd{`$\occupation{\stateset{\mathrm{C}_{1}}{}}{},t_{1}$'}{}\,\right]},
\label{5-35}
\end{align}
the conditional probability in Eq.\eqref{5-33} can be cast into a form:
\begin{align}
& {\mathrm{Pr}}(\mbox{`$\occupation{\stateset{\mathrm{C}_{2}}{}}{},t_{2}$'}|\mbox{`$\occupation{\stateset{\mathrm{C}_{1}}{}}{},t_{1}$'})
\nonumber\\
& = {\mathrm{tr}}\left[\,\zop{`$\occupation{\stateset{\mathrm{C}_{2}}{}}{},t_{2}$'}{}\,\hat{\rho}(\mbox{`$\occupation{\stateset{\mathrm{C}_{1}}{}}{},t_{1}$'})\,\zopd{`$\occupation{\stateset{\mathrm{C}_{2}}{}}{},t_{2}$'}{}\,\right].
\label{5-36}
\end{align}
This result, known as the L\"{u}ders projection postulate~\cite{Lueders1951}, provides the standard prescription for updating density operators in projective measurements.
Repeatedly applying Eq.\eqref{5-35} for multiple propositions and using the Unitarity assumption, we again arrive at our definition of unitary quantum trajectories (Eq.\eqref{4-86})~\footnote{Since our quantum trajectories are generally not Markovian, calculating the probability of the next outcome may require all previous outcomes. Nevertheless, simple operator algebra reveals that two-level atoms observed through photodetectors is first-order Markovian.}.

\subsubsection{Entangled non-interacting subsystems}
The ability to represent deterministic relations provides a natural description of entanglement.
To see this, consider a system consisting of two non-interacting subsystems held by Alice and Bob.
Suppose that these subsystems are characterized by stationary states $\statevectorR{a;{\mathrm{x}}}{\mathrm{A}}$ and $\statevectorR{b;{\mathrm{y}}}{\mathrm{B}}$, where ${\mathrm{x}}$ and ${\mathrm{y}}$ represent the settings for Alice and Bob, and $a$ and $b$ represent the respective outcomes.
Suppose that these subsystems are entangled such that 
\begin{align}
\hat{\rho}(t)
& = |\phi(t)\rangle\langle\phi(t)|
\label{5-41}\\
|\phi(t)\rangle
& = \sum_{(a,b)} c_{(a,b)}(t) \statevectorR{a;{\mathrm{u}}}{\mathrm{A}} \statevectorR{b;{\mathrm{u}}}{\mathrm{B}},
\label{5-42}
\end{align}
where the tuple $(a,b)$ represents a set of outcomes linked by entanglement, $c_{(a,b)}(t)$ denotes its probability amplitude, and ${\mathrm{u}}$ is a given setting.
By definition, the stationary states $\statevectorR{\forall a;{\mathrm{u}}}{\mathrm{A}}$ and $\statevectorR{\forall b;{\mathrm{u}}}{\mathrm{B}}$ are mutually orthogonal for each fixed $\mathrm{u}$.
\par
We can then calculate the joint and marginal probabilities of subsystem propositions using our general assumptions (Eqs.\eqref{4-21}--\eqref{4-22}, Eq.\eqref{4-75}, Eq.\eqref{4-81}).
For $\mathrm{x}\,{=}\,\mathrm{u}$, we have
\begin{align}
& \big\langle \xvar{`$\occupation{\statesingle{a;{\mathrm{x}}}{\mathrm{A}}}{},t$'$\wedge$`$\occupation{\statesingle{b;{\mathrm{y}}}{\mathrm{B}}}{},t$'}{leftright} \big\rangle
= |c_{(a,b)}(t)|^{2} \left| \statevectorLR{b;{\mathrm{x}}}{b;{\mathrm{y}}}{\mathrm{B}} \right|^{2}
\label{5-43}\\
& \big\langle \xvar{`$\occupation{\statesingle{a;{\mathrm{x}}}{\mathrm{A}}}{},t$'}{leftright} \big\rangle
= \big\langle \xvar{`$\occupation{\statesingle{b;{\mathrm{y}}}{\mathrm{B}}}{},t$'}{leftright} \big\rangle
= |c_{(a,b)}(t)|^{2},
\label{5-44}
\end{align}
while for $\mathrm{y}\,{=}\,\mathrm{u}$, we obtain similar expressions where $\left| \statevectorLR{a;{\mathrm{x}}}{a;{\mathrm{y}}}{\mathrm{A}} \right|^{2}$ replaces the last factor in Eq.\eqref{5-43}.
As a consequence, for any entangled pair $(a,b)$ satisfying
\begin{align}
\left| \statevectorLR{a;{\mathrm{x}}}{a;{\mathrm{y}}}{\mathrm{A}} \right|^{2} = 1
~~\mbox{or}~~
\left| \statevectorLR{b;{\mathrm{x}}}{b;{\mathrm{y}}}{\mathrm{B}} \right|^{2} = 1,
\label{5-47}
\end{align}
we obtain the deterministic relation
\begin{align}
& \big\langle \xvar{`$\occupation{\statesingle{a;{\mathrm{x}}}{\mathrm{A}}}{},t$'$\wedge$`$\occupation{\statesingle{b;{\mathrm{y}}}{\mathrm{B}}}{},t$'}{leftright} \big\rangle
\nonumber\\
& ~~ = \big\langle \xvar{`$\occupation{\statesingle{a;{\mathrm{x}}}{\mathrm{A}}}{},t$'}{leftright} \big\rangle
= \big\langle \xvar{`$\occupation{\statesingle{b;{\mathrm{y}}}{\mathrm{B}}}{},t$'}{leftright} \big\rangle
\label{5-45}
\end{align}
together with its equivalent expression
\begin{align}
\xjvar{`$\occupation{\statesingle{a;{\mathrm{x}}}{\mathrm{A}}}{},t$'}{leftright} 
= \xjvar{`$\occupation{\statesingle{b;{\mathrm{y}}}{\mathrm{B}}}{},t$'}{leftright}
,~~\forall j.
\label{5-46}
\end{align}
Below, we provide three specific examples of deterministic relations arising from entanglement.

\subsection{EPR correlations and Bell's theorems}
\label{sDiscussionD}
For a system consisting of two spin-1/2 particles, settings are unit vectors ${\bm x}$ and ${\bm y}$ specifying measurement axes, and outcomes $a$ and $b$ are spin directions taking one of $\{+,-\}$.
When the system is in a singlet state with a source axis ${\bm u}$,
\begin{align}
|\phi\rangle
= \frac{1}{\sqrt{2}} \left( \statevectorR{+;{\bm u}}{\mathrm{A}} \statevectorR{-;{\bm u}}{\mathrm{B}}
- \statevectorR{-;{\bm u}}{\mathrm{A}} \statevectorR{+;{\bm u}}{\mathrm{B}} \right),
\label{5-51}
\end{align}
the joint and marginal probabilities for the outcomes $(a,b)\,{=}\,(+,-),(-,+)$ agree with the standard expressions in quantum theory~\cite{Peres}:
\begin{align}
& \big\langle \xvar{`$\occupation{\statesingle{a;{\bm x}}{\mathrm{A}}}{},t$'$\wedge$`$\occupation{\statesingle{b;{\bm y}}{\mathrm{B}}}{},t$'}{leftright} \big\rangle
= \frac{1-{\bm x}\cdot{\bm y}}{4}
\label{5-52}\\
& \big\langle \xvar{`$\occupation{\statesingle{a;{\bm x}}{\mathrm{A}}}{},t$'}{leftright} \big\rangle
= \big\langle \xvar{`$\occupation{\statesingle{b;{\bm y}}{\mathrm{B}}}{},t$'}{leftright} \big\rangle
= \frac{1}{2}.
\label{5-53}
\end{align} 
Consequently, for the combination of axes ${\bm x}\,{=}\,-{\bm y}$ and outcomes $(a,b)\,{=}\,(+,-),(-,+)$, we obtain the deterministic relation:
\begin{align}
\xjvar{`$\occupation{\statesingle{a;{\bm x}}{\mathrm{A}}}{}$'}{leftright}
= \xjvar{`$\occupation{\statesingle{b;{\bm y}}{\mathrm{B}}}{}$'}{leftright} 
,~~ \forall j.
\label{5-54}
\end{align}
\par
In this system, Bell's theorems provide significant boundary cases that warrant detailed analysis.
We first show that our logical variables are nonlocal HVs that do not satisfy Bell's assumption.
In his 1964 paper, Bell~\cite{Bell} considered a HV satisfying the following conditions:
\begin{align}
p_{\bm{xy}}(a,b) 
& = \int p_{\bm{xy}}(a,b|\lambda) \, p_{\bm{xy}}(\lambda) d\lambda
~~\mbox{(HV)}
\label{5-61}\\
p_{\bm{xy}}(a,b|\lambda) 
& = p_{\bm{x}}(a|\lambda) \, p_{\bm{y}}(b|\lambda)
~~\mbox{(locality)},
\label{5-62}
\end{align}
where $p_{\bm{xy}}(a,b)$ denotes the joint probability that Alice and Bob obtain outcomes $a$ and $b$ using settings $\bm{x}$ and $\bm{y}$; $p_{\bm{xy}}(a,b|\lambda)$ is the conditional probability given the value of $\lambda$; $p_{\bm{xy}}(\lambda)\,{=}\,p(\lambda)$ is the probability distribution of $\lambda$.
In our formulation, the Duality assumption yields an equation analogous to Eq.\eqref{5-61}:
\begin{align}
p_{\bm{xy}}(a,b)
= \lim_{\mathfrak{N}\to\infty} \frac{1}{\mathfrak{N}} \sum_{j=1}^{\mathfrak{N}} \xjvar{`$\occupation{\statesingle{a;\bm{x}}{\mathrm{A}}}{}$'${\wedge}$`$\occupation{\statesingle{b;\bm{y}}{\mathrm{B}}}{}$'}{},\!\!
\label{5-63}
\end{align}
while Boolean logic and the assumption of non-interacting subsystems guarantee the factorizability of logical variables and the corresponding observables:
\begin{align}
& \xjvar{`$\occupation{\statesingle{a;\bm{x}}{\mathrm{A}}}{}$'${\wedge}$`$\occupation{\statesingle{b;\bm{y}}{\mathrm{B}}}{}$'}{}
\nonumber\\
& ~~ = \xjvar{`$\occupation{\statesingle{a;\bm{x}}{\mathrm{A}}}{}$'}{} \xjvar{`$\occupation{\statesingle{b;\bm{y}}{\mathrm{B}}}{}$'}{}
,~~ \forall j
\label{5-64}\\
& \xop{`$\occupation{\statesingle{a;\bm{x}}{\mathrm{A}}}{}$'${\wedge}$`$\occupation{\statesingle{b;\bm{y}}{\mathrm{B}}}{}$'}{}
\nonumber\\
& ~~ = \xop{`$\occupation{\statesingle{a;\bm{x}}{\mathrm{A}}}{}$'}{} \xop{`$\occupation{\statesingle{b;\bm{y}}{\mathrm{B}}}{}$'}{}.
\label{5-68}
\end{align}
These equations, however, do not imply the factorizability of probabilities---the expectation values of logical variables that satisfy the Born rule:
\begin{align}
p_{\bm{xy}}(a,b)
& = {\mathrm{tr}}\big[ \,\xop{`$\occupation{\statesingle{a;\bm{x}}{\mathrm{A}}}{}$'${\wedge}$`$\occupation{\statesingle{b;\bm{y}}{\mathrm{B}}}{}$'}{}\,\hat{\rho}\,\big]
\label{5-65}\\
p_{\bm{x}}(a)
& = {\mathrm{tr}}\big[ \,\xop{`$\occupation{\statesingle{a;\bm{x}}{\mathrm{A}}}{}$}{}\,\hat{\rho}\,\big]
\label{5-66}\\
p_{\bm{y}}(b)
& = {\mathrm{tr}}\big[ \,\xop{`$\occupation{\statesingle{b;\bm{y}}{\mathrm{B}}}{}$}{}\,\hat{\rho}\,\big]
\label{5-67}
\end{align}
(cf. Eq.\eqref{5-52}).
This shows that our logical variables narrowly escape the locality assumption in Eq.\eqref{5-62}.
\par
To see how our formulation violates the classical bound in the CHSH inequality, consider two spin measurements with directions ${\bm x}_{1}$ and ${\bm x}_{2}$ for Alice and ${\bm y}_{1}$ and ${\bm y}_{2}$ for Bob.
Here, the spin operators can be expressed as 
\begin{align}
\hat{A}_{{\bm x}_{i}} & = 2 \xop{`$\occupation{\statesingle{+;{\bm x}_{i}}{\mathrm{A}}}{}$'}{} - 1
,~~i=1,2
\label{5-71}\\
\hat{B}_{{\bm y}_{i}} & = 2 \xop{`$\occupation{\statesingle{+;{\bm y}_{i}}{\mathrm{B}}}{}$'}{} - 1
,~~i=1,2,
\label{5-72}
\end{align}
where the projectors to the spin-up states have the form:
\begin{align}
\xop{`$\occupation{\statesingle{+;{\bm x}_{i}}{\mathrm{A}}}{}$'}{}
& = \statevectorR{+;{\bm x}_{i}}{\mathrm{A}}\statevectorL{+;{\bm x}_{i}}{\mathrm{A}}
,~~i=1,2
\label{5-73}\\
\xop{`$\occupation{\statesingle{+;{\bm y}_{i}}{\mathrm{B}}}{}$'}{}
& = \statevectorR{+;{\bm y}_{i}}{\mathrm{B}}\statevectorL{+;{\bm y}_{i}}{\mathrm{B}}
,~~i=1,2.
\label{5-74}
\end{align}
We define the corresponding spin values using logical variables as follows:
\begin{align}
{A}^{(j)}_{{\bm x}_{i}} & = 2 \xjvar{`$\occupation{\statesingle{+;{\bm x}_{i}}{\mathrm{A}}}{}$'}{} - 1
,~~i=1,2
,~~\forall j
\label{5-75}\\
{B}^{(j)}_{{\bm y}_{i}} & = 2 \xjvar{`$\occupation{\statesingle{+;{\bm y}_{i}}{\mathrm{B}}}{}$'}{} - 1
,~~i=1,2
,~~\forall j.
\label{5-76}
\end{align}
Conventionally, mathematical HV theories have assumed that the sum and product of these spin variables always exist if original observables have definite values.
This assumption yields the CHSH value:
\begin{align}
S^{(j)}
& \equiv {A}^{(j)}_{{\bm x}_{1}} {B}^{(j)}_{{\bm y}_{1}} + {B}^{(j)}_{{\bm y}_{1}} {A}^{(j)}_{{\bm x}_{2}} + {A}^{(j)}_{{\bm x}_{2}} {B}^{(j)}_{{\bm y}_{2}} - {B}^{(j)}_{{\bm y}_{2}} {A}^{(j)}_{{\bm x}_{1}}
\nonumber\\
& = {A}^{(j)}_{{\bm x}_{1}} \left({B}^{(j)}_{{\bm y}_{1}}- {B}^{(j)}_{{\bm y}_{2}} \right) 
+ {A}^{(j)}_{{\bm x}_{2}} \left({B}^{(j)}_{{\bm y}_{1}} + {B}^{(j)}_{{\bm y}_{2}} \right) 
\nonumber\\
& = \pm 2
,\quad \forall j,
\label{5-77}
\end{align}
which leads to the classical bound $|\langle S \rangle|\,{\le}\,2$ that conflicts with experiments.
\par
In contrast, our logical variables are correlated through the exclusivity.
Correlated logical variables must satisfy the Compatibility assumption, and arithmetic operations between uncorrelated logical variables must be distinguished by different trial labels (see Section \ref{sDiscussionB}).
Specifically, when the four logical variables in Eqs.\eqref{5-75}--\eqref{5-76} are non-deterministic, the Compatibility assumption is satisfied if and only if the corresponding projectors commute.
It follows that logical variables must have different trial labels when $\statevectorLR{+;{\bm x}_{1}}{+;{\bm x}_{2}}{\mathrm{A}}\,{\neq}\,0$ or $\statevectorLR{+;{\bm y}_{1}}{+;{\bm y}_{2}}{\mathrm{B}}\,{\neq}\,0$.
This implies that each product ${A}^{(j)}_{{\bm x}_{i}} {B}^{(j)}_{{\bm y}_{i'}}$ is correlated, while their sum must be uncorrelated.
Thus, instead of Eq.\eqref{5-77}, we obtain
\begin{align}
{A}^{(j)}_{{\bm x}_{1}} {B}^{(j)}_{{\bm y}_{1}} + {B}^{(j')}_{{\bm y}_{1}} {A}^{(j')}_{{\bm x}_{2}} + {A}^{(j'')}_{{\bm x}_{2}} {B}^{(j'')}_{{\bm y}_{2}} - {B}^{(j''')}_{{\bm y}_{2}} {A}^{(j''')}_{{\bm x}_{1}},
\label{5-78}
\end{align}
where $j$, $j'$, $j''$, $j'''$ represent different trials.
Evaluating the expectation value of each term using the Duality assumption and the Born rule, our framework reproduces the standard expression in quantum theory:
\begin{align}
|\langle S \rangle|
& = \left| \langle {A}_{{\bm x}_{1}} {B}_{{\bm y}_{1}} \rangle + \langle {B}_{{\bm y}_{1}} {A}_{{\bm x}_{2}} \rangle + \langle {A}_{{\bm x}_{2}} {B}_{{\bm y}_{2}} \rangle - \langle {B}_{{\bm y}_{2}} {A}_{{\bm x}_{1}} \rangle 
\right|
\nonumber\\
& = \left| {\bm x}_{1}\cdot{\bm y}_{1} +  {\bm y}_{1}\cdot{\bm x}_{2} + {\bm x}_{2}\cdot{\bm y}_{2} - {\bm y}_{2}\cdot{\bm x}_{1}\right|.
\label{5-79}
\end{align}
This expression has the correct upper bound $2\sqrt{2}$ (Tsirelson's bound)~\cite{Nielsen2000}.

\subsection{Original EPR problem}
\label{sDiscussionE}
The original EPR problem addresses simultaneous values of positions and momenta in individual trials~\cite{EPR}.
When two charged particles are in an eigenstate of the relative position operator, the two-particle wavefunction is entangled both in positions and momenta:
\begin{align}
|\phi\rangle
& = \iiint c_{({\bm r}_{\mathrm{A}},{\bm r}_{\mathrm{B}})}  
\statevectorR{{\bm r}_{\mathrm{A}}}{\mathrm{A}}
\statevectorR{{\bm r}_{\mathrm{B}}}{\mathrm{B}} d{\bm r}'_{\mathrm{B}}
\label{5-81}\\
& = \iiint c_{({\bm p}_{\mathrm{A}},{\bm p}_{\mathrm{B}})}  
\statevectorR{{\bm p}_{\mathrm{A}}}{\mathrm{A}}
\statevectorR{{\bm p}_{\mathrm{B}}}{\mathrm{B}} d{\bm p}'_{\mathrm{A}},
\label{5-82}
\end{align}
where we assume that the center-of-mass is at rest, and $\statevectorR{{\bm r}_{\mathrm{X}}}{\mathrm{X}}$ and $\statevectorR{{\bm p}_{\mathrm{X}}}{\mathrm{X}}$ (${\mathrm{X}}\,{=}\,{\mathrm{A}},{\mathrm{B}}$) denote the position and momentum eigenstates of each particle, respectively.
The coefficients are given by
\begin{align}
c_{({\bm r}_{\mathrm{A}},{\bm r}_{\mathrm{B}})}
& = \frac{1}{\sqrt{V}} \delta({\bm r}_{\mathrm{A}}-{\bm r}_{\mathrm{B}}'-{\bm r}_{0}) \delta({\bm r}_{\mathrm{B}}-{\bm r}'_{\mathrm{B}})
\label{5-83}\\
c_{({\bm p}_{\mathrm{A}},{\bm p}_{\mathrm{B}})}
& = \frac{1}{\sqrt{V}} e^{-\frac{\mathrm{i}}{\hbar} {\bm p}_{\mathrm{A}}\cdot{\bm r}_{0}} \delta({\bm p}_{\mathrm{A}}-{\bm p}'_{\mathrm{A}}) \delta({\bm p}_{\mathrm{B}}+{\bm p}'_{\mathrm{A}}),
\label{5-84}
\end{align}
where ${\bm r}_{0}$ denotes a relative position.
This entanglement leads to deterministic relations in individual trials:
\begin{align}
{\bm r}_{\mathrm{B}}^{(j)}(t) = {\bm r}_{\mathrm{A}}^{(j)}(t) + {\bm r}_{0}
,~~ 
{\bm p}_{\mathrm{B}}^{(j)}(t) = -{\bm p}_{\mathrm{A}}^{(j)}(t),
\label{5-85}
\end{align}
where ${\bm r}^{(j)}_{\mathrm{X}}(t)$ and ${\bm p}^{(j)}_{\mathrm{X}}(t)$ denote the instantaneous values of positions and momenta for the $j$-th trial.
These values, however, must be defined with special care.
\par
Specifically, we define these values using proxy representations.
That is, we map the occupation of stationary states onto the occupation of observable eigenstates.
The former represents a realistic system with apparatus, while the latter represents an idealized system without apparatus.
If we require the same truth values and probabilities for these proxy representations, they preserve our Compatibility assumption.
Furthermore, these representations must be coarse-grained because stationary states are not localized in position and momentum space.
We therefore consider propositions `$\occupation{\stateset{{\bm r}_{\mathrm{X}}; \Delta{\bm r}}{\mathrm{X}}}{}$' and `$\occupation{\stateset{{\bm p}_{\mathrm{X}}; \Delta{\bm p}}{\mathrm{X}}}{}$' that represent the occupation of the position/momentum eigenstates, where ${\bm r}_{\mathrm{X}}$ and ${\bm p}_{\mathrm{X}}$ denote center coordinates, and $\Delta{\bm r}$ and $\Delta{\bm p}$ denote cell sizes.
\par
Then, the instantaneous values of positions and momenta can be expressed as
\begin{align}
& {\bm r}^{(j)}_{\mathrm{X}}(t) 
\equiv \sum_{{\bm r}_{\mathrm{X}}} {\bm r}_{\mathrm{X}} \, \xjvar{`$\occupation{\stateset{{\bm r}_{\mathrm{X}};\Delta{\bm r}}{\mathrm{X}}}{}$'}{leftright}
,~~ \forall j
\label{5-88}\\
& {\bm p}^{(j)}_{\mathrm{X}}(t) 
\equiv \sum_{{\bm p}_{\mathrm{X}}} {\bm p}_{\mathrm{X}} \, \xjvar{`$\occupation{\stateset{{\bm p}_{\mathrm{X}};\Delta{\bm p}}{\mathrm{X}}}{}$'}{leftright}
,~~ \forall j,
\label{5-89}
\end{align}
where the sums run over all cells, and the projectors are given by
\begin{align}
\zop{`$\occupation{\stateset{{\bm r}_{\mathrm{X}}\pm\Delta{\bm r}}{\mathrm{X}}}{}$'}{}
& = \iiint_{{\bm r}_{\mathrm{X}}\pm\Delta{\bm r}} \!\!\!\! \statevectorR{{\bm r}_{\mathrm{X}}}{\mathrm{X}} \statevectorL{{\bm r}_{\mathrm{X}}}{\mathrm{X}}
\,d{\bm r}_{\mathrm{X}}
\label{5-86}\\
\zop{`$\occupation{\stateset{{\bm p}_{\mathrm{X}}\pm\Delta{\bm p}}{\mathrm{X}}}{}$'}{}
& = \iiint_{{\bm p}_{\mathrm{X}}\pm\Delta{\bm p}} \!\!\!\! \statevectorR{{\bm p}_{\mathrm{X}}}{\mathrm{X}} \statevectorL{{\bm p}_{\mathrm{X}}}{\mathrm{X}}
\,d{\bm p}_{\mathrm{X}}.
\label{5-87}
\end{align}
The key insight for classical description is that the logical variables must satisfy the Compatibility assumption, while the projectors do not commute.
Thus, the propositions `$\occupation{\stateset{{\bm r}_{\mathrm{X}}; \Delta{\bm r}}{\mathrm{X}}}{}$' and `$\occupation{\stateset{{\bm p}_{\mathrm{X}}; \Delta{\bm p}}{\mathrm{X}}}{}$' must be deterministic:
\begin{align}
1 & = \left\langle \xvar{`$\occupation{\stateset{{\bm r}_{\mathrm{X}}\pm\Delta{\bm r}}{\mathrm{X}}}{},t$'}{} \right\rangle_{\mathrm{cl}}
\label{5-91}\\
1 & = \left\langle \xvar{`$\occupation{\stateset{{\bm p}_{\mathrm{X}}\pm\Delta{\bm p}}{\mathrm{X}}}{},t$'}{} \right\rangle_{\mathrm{cl}}.
\label{5-92}
\end{align}
\par
Another important aspect of classical descriptions is that the density operator must be conditioned to trace the time variation of a specific trial.
We therefore define a classical description for the $j$-th trial by the requirement
\begin{align}
& \left\langle \xvar{`$\occupation{\stateset{{\bm r}_{\mathrm{X}}\pm\Delta{\bm r}}{\mathrm{X}}}{},t$'}{} \right\rangle_{\mathrm{cl}}
\nonumber\\
& = {\mathrm{tr}}\left[\,\zop{`$\occupation{\stateset{{\bm r}_{\mathrm{X}}\pm\Delta{\bm r}}{\mathrm{X}}}{}$'}{}
\,\hat{\rho}_{\mathrm{cl}}(t)\,\zopd{`$\occupation{\stateset{{\bm r}_{\mathrm{X}}\pm\Delta{\bm r}}{\mathrm{X}}}{}$'}{}\,\right]
\label{5-93}\\
& = \xjvar{`$\occupation{\stateset{{\bm r}_{\mathrm{X}}\pm\Delta{\bm r}}{\mathrm{X}}}{},t$'}{},
\label{5-94}
\end{align}
and the corresponding expression for `$\occupation{\stateset{{\bm p}_{\mathrm{X}}\pm\Delta{\bm p}}{\mathrm{X}}}{},t$'.
Eqs.\eqref{5-91}--\eqref{5-92} then require that the cell sizes $\Delta{\bm r}$ and $\Delta{\bm p}$ cover the entire spread of the conditional density operator, which leads to the complementarity principle: $\Delta{r}_{x}\Delta{p}_{x}\,{\gg}\,\hbar$, $\Delta{r}_{y}\Delta{p}_{y}\,{\gg}\,\hbar$, and $\Delta{r}_{z}\Delta{p}_{z}\,{\gg}\,\hbar$.
The trial-specific density operator satisfies
\begin{align}
\hat{\rho}_{\mathrm{cl}}(t)
& = \hat{U}^{\dagger}(\delta{t}_{\mathrm{cl}})
\zop{`$\occupation{\stateset{{\bm r}^{(j)}_{\mathrm{X}}(t)\pm\Delta{\bm r}}{\mathrm{X}}}{}$'}{}
\hat{\rho}_{\mathrm{cl}}(t-\delta{t}_{\mathrm{cl}})
\nonumber\\
& ~~~~ \zop{`$\occupation{\stateset{{\bm r}^{(j)}_{\mathrm{X}}(t)\pm\Delta{\bm r}}{\mathrm{X}}}{}$'}{} \hat{U}(\delta{t}_{\mathrm{cl}})
\label{5-95}\\
& = \hat{U}^{\dagger}(\delta{t}_{\mathrm{cl}})
\hat{\rho}_{\mathrm{cl}}(t-\delta{t}_{\mathrm{cl}}) \hat{U}(\delta{t}_{\mathrm{cl}}).
\label{5-96}
\end{align}
Here, the position proposition in Eq.\eqref{5-95} can be replaced by the corresponding momentum proposition.
The small time interval $\delta{t}_{\mathrm{cl}}$ can be chosen arbitrarily, since the classical description merely approximates the realistic system described by stationary states.
\par
In classical descriptions, the instantaneous values in Eqs.\eqref{5-88}--\eqref{5-89} consist of a single term with ${\bm r}_{\mathrm{X}}\,{=}\,{\bm r}^{(j)}_{\mathrm{X}}(t)$ and ${\bm p}_{\mathrm{X}}\,{=}\,{\bm p}^{(j)}_{\mathrm{X}}(t)$, and the instantaneous values equal the expectation values:
\begin{align}
{\bm r}^{(j)}_{\mathrm{X}}(t) = \left\langle {\bm r}_{\mathrm{X}}(t) \right\rangle_{\mathrm{cl}}
,~~
{\bm p}^{(j)}_{\mathrm{X}}(t) = \left\langle {\bm p}_{\mathrm{X}}(t) \right\rangle_{\mathrm{cl}}.
\label{5-99}
\end{align}
Consequently, the classical equations of motion follow from the Ehrenfest theorem~\cite{Messiah} (cf. Eq.\eqref{5-96}), although they merely lead to trivial linear uniform motion in this example.
Meanwhile, the entanglement yields the deterministic relations:
\begin{align}
& \xjvar{`$\occupation{\stateset{{\bm r}_{\mathrm{A}};\Delta{\bm r}}{\mathrm{A}}}{}$'}{leftright}
= \xjvar{`$\occupation{\stateset{{\bm r}_{\mathrm{B}};\Delta{\bm r}}{\mathrm{B}}}{}$'}{leftright} 
,~~ \forall j
\label{5-101}\\
& \xjvar{`$\occupation{\stateset{{\bm p}_{\mathrm{A}};\Delta{\bm p}}{\mathrm{A}}}{}$'}{leftright}
= \xjvar{`$\occupation{\stateset{{\bm p}_{\mathrm{B}};\Delta{\bm p}}{\mathrm{B}}}{}$'}{leftright} 
,~~ \forall j,
\label{5-102}
\end{align}
which restores the EPR relations in Eq.\eqref{5-85}.
Our framework thus circumvents the paradox~\cite{EPR,EPR-Bohr} through the Compatibility assumption, which establishes complementarity in individual trials and requires constant updating of the statistical ensemble.
\par
If we do not require proxy representations for momenta, the position uncertainty $\Delta{\bm r}$ can be arbitrarily small.
Then, the logical variables $\xjvar{`$\occupation{\stateset{{\bm r}_{\mathrm{X}};\Delta{\bm r}}{\mathrm{X}}}{}$'}{}$ behave stochastically, and the Born rule and the projector in Eq.\eqref{5-86} yield the probabilistic interpretation of wavefunctions.

\subsection{Light quantum hypothesis}
\label{sDiscussionF}
Our framework, postulating both HVs and stationary states, reconciles Einstein's interpretation of quantum theory~\cite{Einstein1949,Ballentine} with Bohr's model of physical reality~\cite{Bohr1913,Bohr1950}.
Consequently, it reveals the limitations of Einstein's model of physical reality, as well as Bohr's interpretation of quantum mechanics.
Our HV formulation expresses the light quantum hypothesis~\cite{Einstein1905} in the form:
\begin{align}
\xjvar{`ionized,\,$t$'}{} = \xjvar{`absorbed,\,$t$'}{}
,~~ \forall j,
\label{5-111}
\end{align}
where the symbol `absorbed,\,$t$' represents the field proposition `A photon is absorbed by time $t$'.
Below, we demonstrate that this equation holds for the $n\,{=}\,1$ Fock state, but not for coherent states.
To avoid the complexity arising from relaxations in multiparticle systems, we focus on the simplest problem where a hydrogen atom is initially in the 1s state $\statevectorR{\mbox{1s}}{\mathrm{A}}$.
\par
When the radiation field is initially in the $n\,{=}\,1$ Fock state $\statevectorR{1}{\mathrm{R}}$, the radiation field and atomic system at time $t$ are entangled as a result of unitary evolution:
\begin{align}
|\psi(t)\rangle
& = \hat{U}(t) \statevectorR{\mbox{1}}{\mathrm{R}}\statevectorR{\mbox{1s}}{\mathrm{A}}
\label{5-112}\\
& \cong c_{\mathrm{1s}}(t) \statevectorR{\mbox{1}}{\mathrm{R}}\statevectorR{\mbox{1s}}{\mathrm{A}}
+ \iiint c_{\bm p}(t) \statevectorR{\mbox{0}}{\mathrm{R}}\statevectorR{\bm p}{\mathrm{A}}\,d{\bm p}.
\label{5-113}
\end{align}
Here, $\statevectorR{\bm p}{\mathrm{A}}$ denotes a member of the scattering states $\stateset{E>0}{\mathrm{A}}$ specified by a relative momentum ${\bm p}$, and $c_{\mathrm{1s}}(t)$ and $c_{\bm p}(t)$ denote complex probability amplitudes.
We employ standard approximations in quantum optics~\cite{Loudon}.
Then, the general discussion in Section \ref{sDiscussionC} shows that this entanglement leads to a deterministic relation involving atomic states and field states.
Thus, a straightforward definition
\begin{align}
\mbox{`absorbed,\,$t$'}
= \mbox{`$\occupation{\statesingle{0}{\mathrm{R}}}{},t$'${\wedge}$`$\occupation{\statesingle{1}{\mathrm{R}}}{},0$'}
,\quad t \ge 0
\label{5-114}
\end{align}
reproduces the light quantum hypothesis, in agreement with Dirac's explanation~\cite{Dirac1927} that established the modern foundation of the photon concept.
\par
The situation changes when the radiation field is initially in a coherent state,
\begin{align}
\statevectorR{\tilde{\alpha}}{\mathrm{R}} 
= e^{-\frac{|\tilde{\alpha}|^{2}}{2}} \sum_{n=0}^{\infty} \frac{\tilde{\alpha}^{n}}{\sqrt{n!}}\,
\statevectorR{n}{\mathrm{R}},
\label{5-115}
\end{align}
where $\statevectorR{n}{\mathrm{R}}$ ($n\,{=}\,0,1,\ldots$) denotes the $n$-th Fock state.
In this case, the state vector becomes
\begin{align}
& |\psi(t)\rangle
\equiv \hat{U}(t) \statevectorR{\tilde{\alpha}}{\mathrm{R}}\statevectorR{\mbox{1s}}{\mathrm{A}}
\label{5-116}\\
& \cong \sum_{n=0}^{\infty} \left( c'_{n,\mathrm{1s}}(t) \statevectorR{\mbox{1s}}{\mathrm{A}}
+ c'_{n,{\bm p}}(t) \iiint \statevectorR{\bm p}{\mathrm{A}}\,d{\bm p} \right) \statevectorR{n}{\mathrm{R}},
\label{5-117}
\end{align}
which no longer exhibits entanglement.
Thus, the light quantum hypothesis does not hold for coherent states.
The difficulty is even more obvious in the statistical domain, where the light quantum hypothesis requires
\begin{align}
{\mathrm{tr}}[\,\xop{`ionized,\,$t$'}{}\,\hat{\rho}\,]
\overset{?}{=} {\mathrm{tr}}[\,\xop{`absorbed,\,$t$'}{}\,\hat{\rho}\,].
\label{5-118}
\end{align}
This equality is not satisfied without entanglement because the operators $\xop{`ionized,$t$'}{}$ and $\xop{`absorbed,$t$'}{}$ act on different Hilbert spaces.
In this respect, the light quantum hypothesis is not a universal law for physical reality, but a deterministic relation that holds for specific density operators.
\par
It is worth noting that today's explanation of blackbody radiation no longer depends on the light quantum hypothesis.
This is because non-relativistic QED~\cite{Dirac1927,Fermi1932} has quantized both the atomic system and radiation field.
In quantum optics, photons are defined as what field operators create in the Hilbert space, but physical photons with only two polarization components have no well-defined wavefunction~\cite{Mandel,Scully}.
Accordingly, photons in modern optical theories effectively represent quantum jumps in photodetectors~\footnote{We quote Glauber's words~\cite{Glauber2006}: ``I don't know anything about photons, but I know one when I see one.'' Other criticisms of the photon concept can be found in Lamb~\cite{Antiphoton} and other semi-popular literature~\cite{Muthukrishnan,Knight1983}.}.
Furthermore, if we extend our formulation using the relativistic Hamiltonian~\cite{Messiah,Milonni1993} for a fixed inertial frame, the entanglement in the final state of Compton scattering leads to a deterministic relation ${\bm p}+\hbar{\bm k}={\bm p}_{0}+\hbar{\bm k}_{0}$\footnote{Specifically, the wavefunction has the form $|\phi\rangle = \iiint c_{({\bm p},s,{\bm k},\lambda)} | {\bm p},s \rangle_{\mathrm{A}}| 1 \rangle_{{\bm k}\lambda} d{\bm p}$ with $c_{({\bm p},s,{\bm k},\lambda)} \propto \delta({\bm p}+\hbar{\bm k}-{\bm p}_{0}-\hbar{\bm k}_{0})$, where ${\bm p}$ and $s$ denote the momentum and spin of a scattered electron, ${\bm k}$ and $\lambda$ denote the wavenumber and polarization of the scattered field, and ${\bm p}_{0}$ and ${\bm k}_{0}$ are initial momentum and wavevector, respectively. Here, the stationary states $|{\bm p},s \rangle_{\mathrm{A}}$ and $|1\rangle_{{\bm k}\lambda}$ are defined using the relativistic version of the non-interacting Hamiltonian.}.
Although this relation~\cite{Compton1925,Bothe1} has been viewed as providing evidence for the light quantum hypothesis, our analysis suggests that individual trials in Compton scattering can be described in terms of the occupation of stationary states.

\subsection{Numerical simulations of the entire optical channels}
\label{sDiscussionG}
Our framework enables numerical simulation of the entire optical channel, including quantum jumps in photodetectors and free space propagation from light source to photodetectors.
Before proceeding to numerical demonstrations, we extend the simple model of light absorption in Eq.\eqref{5-23}.
These analyses illustrate how our framework naturally describes critical details of quantum coherence: classical post-detection noise, detector correlations, and the non-exclusivity of photoionization sites.

\subsubsection{Practical photodetector model}
We model classical post-detection noise through true-negative (`TN') and false-positive (`FP') errors that occur with probabilities $p_{\mathrm{TN}}$ and $p_{\mathrm{FP}}$, respectively.
These propositions are compatible with any proposition due to the projectors that commute with any operator:
\begin{align}
\zop{`TN'}{} = \sqrt{p_{\mathrm{TN}}}\,\hat{1}
,~~ \zop{`FP'}{} = \sqrt{p_{\mathrm{FP}}}\,\hat{1}.
\label{5-131}
\end{align}
We use these propositions to define proposition `$\mathrm{output},t$' that represents whether any output signal is generated during $t$.
Neglecting time delays and excess noise ($p_{\mathrm{FP}}\,{=}\,0$) for simplicity, we obtain
\begin{align}
\mbox{`$\mathrm{output},t$'}
= \big( \mbox{`$\mathrm{ionized},t$'} \wedge \lnot \mbox{`TN'} \big) \vee \mbox{`FP'},
\label{5-132}
\end{align}
and the detection probability can be expressed as 
\begin{align}
{\mathrm{Pr}}(\mbox{`$\mathrm{output},t$'})
= {\mathrm{tr}}[\,\xop{`$\mathrm{output},t$'}{}\hat{\rho}\,].
\label{5-133}
\end{align}
In this expression, the following operator describes the probability of each detection:
\begin{align}
\xop{`$\mathrm{output},t$'}{}
& \equiv \zopd{`$\mathrm{output},t$'}{} \zop{`$\mathrm{output},t$'}{}
\label{5-134}\\
& = \eta_{\mathrm{amp}} \sum_{\statevectorR{\phi}{\mathrm{A}}}^{\stateset{E>0}{\mathrm{A}}} \hat{\pi}^{\dagger}(\phi,t) \hat{\pi}(\phi,t),
\label{5-135}
\end{align}
where $\hat{\pi}(\phi,t)\,{\equiv}\,\hat{U}(t)\statevectorR{\mathrm{ref}}{\mathrm{A}} \statevectorL{\phi}{\mathrm{A}}\hat{U}^{\dagger}(t)$ denotes a de-excitation operator for a scattering state $\statevectorR{\phi}{\mathrm{A}}$, with $\statevectorR{\mathrm{ref}}{\mathrm{A}}$ as a reference state taken from the bound states $\stateset{E<0}{\mathrm{A}}$.
Evaluating $\hat{\pi}(\phi,t)$ in the Heisenberg picture yields standard formulas for detection probability with a signal loss coefficient $\eta_{\mathrm{amp}}\,{\equiv}\,1\,{-}\,p_{\mathrm{TN}}$.
\par
The proposition `$\mathrm{output},t$' reproduces the central result of the quantum theory of coherence~\cite{Glauber1963} in a remarkably simple manner.
Considering multiple photodetectors labeled by $i$ ($i\,{=}\,1,{\ldots},I$), our framework represents joint detection across all photodetectors by the compound proposition:
\begin{align}
\bigwedge_{i=1}^{I} \mbox{`$\mathrm{output}_{i}$,\,$t_{i}$'}.
\label{5-141}
\end{align}
Evaluating its probability using the Born rule and Boolean logic, we obtain the joint detection probability by $I$ photodetectors:
\begin{align}
& t_{1} \cdots t_{N}\,{\mathrm{tr}}\Bigg[\,\sum_{\xi'_{N}}^{\{x,y,z\}} \sum_{\xi''_{N}}^{\{x,y,z\}} \cdots \sum_{\xi'_{1}}^{\{x,y,z\}} \sum_{\xi''_{1}}^{\{x,y,z\}} 
\nonumber\\
& ~~~~ \hat{\kappa}_{\xi'_{N}\xi''_{N}} \cdots \hat{\kappa}_{\xi'_{1}\xi''_{1}}
\left({\tilde{A}}^{(-)}_{{\mathrm{T}}\xi''_{1}}({\bm r}_{1},t_{1}) \cdots {\tilde{A}}^{(-)}_{{\mathrm{T}}\xi''_{I}}({\bm r}_{I},t_{I}) \right)
\nonumber\\
& ~~~~ \left({\tilde{A}}^{(+)}_{{\mathrm{T}}\xi'_{I}}({\bm r}_{I},t_{I}) \cdots {\tilde{A}}^{(+)}_{{\mathrm{T}}\xi'_{1}}({\bm r}_{1},t_{1}) \right)\hat{\rho}\,\Bigg]
,~~ t_{1} < \ldots < t_{I}.
\nonumber\\
& \label{5-142}
\end{align}
Here, ${\bm r}_{i}$ denotes the location of the $i$-th detector, ${\tilde{A}}_{\mathrm{T}\xi_{i}}^{(\pm)}({\bm r}_{i},t_{i})$ denotes the positive and negative oscillating parts of the transverse vector potential in the interaction picture, and $\hat{\kappa}_{\xi'_{i}\xi''_{i}}$ denotes an operator coefficient in Mandel's formula.
Owing to the time ordering symbol $\mathcal{T}$ in the transformation rules (Eq.\eqref{4-75}), this expression has the correct operator ordering to reproduce the standard expressions of quantum coherence functions~\cite{Mandel}.
\par
Counting statistics reveal how uncorrelated---or incoherent---sums are justified in the analysis of quantum coherence.
To see this, consider photosensitive materials with $L$ local photoionization sites.
Since photoionization at each site is an irreversible event that occurs only once, the count of the $i$-th photodetector can be expressed as
\begin{align}
{\mathcal N}_{i}^{(j)}(t) 
= \sum_{l=1}^{L} \xjvar{`$\mathrm{output}_{i,l},t$'}{}
,~~\forall j,
\label{5-151}
\end{align}
where the proposition `$\mathrm{output}_{i,l},t$' represents the presence of the output signal from the $l$-th site in the $i$-th photodetector.
If we model photoionization sites as non-interacting subsystems~\footnote{Physically, this assumption is grounded in the localization of stationary states~\cite{Knoster2008,Toyozawa1982} and rapid decoherence caused by the local environment~\cite{Moener,Orrit}.}, Eq.\eqref{5-151} has the physically correct upper bound $L$, even though the logical variables share the same trial index.
This is because the assumption of non-interacting subsystems factorizes the Hilbert space into a tensor product of $L$ subspaces, making subsystem propositions non-exclusive rather than statistically independent.
The calculation of the expectation values $\langle{\mathcal N}_{i}(t)\rangle={\mathrm{tr}}[\hat{\mathcal N}_{i}(t) \hat{\rho}]$ is straightforward, where the operator $\hat{\mathcal N}_{i}(t)$ has the form:
\begin{align}
& \hat{\mathcal N}_{i}(t) 
\equiv \sum_{l=1}^{L} \xop{`$\mathrm{output}_{i,l},t$'}{}
\label{5-153}\\
& \xop{`$\mathrm{output}_{i,l},t$'}{}
= \eta_{\mathrm{amp}} \sum_{l=1}^{L}\sum_{\statevectorR{\phi}{i,l}}^{\stateset{E>0}{i,l}} \hat{\pi}_{i,l}^{\dagger}(\phi,t) \hat{\pi}_{i,l}(\phi,t).
\label{5-154}
\end{align}

\subsubsection{Counting statistics}
We can now explicitly write down the compound proposition that yields the probability distribution of detector counts ${\mathcal N}_{i}$.
This proposition, denoted by `${\mathcal N}_{i},t$', represents whether the detector generates ${\mathcal N}_{i}$ output signals during $t$.
If we divide the time interval $t$ into $M$ subintervals, we can construct the proposition `${\mathcal N}_{i},t$' by enumerating possible ways to assign ${\mathcal N}_{i}$ photoionization sites to $M$ subintervals.
Let the subintervals be defined by $s_{m}\,{\equiv}\,(m/M)t$ ($m\,{=}\,1,\ldots,M$), and let tuple $(m_{1},m_{2},\ldots,m_{{\mathcal N}})$ represent any combination of subinterval indices.
Then a simple calculation yields
\begin{align}
& \mbox{`${\mathcal N}_{i},t$'}
= \bigvee_{(m_{1},m_{2},\ldots,m_{{\mathcal N}})}^{{m_{1}}<{m_{2}}<\cdots<{m_{\mathcal N}}} 
\Bigg( \bigwedge_{m}^{(m_{1},m_{2},\ldots,m_{{\mathcal N}})} \!\!\!\!\!\!\!\!\!
\mbox{`$\mathrm{output}_{i},[s_{m},s_{m+1})$'}
\nonumber\\
& ~~~~~~~~~~~~~~~~~
\wedge \!\!\!\!\!\!\!\!\! \bigwedge_{m}^{{\notin}\,(m_{1},m_{2},\ldots,m_{{\mathcal N}})} \!\!\!\!\!\!\!\!\!
\lnot\,\mbox{`$\mathrm{output}_{i},[s_{m},s_{m+1})$'}\Bigg),
\label{5-155}
\end{align}
where the proposition `$\mathrm{output}_{i},[s_{m},s_{m+1})$' is given by
\begin{align}
& \mbox{`$\mathrm{output}_{i},[s_{m},s_{m+1})$'}
\nonumber\\
& ~~ \equiv \bigvee_{l=1}^{L} \mbox{`$\mathrm{output}_{i,l},s_{m+1}$'}
\wedge \lnot \mbox{`$\mathrm{output}_{i,l},s_{m}$'}.
\label{5-156}
\end{align}
Note that $M$ is infinitely large, while $L$ is large but finite.
We confirmed that this proposition reproduces the standard formulas for counting statistics~\cite[Eqs.(14.8-6)--(14.8-7)]{Mandel}:
\begin{align}
\langle x(\mbox{`${\mathcal N}_{i},t$'}) \rangle
& = {\mathrm{tr}}\left[\,\xop{`${\mathcal N}_{i},t$'}{leftright} \hat{\rho}\,\right]
\label{5-161}\\
\xop{`${\mathcal N}_{i},t$'}{leftright}
& \cong \frac{\hat{\mathcal{N}_{i}}(t)^{{\mathcal N}_{i}}}{{\mathcal N}_{i}!} e^{-\hat{\mathcal{N}_{i}}(t)},
\label{5-162}
\end{align}
where the Heisenberg-picture operator $\hat{\mathcal{N}_{i}}(t)$ can be expressed using field operators as
\begin{align}
\hat{\mathcal{N}}_{i}(t)
& \cong t\,\frac{\eta_{\mathrm{det}}}{\hbar\omega_{\mathrm{o}}}\,S\,\varepsilon_{0} c\,{\tilde{\bm E}}^{(-)}_{{\mathrm{T}}}({\bm r}_{i},t) \cdot {\tilde{\bm E}}^{(+)}_{{\mathrm{T}}}({\bm r}_{i},t).
\label{5-163}
\end{align}
Here, ${\tilde{\bm E}}^{(\pm)}_{{\mathrm{T}}}({\bm r}_{i},t) \cong {\pm}\,{\mathrm{i}}\omega_{\mathrm{o}}{\tilde{\bm A}}^{(\pm)}_{{\mathrm{T}}}({\bm r}_{i},t)$ denote the positive and negative oscillating parts of the transverse electric field operators, $S$ is a photosensitive area, and the quantum efficiency ${\eta}_{\mathrm{det}}$ is proportional to the statistical average of the expectation value of $\hat{\kappa}_{\xi'_{i}\xi''_{i}}$ (see Eq.\eqref{5-142}).

\subsubsection{Double-slit experiment}
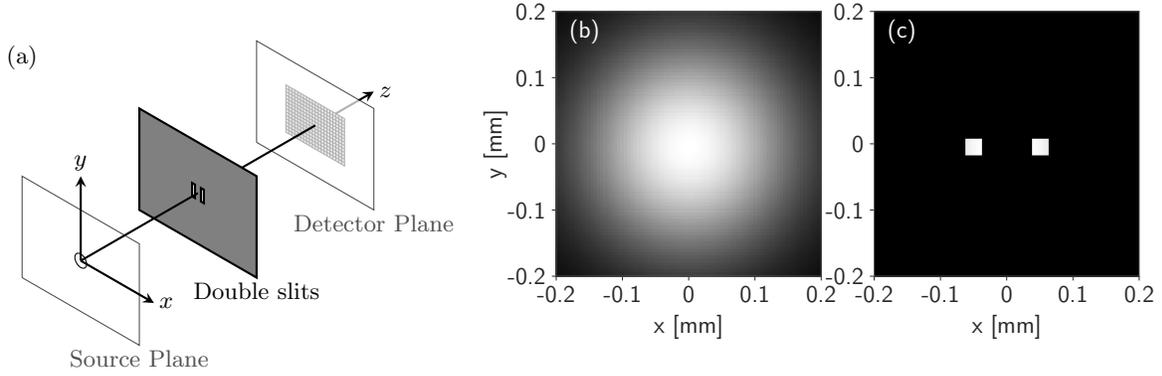
\begin{figure*}[ht]
\centering
\begin{tikzpicture}[x={(0.866cm,-0.5cm)}, y={(0cm,1cm)}, z={(0.866cm,0.5cm)}, scale=0.45,>=stealth, inner sep=0pt, outer sep=2pt, axis/.style={thick,->}, plane/.style={fill=white, opacity=0.7},]
  \coordinate (O) at (0, 0, 0);
  \colorlet{lightred}{red!80!black}
  \node (caption) at (-2, 5, 0) {\small (a)};
  \draw[axis] (0, 0, 8) -- (0, 0, 10) node [right] {\small $z$};
  \begin{scope}
    \draw[fill=lightgray] (-1, -0.7, 8) -- (-1, 0.7, 8) -- (1, 0.7, 8) -- (1, -0.7, 8) -- cycle;
    \foreach \k in {1,2,...,19}{
      \draw[black] (-1+\k*0.1,-0.7,8)--(-1+\k*0.1,0.7,8);
    }
    \foreach \k in {1,2,...,13}{
      \draw[black] (-1,-0.7+\k*0.1,8)--(1,-0.7+\k*0.1,8);
    }
  \end{scope}
  \filldraw[plane] (-2,-1.5,8) -- (-2,1.5,8) -- (2,1.5,8) -- (2,-1.5,8) node[below,sloped]{\small Detector Plane} -- (-2,-1.5,8);
  \draw[thick] (0, 0, 4) -- (0, 0, 8);
  \begin{scope}[thick]
    \draw[fill=gray] (-2, -1.5, 4) -- (-2, 1.5, 4) -- (2, 1.5, 4) -- (2, -1.5, 4) node[below,sloped]{\small Double slits} -- cycle;
    \draw[fill=white] (-0.2,-0.2, 4) -- (-0.2,0.2, 4) -- (-0.1, 0.2, 4) -- (-0.1, -0.2, 4) -- cycle;
    \draw[fill=white] (0.2,-0.2, 4) -- (0.2,0.2, 4) -- (0.1, 0.2, 4) -- (0.1, -0.2, 4) -- cycle;
  \end{scope}
  \filldraw[plane] (-2, -1.5, 0) -- (-2,1.5, 0) -- (2,1.5, 0) -- (2,-1.5, 0) node[below,sloped]{\small Source Plane} -- (-2, -1.5, 0);
  \draw (O) circle [radius=0.2];
  \draw[thick] (O) -- (0,   0, 4);
  \draw[axis] (O) -- +(2.5, 0, 0) node [right] {\small $x$};
  \draw[axis] (O) -- +(0,   2.5, 0) node [above] {\small $y$};
  \end{tikzpicture}   
\includegraphics[width=10cm,keepaspectratio]{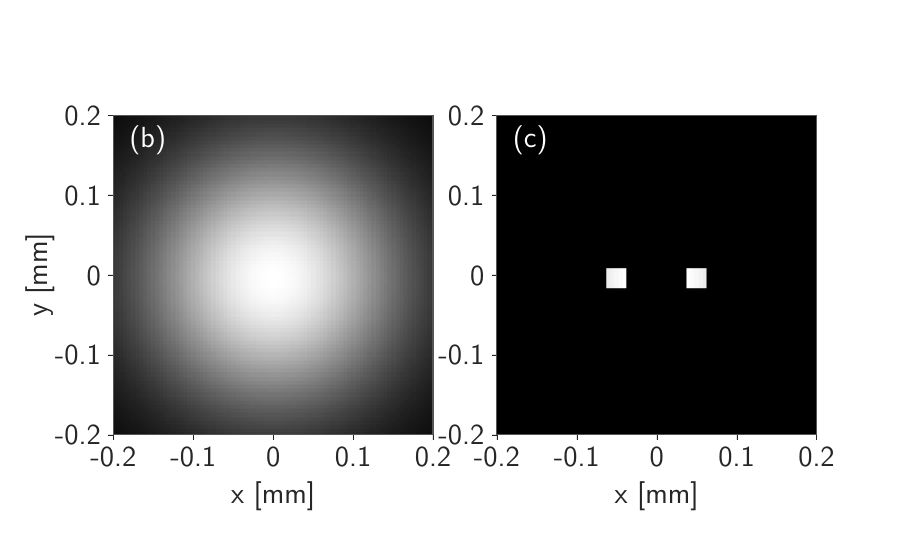}
\caption{Simulation setup: (a) Overall configuration, (b) intensity profile before the slit, and (c) intensity profile after the slit.}
\label{figDoubleSlit2} 
\end{figure*}
\begin{figure*}[ht]
\centering
\adjustbox{trim=0mm 12.3mm 0mm 0mm, clip}{\includegraphics[width=\textwidth,keepaspectratio]{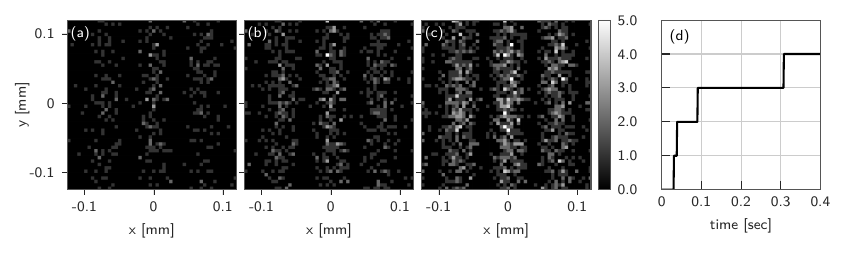}}
\adjustbox{trim=0mm 0mm 0mm 1mm, clip}{\includegraphics[width=\textwidth,keepaspectratio]{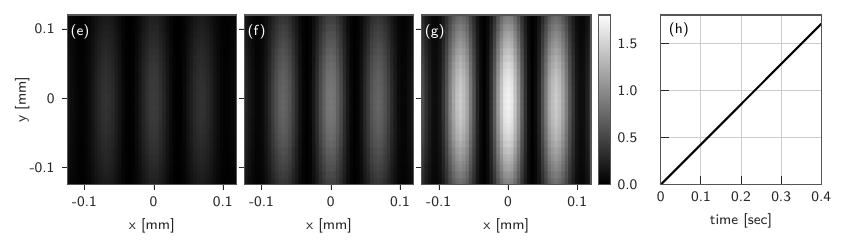}}
\caption{Simulation results. Upper: photodetector counts for individual trials at (a) $t=0.1\,{\mathrm{sec}}$, (b) $t=0.2\,{\mathrm{sec}}$, (c) $t=0.4\,{\mathrm{sec}}$, and (d) temporal variation at $(x,y)\,{=}\,(0,0)$. Lower: expectation values of photodetector counts at the corresponding times and locations. Individual trial counts are discrete integers, while expectation values are continuous real numbers. Standard Python libraries (numpy.fft and scipy.stats.poisson.rvs) are used for the calculations.}
\label{figDoubleSlit3}
\end{figure*}
As a visual example, we present a simple numerical calculation of the double-slit experiment using \emph{coherent light}.
For a multi-mode coherent state emitted from a laser cavity, conventional approaches do not predict quantum jumps in emission processes, while remaining agnostic about quantum jumps in absorption processes (See Section \ref{sDiscussionB}).
In contrast, our formulation allows us to directly simulate the entire quantum optical channel.
\par
The setup is shown in Fig.\ref{figDoubleSlit2}.
The incident light is modeled as a $y$-polarized monochromatic field with wavelength $0.710\mu{\mathrm{m}}$.
The spatial profile is a Gaussian with the beam waist $50\,\mu{\mathrm{m}}\,{\times}\,50\,\mu{\mathrm{m}}$ at the source plane ($z_{\mathrm{src}}\,{=}\,0{\mathrm{mm}}$).
At the slit plane ($z_{\mathrm{slit}}\,{=}\,50{\mathrm{mm}}$), the field before the slits is computed using an analytic formula for Gaussian beams~\cite{Yariv}, then multiplied by a binary transmittance $T(x,y)$ to obtain the field after the slits.
The double slit consists of two $20\,\mu{\mathrm{m}}\,{\times}\,20\,\mu{\mathrm{m}}$ square apertures with a center-to-center separation of $100\,\mu{\mathrm{m}}$.
For linear optical elements, this kind of semiclassical description is commonly employed in cavity QED, although formal derivations~\cite{Milonni2004,Glauber1991} are relatively limited.
The free-space propagation is modeled by the angular spectrum method~\cite{Mandel}, which creates a near-field interference pattern at the detector plane ($z_{\mathrm{det}}\,{=}\,60{\mathrm{mm}}$).
Each photodetector, with $5\,\mu{\mathrm{m}}\,{\times}\,5\,\mu{\mathrm{m}}$ lateral dimension and arranged in a 50$\,{\times}\,$50 two-dimensional array, responds to the light independently with a probability proportional to the local field amplitude according to Eqs.\eqref{5-161}--\eqref{5-163}.
For coherent states, the optical equivalence theorem~\cite{Mandel} justifies a semiclassical evaluation of the expectation values of normally-ordered operator products.
Thus, Eq.\eqref{5-161} reduces to the Poisson distribution:
\begin{align}
& \left\langle \xvar{`${\mathcal N}_{i},t$'}{} \right\rangle
\cong \frac{\langle {\mathcal{N}}_{i}(t) \rangle^{{\mathcal N}_{i}}}{{\mathcal N}_{i}!} \exp(-\langle {\mathcal{N}}_{i}(t) \rangle)
\label{5-171}\\
& \langle{\mathcal{N}}_{i}(t)\rangle
= t ~ \frac{\eta_{\mathrm{det}} W_{\mathrm{total}}}{\hbar\omega_{\mathrm{o}}} \left| \mathcal{E}(x_{i},y_{i},z_{\mathrm{det}},\omega_{\mathrm{o}}) \right|^{2},
\label{5-172}
\end{align}
where ${\bm r}_{i}\,{=}\,(x_{i},y_{i},z_{\mathrm{det}})$ is the $i$-th photodetector location, and ${\mathcal{E}}(x_{i},y_{i},z_{\mathrm{det}},\omega_{\mathrm{o}})$ is the normalized amplitude with a well-defined phase~\cite{Mandel}.
The total power at the detector plane and the quantum efficiency are arbitrarily chosen as $W_{\mathrm{total}}\,{=}\,1\,{\times}\,10^{-14}\,{\mathrm{W}}$ and $\eta_{\mathrm{det}}\,{=}\,0.1$, respectively. 
\par
The calculation results are shown in Fig.\ref{figDoubleSlit3}. 
Figures (a)--(c) show how the detector counts increase discontinuously in individual trials, while (e)--(g) show how the expectation values increase linearly according to Eq.\eqref{5-172}.
(d) and (h) show the temporal variations at a fixed location $(x,y)\,{=}\,(0,0)$.
Conventionally, continuous wave descriptions provided by both classical and quantum electrodynamics failed to represent the time variations of integer functions in (a)--(d)~\cite{Einstein1905}.
Consequently, the discrete jumps have been qualitatively explained via the photon concept~\cite{Einstein1905,Dirac1927}.
In contrast, our framework represents the discontinuities based on stationary state occupations without invoking both photons and environmental decoherence.
While we have focused on the detection of coherent states in this paper, our framework also describes detector correlations, as indicated by the ability to calculate joint probabilities.
A forthcoming paper will describe how to simulate correlated phenomena, such as the photon anti-bunching of the resonance fluorescence from a two-level atom.

\section{Conclusion}
\label{sConclusion}
In this study, we have reexamined the fundamental question of whether quantum theory can describe integer functions.
This question, originally raised by Einstein in 1905, created the photon concept and subsequent controversies over quantum observations.
These controversies recently led to decoherence-based descriptions of quantum jumps.
As an alternative, we developed a theory of logical variables, which explicitly addresses unexplored correlations in individual trials.
Our analysis started by acknowledging the duality of observations: that binary logical variables exist experimentally, though differential laws permit no such solutions.
We then identified assumptions that reconcile Boolean logic with quantum theory, establishing Boolean logic as laws governing individual trials.
Specifically, we focused on the occupation of stationary states and optimized the compatibility criterion to enable a direct description of deterministic relations in individual trials.
We found that the exclusivity of state occupations correlates logical variables and, combined with the new compatibility condition, justifies both the hidden variable formulation and Boolean logic.
This correlation imposes nontrivial constraints on arithmetic operations between logical variables, suggesting a novel mechanism for measurement compatibility and contextuality.
\par
More specifically, the fundamental building blocks of our theory are compound propositions that represent specific time sequences of state occupations.
Their probabilities are calculated using chain operators and density operators, while their truth values are represented by binary logical variables that specify the correlation scope through trial indices.
Our eightfold assumptions describe how to apply the Born rule and Boolean logic to these chain operators and logical variables.
We also presented three concrete examples on Bell's theorems, the original EPR paradox, and the light quantum hypothesis.
We demonstrated that (i) our nonlocal logical variables avoid the classical CHSH bound by prohibiting incompatible arithmetic operations; (ii) mapping stationary state occupations to position/momentum eigenstate occupations provides proxy representations that permit classical descriptions; and (iii) the light quantum hypothesis is a state-specific relation rather than a general law for physical reality.
We also provided a simulation of the double-slit experiment to show the ability to simulate the entire quantum optical channel.
This example also illustrates how our theory describes quantum jumps in the absorption of coherent light---a phenomenon that conventional approaches have struggled to describe.
\par
From a fundamental viewpoint, our approach is unique in postulating both hidden variables and stationary states.
This means that our framework integrates Einstein's interpretation of quantum theory (allowing hidden variables) with Bohr's model of physical reality (postulating stationary states).
In our framework, quantum jumps occur stochastically within the unitary evolution of the entire system including measurement apparatus.
As a result, our framework eliminates the need for external observers, describing quantum jumps and wave-particle duality based only on transitions between stationary states.
A remaining challenge is to develop a unified statistical description of inter-state transitions and external decoherence.
Ultimately, our hidden variable formulation raises a new question about what physical reality is.
Further discussion of the definition of stationary states is necessary to refine our phenomenological analysis.
We believe this framework, though currently limited, offers a unique perspective in exploring the fundamental nature of physical reality.

\bibliographystyle{unsrt}
\bibliography{main}

\clearpage
\appendix
\section*{Supplemental Material}
The supplemental materials below provide the details of mathematical derivations and discuss subtleties worth examining.

\section{Definition of statistical concepts}
\label{sStatistics}
This section provides a concise summary of key statistical concepts used throughout the paper.

\paragraph*{Exclusivity:}
Two propositions `$\mathrm{C}_{1},\,t_{1}$' and `$\mathrm{C}_{2},t_{2}$' are considered mutually exclusive when their joint probability equals zero:
\begin{align}
{\mathrm{Pr}}(\mbox{`$\mathrm{C}_{1},\,t_{1}$'}\,{\wedge}\,\mbox{`$\mathrm{C}_{2},t_{2}$'})
\equiv \big\langle\xvar{`$\mathrm{C}_{1},t_{1}$'$\wedge$`$\mathrm{C}_{2},t_{2}$'}{}\big\rangle
= 0.
\label{eeStatistics5}
\end{align}
This implies that their conjunction is always false:
\begin{align}
\xjvar{`$\mathrm{C}_{1},t_{1}$'$\wedge$`$\mathrm{C}_{2},t_{2}$'}{}=0
,~~\forall j.
\label{eeStatistics6}
\end{align}
Consequently, the logical variables of mutually exclusive propositions are always additive.
\begin{align}
& \xjvar{`$\mathrm{C}_{1},t_{1}$'$\vee$`$\mathrm{C}_{2},t_{2}$'}{}
= \xjvar{`$\mathrm{C}_{1},t_{1}$'}{} + \xjvar{`$\mathrm{C}_{2},t_{2}$'}{}
\nonumber\\
& ,~~\forall j.
\label{eeStatistics7}
\end{align}

\paragraph*{Independence:}
Two propositions `$\mathrm{C}_{1},t_{1}$' and `$\mathrm{C}_{2},t_{2}$' are independent if their conditional probabilities equal their marginal probabilities:
\begin{align}
{\mathrm{Pr}}(\mbox{`$\mathrm{C}_{1},t_{1}$'}|\mbox{`$\mathrm{C}_{2},t_{2}$'})
& = {\mathrm{Pr}}(\mbox{`$\mathrm{C}_{1},t_{1}$'})
\label{eeStatistics8}\\
{\mathrm{Pr}}(\mbox{`$\mathrm{C}_{2},t_{2}$'}|\mbox{`$\mathrm{C}_{1},t_{1}$'})
& = {\mathrm{Pr}}(\mbox{`$\mathrm{C}_{2},t_{2}$'}),
\label{eeStatistics9}
\end{align}
where the conditional probabilities are defined in Eq.\eqref{5-33}.
For independent propositions, the joint probability factorizes into the product of marginal probabilities as
\begin{align}
{\mathrm{Pr}}(\mbox{`$\mathrm{C}_{1},t_{1}$'}\,{\wedge}\,\mbox{`$\mathrm{C}_{2},t_{2}$'})
= {\mathrm{Pr}}(\mbox{`$\mathrm{C}_{1},t_{1}$'})\,{\mathrm{Pr}}(\mbox{`$\mathrm{C}_{2},t_{2}$'}).
\label{eeStatistics10}
\end{align}

\paragraph*{Deterministic propositions:}
A proposition `$\mathrm{C}_{1},t_{1}$' is classified as deterministic when its probability always equals either zero or one:
\begin{align}
\langle \xvar{`$\mathrm{C}_{1},t_{1}$'}{} \rangle \in \{0, 1\}.
\label{eeStatistics11}
\end{align}
This also implies that their truth value is invariably true or false for the given density operator.

\paragraph*{Deterministic relations:}
A deterministic relation exists between propositions `$\mathrm{C}_{1},t_{1}$' and `$\mathrm{C}_{2},t_{2}$' when their truth values are identical in all experimental trials:
\begin{align}
\xjvar{`$\mathrm{C}_{1},t_{1}$'}{} = \xjvar{`$\mathrm{C}_{2},t_{2}$'}{}
,\quad \forall{j}.
\label{eeStatistics12}
\end{align}
By the definition of conditional probabilities, this equation is equivalent to
\begin{align}
1 = {\mathrm{Pr}}(\mbox{`$\mathrm{C}_{1},t_{1}$'}|\mbox{`$\mathrm{C}_{2},t_{2}$'}) = {\mathrm{Pr}}(\mbox{`$\mathrm{C}_{2},t_{2}$'}|\mbox{`$\mathrm{C}_{1},t_{1}$'}),
\label{eeStatistics13}
\end{align}
which can be alternatively expressed as
\begin{align}
\langle \xvar{`$\mathrm{C}_{1},t_{1}$'$\wedge$`$\mathrm{C}_{2},t_{2}$'}{} \rangle
= \langle \xvar{`$\mathrm{C}_{1},t_{1}$'}{} \rangle
= \langle \xvar{`$\mathrm{C}_{2},t_{2}$'}{} \rangle.
\label{eeStatistics14}
\end{align}
Note that deterministic relations do not necessarily require that the propositions involved be deterministic.

\section{Simple model for handling approximate propositions}
\label{sApproximate}
To introduce approximate propositions, a special consideration is necessary because logical variables always have values $0$ or $1$.
We therefore develop a simple model for handling approximate propositions in this section.
Consider propositions `$\mathrm{C}_{1},t_{1}$' and `$\mathrm{C}_{2},t_{2}$' that satisfy a deterministic relation,
\begin{align}
\xjvar{`$\mathrm{C}_{1},t_{1}$'}{leftright}
& = \xjvar{`$\mathrm{C}_{2},t_{2}$'}{leftright}
,\quad \forall j,
\label{eeApproximate1}
\end{align}
and let proposition `$\mathrm{C}_{2},t_{2}$' be approximated by a proposition `${\mathrm{C}}'_{2},t_{2}$' as
\begin{align}
\mbox{`$\mathrm{C}_{2},t_{2}$'}
& = \mbox{$\big(\mbox{`TT'$\wedge$`${\mathrm{C}}'_{2},t_{2}$'}\big)\,{\vee}\, 
\big(\mbox{`TF'$\wedge\lnot$`${\mathrm{C}}'_{2},t_{2}$'}\big)$}.
\label{eeApproximate2}
\end{align}
Here, `TT' represents that `$\mathrm{C}_{2},t_{2}$' is true when `${\mathrm{C}}'_{2},t_{2}$' is true, and `TF' represents that `$\mathrm{C}_{2},t_{2}$' is true when `${\mathrm{C}}'_{2},t_{2}$' is false.
We require that both `TT' and `TF' are compatible with any proposition by assuming
\begin{align}
\zop{`TT'}{} = \sqrt{p_{\mathrm{TT}}} \hat{1}
,~~ \zop{`TF'}{} = \sqrt{p_{\mathrm{TF}}} \hat{1},
\label{eeApproximate3}
\end{align}
where $1\,{-}\,p_{\mathrm{TT}}$ and $p_{\mathrm{TF}}$ represent error probabilities.
Using Boolean logic and these relations, the deterministic relation in Eq.\eqref{eeApproximate1} takes the form:
\begin{align}
\xjvar{`$\mathrm{C}_{1},t_{1}$'}{}
& = \left(\xjvar{`TT'}{} - \xjvar{`TF'}{}\right) \, \xjvar{`${\mathrm{C}}'_{2},t_{2}$'}{} 
\nonumber\\
& ~~~~ + \xjvar{`TF'}{},~~ \forall{j}.
\label{eeApproximate6}
\end{align}
\par
The validity of the approximation in Eq.\eqref{eeApproximate2} must be evaluated in the statistical domain.
For this purpose, we rewrite the deterministic relation in Eq.\eqref{eeApproximate1} as
\begin{align}
\left\langle \xvar{`$\mathrm{C}_{1},t_{1}$'$\wedge$`$\mathrm{C}_{2},t_{2}$'}{} \right\rangle
= \left\langle \xvar{`$\mathrm{C}_{1},t_{1}$'}{} \right\rangle
= \left\langle \xvar{`$\mathrm{C}_{2},t_{2}$'}{} \right\rangle.
\label{eeApproximate11}
\end{align}
Taking the conjunction of Eq.\eqref{eeApproximate2} and `$\mathrm{C}_{1},t_{1}$' and calculating the expectation value, we have
\begin{align}
& \left\langle \xvar{`$\mathrm{C}_{1},t_{1}$'$\wedge$`$\mathrm{C}_{2},t_{2}$'}{} \right\rangle
\nonumber\\
& = \left\langle \xvar{`$\mathrm{C}_{1},t_{1}$'$\wedge\left(\big(\mbox{`TT'$\wedge$`${\mathrm{C}}'_{2},t_{2}$'}\big)\,{\vee}\,\big(\mbox{`TF'$\wedge\lnot$`${\mathrm{C}}'_{2},t_{2}$'}\big)\right)$}{} \right\rangle
\label{eeApproximate12}\\
& = \left( p_{\mathrm{TT}} - p_{\mathrm{TF}} \right) \left\langle \xvar{`$\mathrm{C}_{1},t_{1}$'$\wedge$`${\mathrm{C}}'_{2},t_{2}$'}{} \right\rangle
+ p_{\mathrm{TF}} \left\langle \xvar{`$\mathrm{C}_{1},t_{1}$'}{} \right\rangle.
\label{eeApproximate13}
\end{align}
On the other hand, taking the expectation value of Eq.\eqref{eeApproximate6}, we also obtain
\begin{align}
\left\langle \xvar{`$\mathrm{C}_{1},t_{1}$'}{} \right\rangle
& = \left( p_{\mathrm{TT}} - p_{\mathrm{TF}} \right) \left\langle \xvar{`${\mathrm{C}}'_{2},t_{2}$'}{} \right\rangle
+ p_{\mathrm{TF}}.
\label{eeApproximate14}
\end{align}
Combining Eqs.\eqref{eeApproximate11}, \eqref{eeApproximate13}, and \eqref{eeApproximate14}, we find
\begin{align}
& \left\langle \xvar{`$\mathrm{C}_{1},t_{1}$'$\wedge$`${\mathrm{C}}'_{2},t_{2}$'}{} \right\rangle
\nonumber\\
& = \frac{ 1 - p_{\mathrm{TF}} }{\left( p_{\mathrm{TT}} - p_{\mathrm{TF}} \right) } \left\langle \xvar{`$\mathrm{C}_{1},t_{1}$'}{} \right\rangle 
\label{eeApproximate15}\\
& = \left( 1 - p_{\mathrm{TF}} \right) \left\langle \xvar{`${\mathrm{C}}'_{2},t_{2}$'}{} \right\rangle 
+ \frac{\left(1 - p_{\mathrm{TF}}\right)p_{\mathrm{TF}}}{\left( p_{\mathrm{TT}} - p_{\mathrm{TF}} \right)}.
\label{eeApproximate16}
\end{align}
When the error probabilities are sufficiently small such that $p_{\mathrm{TT}}\,{\cong}\,1$ and $p_{\mathrm{TF}}\,{\cong}\,0$, Eq.\eqref{eeApproximate16} simplifies to
\begin{align}
\langle \xvar{`$\mathrm{C}_{1},t_{1}$'$\wedge$`${\mathrm{C}}'_{2},t_{2}$'}{} \rangle
\cong \langle \xvar{`$\mathrm{C}_{1},t_{1}$'}{} \rangle
\cong \langle \xvar{`${\mathrm{C}}'_{2},t_{2}$'}{} \rangle.
\label{eeApproximate17}
\end{align}
Thus, approximate propositions have reasonable statistical properties if errors are independent of given propositions and error probabilities are sufficiently small.

\section{Compatibility of deterministic propositions}
\label{sAppendix100}
This section presents an elementary derivation of the compatibility condition (Eq.\eqref{4-94}).
Consider a compound proposition `${\bf C},{\bm t}$' consisting of $N$ propositions `$\stateset{{\mathrm{C}}_{n}}{},t_{n}$' ($n\,{=}1,\ldots,N$) and suppose that these propositions are deterministic:
\begin{align}
\left\langle \xvar{`$\occupation{\stateset{\mathrm{C}_{n}}{}}{},t_{n}$'}{} \right\rangle
& = \sum_{\statevectorR{\phi_{n}}{}}^{\stateset{\mathrm{C}_{n}}{}} \statevectorL{\phi_{n}}{} \hat{\rho}(t_{n}) \statevectorR{\phi_{n}}{}
= 1.
\label{eeGroup1}
\end{align}
Our general assumptions then yield
\begin{align}
\left\langle \xvar{$\lnot$`$\occupation{\stateset{\mathrm{C}_{n}}{}}{},t_{n}$'}{} \right\rangle
& = \sum_{\statevectorR{\phi_{n}}{}}^{\notin\stateset{\mathrm{C}_{n}}{}} \statevectorL{\phi_{n}}{} \hat{\rho}(t_{n}) \statevectorR{\phi_{n}}{}
= 0,
\label{eeGroup2}
\end{align}
where each term in the sum is non-negative due to the positivity of density operators.
Thus, Eq.\eqref{eeGroup2} implies
\begin{align}
\statevectorL{\phi_{n}}{} \hat{\rho}(t_{n}) \statevectorR{\phi_{n}}{} = 0
,~~\forall \statevectorR{\phi_{n}}{} \notin \stateset{\mathrm{C}_{n}}{}
\label{eeGroup3}
\end{align}
Applying the Cauchy-Schwarz inequality to $\hat{\rho}(t)^{1/2} |\phi_{n}\rangle$, we further obtain
\begin{align}
|\statevectorL{\phi_{n}}{} \hat{\rho}(t_{n}) \statevectorR{\phi_{m}}{}| 
\le \sqrt{\statevectorL{\phi_{n}}{} \hat{\rho}(t_{n}) \statevectorR{\phi_{n}}{} \statevectorL{\phi_{m}}{} \hat{\rho}(t_{n}) \statevectorR{\phi_{m}}{}},
\label{eeGroup4}
\end{align}
which, on substituting Eq.\eqref{eeGroup3}, yields
\begin{align}
& 0 = \statevectorL{\phi_{n}}{} \hat{\rho}(t_{n}) \statevectorR{\phi'_{n}}{}
= \statevectorL{\phi'_{n}}{} \hat{\rho}(t_{n}) \statevectorR{\phi_{n}}{},
\nonumber\\
& ~~\forall\statevectorR{\phi_{n}}{}\,{\notin}\,\stateset{\mathrm{C}_{n}}{}
,~~\forall\statevectorR{\phi'_{n}}{}\,{\notin}\,\stateset{\mathrm{C}_{n}}{}.
\label{eeGroup5}
\end{align}
\par
Now, denoting the minimum element of $\mathcal{C}$ by $n_{1} \equiv \min{\mathcal{C}}$ and the remaining elements of ${\mathcal{C}}\,{=}\,\{n_{1},{\mathcal{C}}'\}$ by ${\mathcal{C}}'$, the left-hand side of the compatibility condition (Eq.\eqref{4-94}) can be rewritten as follows:
\begin{align}
& {\mathrm{tr}}\left[\,\hat{K}_{\mathcal{C}}\,\hat{\rho}(t)\,\hat{K}^{\dagger}_{\mathcal{C}}\,\right]
\nonumber\\
& = {\mathrm{tr}}\left[\,\hat{K}_{\mathcal{C}'}\,\zop{`$\occupation{\stateset{{\mathrm{C}}_{n_{1}}}{}}{}$'}{}\,\hat{\rho}(t)\,\zopd{`$\occupation{\stateset{{\mathrm{C}}_{n_{1}}}{}}{}$'}{}\,\hat{K}^{\dagger}_{\mathcal{C}'}\,\right],
\label{eeGroup6}\\
& = \sum_{|\phi\rangle}^{\stateset{{\mathrm{C}}_{n_{1}}}{}} \sum_{|\phi'\rangle}^{\stateset{{\mathrm{C}}_{n_{1}}}{}} 
\langle \phi | \hat{K}^{\dagger}_{\mathcal{C}'} \hat{K}_{\mathcal{C}'} | \phi' \rangle 
\langle\,\phi' | \hat{\rho}(t) | \phi \rangle.
\label{eeGroup8}
\end{align}
Extending the ranges of the sums using Eq.\eqref{eeGroup5}, we obtain
\begin{align}
{\mathrm{tr}}\Big[\,\hat{K}_{\mathcal{C}}\,\hat{\rho}(t)\,\hat{K}^{\dagger}_{\mathcal{C}}\,\Big]
& = \sum^{\stateset{{\mathfrak{E}_{n_{1}}}}{}}_{|\phi\rangle} \sum^{\stateset{{\mathfrak{E}_{n_{1}}}}{}}_{|\phi'\rangle}
\langle \phi | \hat{K}^{\dagger}_{\mathcal{C}'} \hat{K}_{\mathcal{C}'} | \phi' \rangle 
\langle \phi' | \hat{\rho}(t) | \phi \rangle
\label{eeGroup9}\\
& = {\mathrm{tr}}\Big[ \hat{K}_{\mathcal{C}'} \hat{\rho}(t) \hat{K}^{\dagger}_{\mathcal{C}'} \Big],
\label{eeGroup10}
\end{align}
where $\stateset{{\mathfrak{E}_{n_{1}}}}{}$ denotes the complete set of stationary states for time $t_{n_{1}}$.
By repeatedly using Eq.\eqref{eeGroup10}, we find that
\begin{align}
{\mathrm{tr}}\Big[ \hat{K}_{\mathcal{C}} \hat{\rho}(t) \hat{K}^{\dagger}_{\mathcal{C}} \Big]
= {\mathrm{tr}}[ \hat{\rho}(t) ] = 1,
\label{eeGroup11}
\end{align}
which shows the left-hand side of Eq.\eqref{4-94} is independent of operator ordering.
Similarly, we obtain ${\mathrm{tr}}[ \hat{K}_{\mathcal{C}} \hat{\rho}(t) \hat{K}^{\dagger}_{\mathcal{C}} ]\,{=}\,0$ irrespective of operator ordering when one of the single-time propositions is invariably false for a given density operator.

\section{Non-interacting subsystems}
\label{sSubsystems}
We define non-interacting subsystems as a system approximated by an effective Hamiltonian of the form:
\begin{align}
\hat{H}_{0} = \sum_{l=1}^{L} \hat{H}^{(l)}_{0},
\label{eeSubSystem1}
\end{align}
where $\hat{H}^{(l)}_{0}$ ($l\,{=}\,1,\ldots,L$) denotes the non-interacting Hamiltonian of the $l$-th subsystem.
For each subsystem, stationary states are defined by the eigenvalue equation
\begin{align}
\hat{H}_{0}^{(l)}\statevectorR{\phi_{l}}{l} = {E}^{(l)}(\phi_{l})\,\statevectorR{\phi_{l}}{l}
,\quad \statevectorLR{\phi_{l}}{\phi'_{l}}{l} = \delta_{\phi_{l},\phi'_{l}},
\label{eeSubSystem2}
\end{align}
where $\statevectorR{\phi_{l}}{l}$ denotes the stationary state of the $l$-th subsystem characterized by quantum number $\phi_{l}$.
The stationary states of the entire system are tensor products,
\begin{align}
|\phi\rangle = \statevectorR{\phi_{1}}{1}{\otimes}\cdots{\otimes}\statevectorR{\phi_{L}}{L}
,\quad \phi \equiv (\phi_{1}, \ldots, \phi_{L}),
\label{eeSubSystem3}
\end{align}
and the corresponding projectors have the form:
\begin{align}
\zop{$\mbox{`$\occupation{\statesingle{\phi}{}}{}$'}$}{}
& = \statevectorR{\phi}{} \statevectorL{\phi}{}
\label{eeSubSystem4}\\
\zop{$\mbox{`$\occupation{\statesingle{\phi_{l}}{l}}{}$'}$}{}
& = \statevectorR{\phi_{l}}{l} \statevectorL{\phi_{l}}{l}.
\label{eeSubSystem5}
\end{align}
Here, all projectors $\zop{$\mbox{`$\occupation{\statesingle{\phi_{l}}{l}}{}$'}$}{}$ with different $\phi_{l}$ and $l$ commute with each other, and thus all subsystem propositions `$\occupation{\statesingle{\phi_{l}}{l}}{},t$' satisfy the Compatibility assumption.
\par
From these definitions, the following equalities hold for non-interacting subsystems:
\begin{align}
\zop{`$\occupation{\statesingle{\phi}{}}{}$'}{}
& = \prod^{L}_{l=1} \zop{$\mbox{`$\occupation{\statesingle{\phi_{l}}{l}}{}$'}$}{}
= \zop{$\displaystyle \bigwedge^{L}_{l=1} \mbox{`$\occupation{\statesingle{\phi_{l}}{l}}{}$'}$}{leftright}
\label{eeSubSystem6}\\
\xop{`$\occupation{\statesingle{\phi}{}}{}$'}{}
& = \prod^{L}_{l=1} \xop{$\mbox{`$\occupation{\statesingle{\phi_{l}}{l}}{}$'}$}{}
= \xop{$\displaystyle \bigwedge^{L}_{l=1} \mbox{`$\occupation{\statesingle{\phi_{l}}{l}}{}$'}$}{leftright},
\label{eeSubSystem7}
\end{align}
which, by the Born rule and Boolean logic, lead to the equality for the expectation value:
\begin{align}
\left\langle \xvar{`$\occupation{\statesingle{\phi}{}}{}$'}{} \right\rangle
& = \left\langle \xvar{$\displaystyle \bigwedge^{L}_{l=1} \mbox{`$\occupation{\statesingle{\phi_{l}}{l}}{}$'}$}{leftright} \right\rangle
= \left\langle \prod^{L}_{l=1} \xvar{$\mbox{`$\occupation{\statesingle{\phi_{l}}{l}}{}$'}$}{} \right\rangle.
\label{eeSubSystem8}
\end{align}
Based on the Duality assumption, we assume that the corresponding relations also hold for truth values:
\begin{align}
\xjvar{`$\occupation{\statesingle{\phi}{}}{}$'}{}
& = \xjvar{$\displaystyle \bigwedge^{L}_{l=1} \mbox{`$\occupation{\statesingle{\phi_{l}}{l}}{}$'}$}{leftright}
= \prod^{L}_{l=1} \xjvar{$\mbox{`$\occupation{\statesingle{\phi_{l}}{l}}{}$'}$}{}.
\label{eeSubSystem9}
\end{align}
From Eqs.\eqref{eeSubSystem6}--\eqref{eeSubSystem7} and Eq.\eqref{eeSubSystem9}, we obtain the formal identity between propositions:
\begin{align}
\mbox{`$\occupation{\statesingle{\phi}{}}{}$'}{}\,{=}\,\bigwedge^{L}_{l=1} \mbox{`$\occupation{\statesingle{\phi_{l}}{l}}{}$'}.
\label{eeSubSystem10}
\end{align}
\par
We can now see that projectors, observables, and logical variables factor into those in subsystems.
When the density operator is not separable, however, none of these relations require the factorizability of the expectation value:
\begin{align}
\left\langle \xvar{`$\occupation{\statesingle{\phi}{}}{}$'}{} \right\rangle
= \left\langle \prod^{L}_{l=1} \xvar{$\mbox{`$\occupation{\statesingle{\phi_{l}}{l}}{}$'}$}{} \right\rangle
\neq \prod^{L}_{l=1} \left\langle \xvar{$\mbox{`$\occupation{\statesingle{\phi_{l}}{l}}{}$'}$}{} \right\rangle.
\label{eeSubSystem11}
\end{align}
In other words, subsystem propositions `$\occupation{\statesingle{\phi_{l}}{l}}{}$' may not be statistical independent even when operators are local and subsystems do not interact.
This shows that our logical variables are generally nonlocal in Bell's sense.

\section{EPR state vector}
\label{sEPRwavefunction}
This section shows the explicit forms of the wavefunctions corresponding to Eqs.\eqref{5-81}--\eqref{5-84}.
In the original EPR paper, they considered a two-particle wavefunction of the form:
\begin{align}
& \phi({\bm r}_{\mathrm{A}},{\bm r}_{\mathrm{B}})
= \phi({\bm r},{\bm R})
= \frac{1}{\sqrt{V}} \delta({\bm r}\,{-}\,{\bm r}_{0})
\label{eeAppEPR1}\\
& = \frac{1}{\sqrt{V}} \iiint \delta({\bm r}_{\mathrm{A}}\,{-}\,{\bm r}'_{\mathrm{B}} - {\bm r}_{0}) 
\delta({\bm r}_{\mathrm{B}}\,{-}\,{\bm r}'_{\mathrm{B}}) d{\bm r}'_{\mathrm{B}},
\label{eeAppEPR2}
\end{align}
where ${\bm r}$ and ${\bm R}$ denote the relative position and the center-of-mass position, respectively.
The factor $1/\sqrt{V}$ stems from the center-of-mass wavefunction $\chi({\bm R}) = e^{-{\mathrm{i}}{\bm P}\cdot{\bm R}/\hbar}/\sqrt{V}$ for ${\bm P}\,{=}\,{\bm p}_{\mathrm{A}}\,{+}\,{\bm p}_{\mathrm{B}}\,{=}\,{\bf 0}$, where ${\bm P}$ denotes the center-of-mass momentum.
In the momentum representation, this wavefunction becomes
\begin{align}
& \phi({\bm p}_{\mathrm{A}},{\bm p}_{\mathrm{B}})
= \frac{1}{\sqrt{V}} e^{-\frac{\mathrm{i}}{\hbar} {\bm p}_{\mathrm{A}}\cdot{\bm r}_{0}}
\delta({\bm p}_{\mathrm{A}}+{\bm p}_{\mathrm{B}})
\label{eeAppEPR3}\\
& = \frac{1}{\sqrt{V}} \iiint 
e^{-\frac{\mathrm{i}}{\hbar} {\bm p}'_{\mathrm{A}}\cdot{\bm r}_{0}}
\delta({\bm p}_{\mathrm{A}}-{\bm p}'_{\mathrm{A}}) \delta({\bm p}_{\mathrm{B}}+{\bm p}'_{\mathrm{A}})
d{\bm p}'_{\mathrm{A}}.
\label{eeAppEPR4}
\end{align}
These equations directly yield Eqs.\eqref{5-81}--\eqref{5-84}.

\section{Photoionization rate}
\label{sAppPhotoionization}
In this section, we present a formal derivation of the standard photoionization rate formula under approximations commonly used in quantum optics.
\par
To this end, suppose that stationary states of the photodetector material can be classified into two subsets $\stateset{E>0}{\mathrm{A}}$ and $\stateset{E<0}{\mathrm{A}}$ such that the minimum eigenenergy of the former is larger than the maximum eigenenergy of the latter.
For every stationary state $\statevectorR{\phi_{+}}{\mathrm{A}}$ in the upper manifold $\stateset{E>0}{\mathrm{A}}$, let de-excitation operators be defined by
\begin{align}
\hat{\pi}(\phi_{+}) 
\equiv \statevectorR{\mathrm{ref}_{-}}{\mathrm{A}}\statevectorL{\phi_{+}}{\mathrm{A}},
\label{ePhotoionization1}
\end{align}
where $\statevectorR{\mathrm{ref}_{-}}{\mathrm{A}}$ is a reference state taken from the lower manifold $\stateset{E<0}{\mathrm{A}}$.
Excitation operators are Hermitian conjugates of the de-excitation operators.
By Eq.\eqref{ePhotoionization1} and the definition of stationary states, all de-excitation and excitation operators commute except for the combination of $\hat{\pi}(\phi_{+})$ and $\hat{\pi}^{\dagger}(\phi'_{+})$ with $\phi_{+}\,{=}\,\phi'_{+}$.
We assume that all subsystem are initially in one of the lower states:
\begin{align}
\hat{0} = \hat{\pi}(\phi_{+}) \hat{\rho}
= \hat{\rho} \hat{\pi}^{\dagger}(\phi_{+})
\label{ePhotoionization2}
\end{align}
and that the photodetector material is only weakly excited at the time of interest:
\begin{align}
{\mathrm{tr}}\left[\hat{\pi}^{\dagger}(\phi_{+})\hat{\pi}(\phi_{+}) \hat{\rho}(t)\right] 
= \statevectorL{\phi_{+}}{\mathrm{A}}\hat{\rho}(t)\statevectorR{\phi_{+}}{\mathrm{A}}
\ll 1.
\label{ePhotoionization3}
\end{align}
Using the definition of proposition `ionized,$t$' (Eqs.\eqref{5-23}--\eqref{5-24}), we obtain the following expressions for the photoionization probability:
\begin{align}
& \langle \xvar{`ionized,$t$'}{} \rangle
\nonumber\\
& = {\rm{tr}}\left[ \zop{`ionized,$t$'}{}\,\hat{\rho}\,\zopd{`ionized,$t$'}{} \right]
\label{ePhotoionization21}\\
& = {\rm{tr}}\left[\,\left( \sum_{\statevectorR{{\phi_{+}}}{\mathrm{A}}}^{\stateset{E>0}{\mathrm{A}}} \statevectorR{{\phi_{+}}}{\mathrm{A}} \statevectorL{{\phi_{+}}}{\mathrm{A}} \right) \hat{\rho}\,\right]
\label{ePhotoionization23}\\
& = {\rm{tr}}\left[\,\left( \sum_{\statevectorR{{\phi_{+}}}{\mathrm{A}}}^{\stateset{E>0}{\mathrm{A}}} \hat{\pi}^{\dagger}(\phi_{+},t)\,\hat{\pi}(\phi_{+},t) \right) \hat{\rho}\,\right],
\label{ePhotoionization24}
\end{align}
which is the starting point for the derivation of standard photoionization rate formula~\cite{Mandel}.
\par
To evaluate Eq.\eqref{ePhotoionization24}, we use the first-order perturbation approximation:
\begin{align}
\hat{U}(t)
& = \hat{U}_{0}(t) + \frac{1}{{\rm{i}}\hbar}\int_{0}^{t} \!\!\hat{U}_{0}(t-t')\hat{H}_{\mathrm{I}}\hat{U}(t')dt'
\label{ePhotoionization4}\\
& \cong \hat{U}_{0}(t) + \hat{U}_{1}(t),
\label{ePhotoionization5}
\end{align}
where the zero-th and first order contributions take the form:
\begin{align}
& \hat{U}_{0}(t) \equiv e^{-\frac{\mathrm{i}}{\hbar}\hat{H}_{0}t}
,~~ \hat{U}_{1}(t) \equiv \hat{U}_{0}(t)\,\frac{1}{{\rm{i}}\hbar}\int_{0}^{t} \!\!\tilde{H}_{\mathrm{I}}(t')dt'.
\label{ePhotoionization6}
\end{align}
The Heisenberg-picture and interaction-picture operators are defined as follows:
\begin{align}
\hat{O}(t) 
& \equiv \hat{U}^{\dagger}(t) \hat{O} \hat{U}(t)
\label{ePhotoionization7}\\
\tilde{O}(t) 
& \equiv \hat{U}_{0}^{\dagger}(t) \hat{O} \hat{U}_{0}(t).
\label{ePhotoionization8}
\end{align}
Applying the long wavelength approximation and neglecting the non-linear term with respect to $\tilde{\bm A}_{\mathrm{T}}$, the interaction Hamiltonian $\tilde{H}_{\mathrm{I}}(t)$ in Eq.\eqref{ePhotoionization6} takes the form:
\begin{align}
\tilde{H}_{\mathrm{I}}(t)
\cong - \sum^{K}_{k} \frac{e_{k}}{m_{k}} {\tilde{\bm p}}_{k}(t) \cdot \tilde{\bm A}_{\mathrm{T}}({\bm r}_{0},t),
\label{ePhotoionization9}
\end{align}
where ${\bm r}_{0}$ represents the location of the photoionization site.
\par
Eqs.\eqref{ePhotoionization5}--\eqref{ePhotoionization9} cast the Heisenberg-picture de-excitation operator into the form:
\begin{align}
& \hat{\pi}(\phi_{+},t)
\nonumber\\
& \cong
\tilde{\pi}(\phi_{+}, t)
+ \tilde{\pi}(\phi_{+}, t)\left(\frac{1}{{\rm{i}}\hbar}\int_{0}^{t} \!\!\tilde{H}_{\mathrm{I}}(t') dt'\right)
\nonumber\\
& ~~~~~~~~~~~~~~ - \left(\frac{1}{{\rm{i}}\hbar}\int_{0}^{t} \!\!\tilde{H}_{\mathrm{I}}(t') dt'\right)\tilde{\pi}(\phi_{+},t)
\label{ePhotoionization11}\\
& = \tilde{\pi}(\phi_{+}, t) + \int_{0}^{t} \tilde{\bm h}(\phi_{+},t,t')\cdot{\tilde{\bm A}_{\mathrm{T}}}({\bm r}_{0},t')dt',
\label{ePhotoionization12}
\end{align}
where the homogeneous term $\tilde{\pi}(\phi_{+}, t)$ is defined by
\begin{align}
\tilde{\pi}(\phi_{+},t)
\equiv \hat{\pi}(\phi_{+}) e^{-{\rm{i}}\omega_{\phi_{+}}t},
\label{ePhotoionization13}
\end{align}
with $\omega_{\phi_{+}}\,{\equiv}\,(E(\phi_{+})-E({\mathrm{ref}}_{-}))/{\hbar}$ as the transition angular frequency.
The coefficient $\tilde{\bm h}(\phi_{+},t,t')$ has the form:
\begin{align}
\tilde{\bm h}(\phi_{+},t,t')
& \equiv \left[\frac{1}{{\rm{i}}\hbar}\sum^{K}_{k=1} \frac{e_{k}}{m_{k}}\tilde{\bm p}_{k}(t'), \tilde{\pi}(\phi_{+},t)\right]_{-}
\label{ePhotoionization14}\\
& \cong - \sum_{\statevectorR{{\phi_{-}}}{\mathrm{A}}}^{\stateset{E<0}{\mathrm{A}}} \left(E(\phi_{+})-E(\phi_{-})\right) \statevectorL{{\phi_{+}}}{\mathrm{A}}\hat{\bm D}\statevectorR{{\phi_{-}}}{\mathrm{A}}
\nonumber\\
& ~~~~  e^{{\rm{i}}(\omega_{\phi_{+}}-\omega_{\phi_{-}})t'} e^{-{\rm{i}}\omega_{\phi_{+}}t}
\,\statevectorR{{\mathrm{ref}}_{-}}{\mathrm{A}} \statevectorL{{\phi_{-}}}{\mathrm{A}},
\label{ePhotoionization15}
\end{align}
where $\hat{\bm D}\,{\equiv}\,\sum^{K}_{k=1} e_{k}\hat{\bm r}_{k}$ is the dipole moment operator.
To obtain Eq.\eqref{ePhotoionization15}, we used the well-known identity
\begin{align}
\frac{1}{{\rm{i}}\hbar}\sum^{K}_{k=1} \frac{e_{k}}{m_{k}}\hat{\bm p}_{k}
= \left[ \hat{H}_{0}, \hat{\bm D} \right]_{-}
\label{ePhotoionization16}
\end{align}
and dropped diagonal and near-diagonal matrix elements of $\hat{\bm D}$, which are non-dominant.
We also neglected the population of the upper manifold against the lower manifold.
These approximations are not specific to our formulation, but approximations for Heisenberg operators (like Eq.\eqref{ePhotoionization12}) are valid only when evaluating probabilities for a specific density operator.
\par
Below, we summarize the standard derivation~\cite{Mandel}[{\S}14.2] to confirm that our assumptions do not conflict with the standard calculation procedure.
Substituting Eq.\eqref{ePhotoionization12} into Eq.\eqref{ePhotoionization24} and differentiating with respect to time, we find that the photoionization rate has the form:
\begin{align}
& \frac{d}{dt}\langle{x}(\mbox{`ionized',\,$t$})\rangle
= \frac{d}{dt}{\rm{tr}}\Bigg[\Bigg(\int_{0}^{t} dt_{1} \int_{0}^{t} dt_{2} \sum_{\xi}^{x,y,z} \sum_{\xi'}^{x,y,z} 
\nonumber\\
& ~~~~~~~~ \hat{\zeta}_{\xi \xi'}(t_{1}-t_{2}) {\tilde{A}_{{\mathrm{T}}\xi}}({\bm r}_{0},t_{2}){\tilde{A}_{{\mathrm{T}}\xi'}}({\bm r}_{0},t_{1})\Bigg) \hat{\rho}\,\Bigg],
\label{ePhotoionization25}
\end{align}
where used Eq.\eqref{ePhotoionization2} to drop the homogeneous term $\tilde{\pi}(\phi_{+},t)$.
We focus on a quasi-monochromatic light centered at an angular frequency $\omega_{\mathrm{o}}$ and assume that an exposure time $t$ is sufficiently longer than the optical period $2\pi/\omega_{\mathrm{o}}$.
To distinguish between terms containing annihilation and creation operators, we divide the field operator into its positive and negative oscillating parts:
\begin{align}
{\tilde{A}}_{\mathrm{T}\xi}({\bm r}_{0},t) = {\tilde{A}}_{\mathrm{T}\xi}^{(+)}({\bm r}_{0},t) + {\tilde{A}}_{\mathrm{T}\xi}^{(-)}({\bm r}_{0},t).
\label{ePhotoionization26}
\end{align}
Substituting Eq.\eqref{ePhotoionization26} into Eq.\eqref{ePhotoionization25}, the product ${\tilde{A}_{{\mathrm{T}}\xi}}({\bm r}_{0},t_{2}){\tilde{A}_{{\mathrm{T}}\xi'}}({\bm r}_{0},t_{1})$ yields the sum of a commutation relation and a normally-ordered product:
\begin{align}
& :{\tilde{A}_{{\mathrm{T}}\xi}}({\bm r}_{0},t_{2}){\tilde{A}_{{\mathrm{T}}\xi'}}({\bm r}_{0},t_{1}):
\label{ePhotoionization28}\\
& \equiv 
{\tilde{A}^{(+)}_{{\mathrm{T}}\xi}}({\bm r}_{0},t_{2}){\tilde{A}^{(+)}_{{\mathrm{T}}\xi'}}({\bm r}_{0},t_{1})
+ {\tilde{A}^{(-)}_{{\mathrm{T}}\xi}}({\bm r}_{0},t_{2}){\tilde{A}^{(-)}_{{\mathrm{T}}\xi'}}({\bm r}_{0},t_{1})
\nonumber\\
& + {\tilde{A}^{(-)}_{{\mathrm{T}}\xi}}({\bm r}_{0},t_{2}){\tilde{A}^{(+)}_{{\mathrm{T}}\xi'}}({\bm r}_{0},t_{1})
+ {\tilde{A}^{(+)}_{{\mathrm{T}}\xi}}({\bm r}_{0},t_{2}){\tilde{A}^{(-)}_{{\mathrm{T}}\xi'}}({\bm r}_{0},t_{1}).
\nonumber
\end{align}
Using the explicit form of $\hat{\zeta}_{\xi\xi'}(t_{1}-t_{2})$, we find that the contribution of the commutation relation---so-called ``vacuum contribution to the light detection probability''---vanishes for $t\,{\gg}\,2\pi/\omega_{\mathrm{o}}$.
Dropping the rapidly oscillating terms at double the optical frequency, we obtain
\begin{align}
& \frac{d}{dt}\langle{x}(\mbox{`ionized',\,$t$})\rangle
\label{ePhotoionization31}\\
& = {\rm{tr}}\left[\left(\sum_{\xi}^{x,y,z} \sum_{\xi'}^{x,y,z} \hat{\kappa}_{\xi \xi'}(\omega_{\mathrm{o}})
{\tilde{A}_{{\mathrm{T}}\xi}}^{(-)}({\bm r}_{0},t){\tilde{A}_{{\mathrm{T}}\xi'}}^{(+)}({\bm r}_{0},t)\right) \hat{\rho}\,\right],
\nonumber
\end{align}
where we defined the operator coefficient $\hat{\kappa}_{\xi \xi'}(\omega_{\mathrm{o}})$ by
\begin{align}
& \hat{\kappa}_{\xi \xi'}(\omega_{\mathrm{o}})
\nonumber\\
& \cong 
\sum_{\statevectorR{\phi_{+}}{\mathrm{A}}}^{\stateset{E>0}{\mathrm{A}}} 
\sum_{\statevectorR{\phi_{-}}{\mathrm{A}}}^{\stateset{E<0}{\mathrm{A}}} 
2\pi\,\delta\left(\omega_{\phi_{+}}-\omega_{\phi_{-}}-\omega_{\mathrm{o}}\right)
\left(E(\phi_{+}) - E(\phi_{-})\right)^{2} 
\nonumber\\
& ~~~~
\statevectorL{\phi_{-}}{\mathrm{A}}\hat{D}_{\xi}\statevectorR{\phi_{+}}{\mathrm{A}}
\statevectorL{\phi_{+}}{\mathrm{A}}\hat{D}_{\xi'}\statevectorR{\phi_{-}}{\mathrm{A}}
\,\statevectorR{\phi_{-}}{\mathrm{A}}\statevectorL{\phi_{-}}{\mathrm{A}}.
\label{ePhotoionization32}
\end{align}
The combination of Eqs.\eqref{ePhotoionization31}--\eqref{ePhotoionization32} agrees with Eq.(14.2-24) of Mandel\&Wolf, except that the double sum is replaced by an energy integral weighted by the atomic density of states.

\section{Photoelectric effect of hydrogen atoms}
\label{sHydrogen}
As a more concrete example, this section examines photoionization rate calculations for the photoelectric effect in hydrogen atoms~\cite{Loudon,Bethe}.
For hydrogen atoms, the non-interacting Hamiltonian $\hat{H}_{0}$ takes the form:
\begin{align}
\hat{H}_{0} = \hat{H}_{{\bm r}\mathrm{S}} + \hat{H}_{\bm R} + \hat{H}_{\mathrm{F}},
\label{eHydrogen1}
\end{align}
where the Hamiltonians for relative motion (${\bm r}$), Zeeman interaction ($\mathrm{S}$), center-of-mass motion ($\bm R$), and radiation field ($\mathrm{F}$) are respectively given by
\begin{align}
\hat{H}_{{\bm r}\mathrm{S}}
& \equiv \frac{1}{2m}\|{\hat{\bm p}}\|^{2}
- \frac{e^{2}}{4\pi\varepsilon_{0}}\frac{1}{\left\|\hat{\bm r}\right\|}
\nonumber\\
& ~~~~ + \frac{g_{\mathrm{p}}e}{2m_{\mathrm{p}}}{\hat{\bm s}}_{\mathrm{p}}\cdot\hat{\bm B}(\hat{\bm r}_{\mathrm{p}}) - \frac{g_{\mathrm{e}}e}{2m_{\mathrm{e}}}{\hat{\bm s}}_{\mathrm{e}}\cdot\hat{\bm B}(\hat{\bm r}_{\mathrm{e}})
\label{eHydrogen2}\\
\hat{H}_{\bm R}
& \equiv \frac{1}{2M}\|{\hat{\bm P}}\|^{2}
\label{eHydrogen4}\\
\hat{H}_{\mathrm{F}} 
& \equiv \iiint\frac{\varepsilon_{0}}{2}\left[\|\hat{\bm E}_{\mathrm{T}}({\bm x})\|^{2}+{c^{2}}\|\hat{\bm B}({\bm x})\|^{2}\right]d{\bm x}.
\label{eHydrogen5}
\end{align}
$\hat{\bm r}_{k}$, $\hat{\bm p}_{k}$, and ${\hat{\bm s}}_{k}$ ($k\,{=}\,\mathrm{e}, \mathrm{p}$) denote the position, momentum, and spin of the electron and proton, respectively.
The stationary states of the entire system have the form:
\begin{align}
|\phi\rangle
= |\phi\rangle^{\!{\bm r}\mathrm{S}} \otimes |\phi\rangle^{\!\bf R} \otimes \prod_{{\bm k}\lambda} |\phi\rangle^{\!{\bm k}\lambda},
\label{eHydrogen6}
\end{align}
where $|\phi\rangle^{\!{\bm r}\mathrm{S}} $, $|\phi\rangle^{\!\bf R}$, and $|\phi\rangle^{\!{\bm k}\lambda}$ denote the orthonormalized eigenstates of $\hat{H}_{{\bm r}\mathrm{S}}$, $\hat{H}_{\bm R}$, and $\hat{H}_{\mathrm{F}}$, respectively.
${\bm k}\,{\in}\,{\mathbb{R}^{3}}$ and $\lambda\,{\in}\,\{1,2\}$ denote the wavenumber and polarization of each field mode.
\par
We now find that Eqs.\eqref{eHydrogen1}--\eqref{eHydrogen5} define non-interacting subsystems.
Moreover, the eigenenergies of scattering states are almost independent of spin directions when external magnetic fields are not applied, although the Zeeman interaction in Eq.\eqref{eHydrogen2} leads to the fine structure of bound states.
Hence, the atomic system proposition `$\occupation{\stateset{E>0}{\mathrm{A}}}{}$' describing the occupation of scattering states can be expressed as
\begin{align}
\mbox{`$\occupation{\stateset{E>0}{\mathrm{A}}}{}$'}
= \mbox{`$\occupation{\stateset{E>0}{{\bm r}}}{}$'},
\label{eHydrogen11}
\end{align}
where the scattering states for the relative motion are defined by
\begin{align}
\stateset{E>0}{{\bm r}}
\equiv & \left\{\,\statevectorR{{\bm p}}{{\bm r}}\,\Big|\,
\hat{H}_{\bm r} \statevectorR{{\bm p}}{{\bm r}} = E^{\bm r}({\bm p}) \statevectorR{{\bm p}}{{\bm r}}
,\,\statevectorLR{{\bm p}}{{\bm p}'}{{\bm r}} = \delta_{{\bm p},{\bm p}'},
\right.
\nonumber\\
& ~~~~~~~~~~ \left. E^{\bm r}({\bm p}) \ge 0 \phantom{\Big|} \right\}.
\label{eHydrogen12}
\end{align}
In this expression, each ${\bm p}$ denotes a relative momentum of the incoming plane wave that generates outgoing partial waves ($\approx 1/\|{\bm r}\|$).
\par
Substituting Eq.\eqref{eHydrogen12} into Eq.\eqref{ePhotoionization23} and taking the partial trace, we obtain
\begin{align}
\frac{d}{dt}\big\langle \xvar{`$\mathrm{ionized},t$'}{} \big\rangle
= \frac{d}{dt}\iiint \statevectorL{{\bm p}}{\bm r} \hat{\rho}_{\bm r}(t) \statevectorR{{\bm p}}{\bm r}\,d{\bm p},
\label{eHydrogen13}
\end{align}
where the reduced density operator is defined by
\begin{align}
& \hat{\rho}_{\bm r}(t)
\equiv {\mathrm{tr}}_{\overline{\bm r}} \left[ \hat{U}(t) \hat{\rho} \hat{U}^{\dagger}(t) \right].
\label{eHydrogen14}
\end{align}
with ${\mathrm{tr}}_{\overline{\bm r}}$ denoting the partial trace over all degrees of freedom except for the relative motion.
When the initial state is separable and pure, the reduced density operator has a relatively simple form in the first-order perturbation approximation.
Consider the simplest case where the atom is initially in the 1s state $\statevectorR{\mbox{1s}}{{\bm r}}$ and the field is initially in the vacuum state, except for a particular mode ${\bm k}\lambda$.
Then the initial density operator has the form:
\begin{align}
\hat{\rho}
= \statevectorR{\mbox{1s}}{{\bm r}}\statevectorL{\mbox{1s}}{{\bm r}} \otimes 
\statevectorR{\mbox{rad}}{{\bm k}\lambda}\statevectorL{\mbox{rad}}{{\bm k}\lambda} \otimes
\hat{\rho}_{\mathrm{r.o.w.}},
\label{eHydrogen21}
\end{align}
where $\statevectorR{\mbox{rad}}{{\bm k}\lambda}$ denotes the initial state of the ${\bm k}\lambda$-mode, and $\hat{\rho}_{\mathrm{r.o.w.}}$ represents the density operator of the rest of the world. 
Substituting Eq.\eqref{eHydrogen21} and Eq.\eqref{ePhotoionization5} into Eq.\eqref{eHydrogen14} and tracing over the field, we obtain
\begin{align}
\hat{\rho}_{\bm r}(t)
& \cong \statevectorR{\psi(t)}{\bm r} \statevectorL{\psi(t)}{\bm r}
\label{eHydrogen22}\\
\statevectorR{\psi(t)}{\bm r}
& \equiv \hat{U}_{\bm r}(t)|\mathrm{1s}\rangle,
\label{eHydrogen23}
\end{align}
where the reduced time evolution operator $\hat{U}_{\bm r}(t)$ for the $n\,{=}\,1$ Fock state $\statevectorR{\mbox{1}}{{\bm k}\lambda}$ is given by
\begin{align}
\hat{U}_{\bm r}(t) 
= {\mathrm{tr}}_{\overline{{\bm r}{\bm k}\lambda}}
\left[\,\statevectorL{1}{{{\bm k}\lambda}} \hat{U}_{0}(t) \statevectorR{1}{{{\bm k}\lambda}} + \statevectorL{0}{{{\bm k}\lambda}} \hat{U}_{1}(t) \statevectorR{1}{{{\bm k}\lambda}}\,\,\right],
\label{eHydrogen24}
\end{align}
while for the coherent state $\statevectorR{\tilde{\alpha}}{{{\bm k}\lambda}}$, it has the form:
\begin{align}
\hat{U}_{\bm r}(t) 
= {\mathrm{tr}}_{\overline{{\bm r}{\bm k}\lambda}}\left[\,\statevectorL{\tilde{\alpha}}{{{\bm k}\lambda}} \hat{U}_{0}(t) + \hat{U}_{1}(t) \statevectorR{\tilde{\alpha}}{{{\bm k}\lambda}}\,\right].
\label{eHydrogen25}
\end{align}
To obtain these expressions, we employed similar approximations, as shown in Section \ref{sAppPhotoionization}.
Notably, the time evolution is semiclassical, since all field operators are replaced by their matrix elements.
\par
When the angular optical frequency $\omega_{\mathrm{o}}\,{\equiv}\,c\|{\bm k}\|$ is sufficiently large so that $\omega_{\mathrm{o}}\,{>}\,|E({\mathrm{1s}})|{/}\hbar$, the state vector $\statevectorR{\psi(t)}{\bm r}$ in Eq.\eqref{eHydrogen23} becomes a coherent superposition of the 1s state and scattering states:
\begin{align}
\statevectorR{\psi(t)}{\bm r}
= c_{\mathrm{1s}}(t) \statevectorR{\mbox{1s}}{\bm r} + \iiint c_{\bm p}(t) \statevectorR{{\bm p}}{\bm r} d{\bm p},
\label{eHydrogen30}
\end{align}
where the c-number coefficients are given by
\begin{align}
c_{\mathrm{1s}}(t)
& = \statevectorL{\mbox{1s}}{\bm r} \hat{U}_{\bm r}(t) \statevectorR{\mbox{1s}}{\bm r}
\label{eHydrogen31}\\
c_{\bm p}(t)
& = \statevectorL{{\bm p}}{\bm r} \hat{U}_{\bm r}(t) \statevectorR{\mbox{1s}}{\bm r}.
\label{eHydrogen32}
\end{align}
Substituting Eq.\eqref{eHydrogen22} and Eq.\eqref{eHydrogen30} into Eq.\eqref{eHydrogen13}, we obtain the following expression for the photoionization rate:
\begin{align}
& \frac{d}{dt}\left\langle \xvar{`$\mathrm{ionized},t$'}{} \right\rangle
\nonumber\\
& = \lim_{\Delta{t}{\to}0} \iiint \frac{1}{\Delta{t}} \left(|c_{\bm p}(t + \Delta{t})|^{2}- |c_{\bm p}(t)|^{2}\right)\,d{\bm p}.
\label{eHydrogen34}
\end{align}
Evaluating $|c_{\bm p}(t + \Delta{t})|^{2}-|c_{\bm p}(t)|^{2}$ using Eq.\eqref{eHydrogen32}, we obtain
\begin{align}
& \frac{d}{dt}\left\langle \xvar{`$\mathrm{ionized},t$'}{} \right\rangle
\nonumber\\
& \approx \mbox{const.} \iiint \frac{\cos^{2}({\bm e}_{\lambda},{\bm p})}{\nu_{o}^{7/2}}
\,\delta\left( \frac{\|{\bm p}\|^{2}}{2m} - E_{\mathrm{1s}} - h\nu_{o} \right) d{\bm p}
\nonumber\\
& ~~~~ \times |c_{\mathrm{1s}}(t)|^{2} \times \mbox{(light power)},
\label{eHydrogen35}
\end{align}
where ${\bm e}_{\lambda}$ is the unit vector indicating the linear polarization vector, and $({\bm e}_{\lambda},{\bm p})$ denotes the angle between ${\bm e}_{\lambda}$ and ${\bm p}$.
Note that Eq.\eqref{eHydrogen35} leads to Einstein's relation for photoelectric effects:
\begin{align}
h\nu_{o} + E_{\mathrm{1s}} = \frac{\|{\bm p}\|^{2}}{2m}.
\label{eHydrogen35a}
\end{align}
A more specific derivation is found in Chap.5 of Loudon's textbook~\cite{Loudon}.
\par
If we take advantages of $|c_{\mathrm{1s}}(t)|^{2} + \iiint |c_{\bm p}(t)|^{2}\,d{\bm p} = 1$, Eq.\eqref{eHydrogen34} also yields
\begin{align}
\frac{d}{dt}\left\langle \xvar{`$\mathrm{ionized},t$'}{} \right\rangle
= - \frac{d}{dt}|c_{\mathrm{1s}}(t)|^{2}.
\label{eHydrogen36}
\end{align}
By comparing Eq.\eqref{eHydrogen35} with Eq.\eqref{eHydrogen36}, we finally obtain the exponential decay of the probability that the atom has not been photoionized:
\begin{align}
& \big\langle \xvar{$\lnot$`$\mathrm{ionized},t$'}{} \big\rangle
= e^{-\gamma{t}},
\label{eHydrogen37}
\end{align}
where the decay constant is given by
\begin{align}
\gamma \approx 
& \mathrm{\,const.} \iiint \frac{\cos^{2}({\bm p},{\bm e}_{\lambda})}{\nu_{o}^{7/2}}\,\delta\left( \frac{\|{\bm p}\|^{2}}{2m} + \Big|E_{\mathrm{1s}}\Big| - h\nu_{o} \right) d{\bm p}
\nonumber\\
& \times \mbox{(light power)}.
\label{eHydrogen38}
\end{align}

\section{Derivation of a useful relation}
\label{sAppUseful}
In this section, we derive a useful relation for the expectation values of normally-ordered products of de-excitation and excitation operators.
This relation plays a key role in deriving joint detection probabilities for multiple photodetectors (Eq.\eqref{5-142}).
\par
\paragraph*{Statement:}
Consider a system consisting of $I$ non-interacting subsystems and classify each subsystem' stationary states into two sets $\stateset{E>0}{i}$ and $\stateset{E<0}{i}$ ($i\,{=}\,1,\ldots,I$), according to the sign of eigenenergy $E$ (see Section \ref{sAppPhotoionization}).
For every upper state $\statevectorR{\phi}{i}$ in $\stateset{E>0}{i}$, let de-excitation operator $\hat{\pi}_{i}(\phi)$ and excitation operator $\hat{\pi}^{\dagger}_{i}(\phi)$ be defined by
\begin{align}
\hat{\pi}_{i}(\phi) \equiv \statevectorR{\mathrm{ref}_{i}}{i}\statevectorL{\phi}{i}
,~~ \hat{\pi}^{\dagger}_{i}(\phi) \equiv \statevectorR{\phi}{i}\statevectorL{\mathrm{ref}_{i}}{i},
\label{eUseful1}
\end{align}
where $\statevectorR{\mathrm{ref}_{i}}{i}$ is a reference state taken from $\stateset{E<0}{i}$.
Due to the definition of stationary states, these operators commute except for the combination of $\hat{\pi}_{i}(\phi)$ and $\hat{\pi}^{\dagger}_{i'}(\phi')$ for $i\,{=}\,i'$ and $\phi\,{=}\,\phi'$.
We assume that all subsystem are initially in one of the lower states $\stateset{E<0}{i}$, which requires
\begin{align}
0 = \hat{\pi}_{i}(\phi) \hat{\rho}
= \hat{\rho} \hat{\pi}_{i}^{\dagger}(\phi)
,~~ \forall \phi \in \stateset{E>0}{i}.
\label{eUseful2}
\end{align}
\par
Now, consider evaluating a trace of the form:
\begin{align}
{\mathcal{Z}}_{N} 
& \equiv {\mathrm{tr}}\left[ \left({\mathcal{T}}\prod_{n=1}^{N}\zop{`$\occupation{{\mathrm{C}}_{n}}{},t_{n}$'}{leftright}\right) \hat{\rho} \left({\mathcal{T}}\prod_{n=1}^{N}\zop{`$\occupation{{\mathrm{C}}_{n}}{},t_{n}$'}{leftright}\right)^{\dagger}\right],
\label{eUseful3}
\end{align}
where the projectors are given by
\begin{align}
\zop{`$\occupation{{\mathrm{C}}_{n}}{},t_{n}$'}{leftright}
= \sum_{\statevectorR{\phi}{i_{n}}}^{\stateset{E>0}{i_{n}}} \hat{\pi}_{i_{n}}^{\dagger}(\phi,t_{n})\,\hat{\pi}_{i_{n}}(\phi,t_{n}),
\label{eUseful4}
\end{align}
and $i_{n}$ denotes the index of a subsystem described by `$\occupation{{\mathrm{C}}_{n}}{},t_{n}$'.
Under the first-order perturbation approximation, the de-excitation operators take the form:
\begin{align}
\hat{\pi}_{i}(\phi,t)
\cong \tilde{\pi}_{i}(\phi, t) + \int_{0}^{t} \tilde{\bm h}_{i}(\phi,t,t')\cdot{\tilde{\bm A}_{\mathrm{T}}}({\bm r}_{i},t')dt',
\label{eUseful5}
\end{align}
where we neglected the non-linear interaction and used the long-wavelength approximation.
The homogeneous term and the operator coefficient are defined by
\begin{align}
& \tilde{\pi}_{i}(\phi,t)
= \hat{\pi}_{i}(\phi) e^{-{\rm{i}}\frac{E_{i}(\phi)-E_{i}({\mathrm{ref}}_{-})}{\hbar}t}
\label{eUseful6}\\
& \tilde{\bm h}_{i}(\phi,t,t')
\equiv \left[(\hat{H}_{0,i} \hat{\bm D}_{i} - \hat{\bm D}_{i} \hat{H}_{0,i}), \tilde{\pi}_{i}(\phi,t)\right]_{-},
\label{eUseful7}
\end{align}
where $\hat{H}_{0,i}$ and $\hat{\bm D}_{i}$ are the non-interacting Hamiltonian and the dipole moment operator for the $i$-th subsystem, respectively (see Eq.\eqref{ePhotoionization16}).
${\bm r}_{i}$ represents the location of the $i$-th subsystem.
\par
Under these conditions, Eq.\eqref{eUseful5} can be replaced by the following simpler expression when evaluating the trace in Eq.\eqref{eUseful3}:
\begin{align}
\hat{\pi}_{i}(\phi, t) 
= \int_{0}^{t} \tilde{\bm h}(\phi,t,t')\cdot{\tilde{\bm A}^{(+)}_{\mathrm{T}}}({\bm r}_{i},t')dt'.
\label{eUseful8}
\end{align}

\paragraph*{Note:}
This expression is applicable to most practical photodetector analyses, but not necessarily for two-level atoms where the first-order perturbation approximation fails when evaluating the long-time behavior.
Below, we suppress the time-ordering symbol in Eq.\eqref{eUseful3} by assuming $t_{1}\,{\le}\,t_{2}\,{\le}\,{\cdots}\,{\le}\,t_{N}$.

\paragraph*{Proof:}
To show that the homogeneous term $\tilde{\pi}_{i}(\phi, t)$ in Eq.\eqref{eUseful5} does not contribute to ${\mathcal{Z}}_{N}$, consider an operator $\hat{\Theta}_{1}, \ldots, \hat{\Theta}_{N}$ defined by the asymptotic equation,
\begin{align}
\hat{\Theta}_{n+1} = \zop{`$\occupation{\stateset{\mathrm{C}_{n}}{}}{},t_{n}$'}{}\,\hat{\Theta}_{n}\,\zopd{`$\occupation{\stateset{\mathrm{C}_{n}}{}}{},t_{n}$'}{}
,~~ \hat{\Theta}_{1}\equiv\hat{\rho},
\label{eUseful11}
\end{align}
and let Eq.\eqref{eUseful3} be expressed as
\begin{align}
{\mathcal{Z}}_{N} = {\mathrm{tr}}\left[\hat{\Theta}_{N}\right].
\label{eUseful12}
\end{align}
Substituting Eq.\eqref{eUseful4} into Eq.\eqref{eUseful11}, we have
\begin{align}
\hat{\Theta}_{n+1} 
& = \sum_{\statevectorR{\phi_{n}}{i_{n}}}^{\stateset{E>0}{i_{n}}} \sum_{\statevectorR{\phi'_{n}}{i_{n}}}^{\stateset{E>0}{i_{n}}} 
\hat{\pi}_{i_{n}}^{\dagger}(\phi_{n},t_{n})\,\hat{\pi}_{i_{n}}(\phi_{n},t_{n})\,\hat{\Theta}_{n}\,
\nonumber\\
& ~~~~~~~~~~~~~~~~~~~~~~~~~ \hat{\pi}_{i_{n}}^{\dagger}(\phi'_{n},t_{n})\,\hat{\pi}_{i_{n}}(\phi'_{n},t_{n}),
\label{eUseful13}
\end{align}
while Eq.\eqref{eUseful1} yields
\begin{align}
\hat{\pi}_{i_{n}}(\phi'_{n},t_{n}) \hat{\pi}_{i_{n}}^{\dagger}(\phi_{n},t_{n})
= \delta_{\phi'_{n},\phi_{n}} \statevectorR{\mathrm{ref}_{i_{n}}}{i_{n}}\statevectorL{\mathrm{ref}_{i_{n}}}{i_{n}}.
\label{eUseful14}
\end{align}
Substituting Eq.\eqref{eUseful13} and Eq.\eqref{eUseful14} into Eq.\eqref{eUseful12} and using the cyclic relation ${\mathrm{tr}}[\hat{A}\hat{B}\hat{C}]\,{=}\,{\mathrm{tr}}[\hat{C}\hat{A}\hat{B}]$, for $n=N-1,\ldots,1$, we obtain
\begin{align}
\hat{\Theta}_{n+1}
= \sum_{\statevectorR{\phi_{n}}{i_{n}}}^{\stateset{E>0}{i_{n}}}
\hat{\pi}_{i_{n}}(\phi_{n},t_{n})\,\hat{\Theta}_{n}\,\hat{\pi}_{i_{n}}^{\dagger}(\phi_{n},t_{n}),
\label{eUseful15}
\end{align}
which further leads to
\begin{align}
\hat{\Theta}_{N}
& = \sum_{\statevectorR{\phi}{i_{N}}}^{\stateset{E>0}{i_{N}}} \cdots \sum_{\statevectorR{\phi}{i_{1}}}^{\stateset{E>0}{i_{1}}} 
 \hat{\pi}_{i_{N}}(\phi_{N},t_{N}) \cdots \hat{\pi}_{i_{1}}(\phi_{1},t_{1})\,\hat{\rho}\,
\nonumber\\
& ~~~~~~ \hat{\pi}_{i_{1}}^{\dagger}(\phi_{1},t_{1}) \cdots \hat{\pi}_{i_{N}}^{\dagger}(\phi_{N},t_{N}).
\label{eUseful16}
\end{align}
Substituting Eq.\eqref{eUseful5} into this equation and using Eq.\eqref{eUseful2} and Eq.\eqref{eUseful6}, we find that the first term of Eq.\eqref{eUseful5} does not contribute to ${\mathcal{Z}}_{N}$.
\par
Comparing Eq.\eqref{eUseful16} with Eq.\eqref{eUseful1}, we find that $\hat{\Theta}_{N}$ will vanish if different propositions refer to the same subsystem (i.e., if $i_{n}\,{=}\,i_{n'}$ for $n\,{\neq}\,n'$).
Otherwise, ${\mathcal{Z}}_{N}$ takes the form:
\begin{align}
{\mathcal{Z}}_{N}
& \cong {\mathrm{tr}}\Bigg[\,\sum_{\xi'_{N}}^{\{x,y,z\}} \sum_{\xi''_{N}}^{\{x,y,z\}} \cdots \sum_{\xi'_{1}}^{\{x,y,z\}} \sum_{\xi''_{1}}^{\{x,y,z\}} 
\nonumber\\
& ~~~~~~~~~~~
\int_{0}^{t_{N}} dt'_{N} \int_{0}^{t_{N}} dt''_{N} \cdots \int_{0}^{t_{1}} dt'_{1} \int_{0}^{t_{1}} dt''_{1} 
\nonumber\\
& ~~~~~~~~~~~~~
\hat{\zeta}_{\xi'_{N} \xi''_{N}}(t'_{N}-t''_{N}) \cdots \hat{\zeta}_{\xi'_{1} \xi''_{1}}(t'_{1}-t''_{1}) 
\nonumber\\
& ~~~~~~~~~~~
\left({\tilde{A}}_{{\mathrm{T}}\xi''_{1}}({\bm r}_{1},t''_{1}) \cdots {\tilde{A}}_{{\mathrm{T}}\xi''_{N}}({\bm r}_{N},t''_{N}) \right)
\nonumber\\
& ~~~~~~~~~~~ \left({\tilde{A}}_{{\mathrm{T}}\xi'_{N}}({\bm r}_{N},t'_{N}) \cdots {\tilde{A}}_{{\mathrm{T}}\xi'_{1}}({\bm r}_{1},t'_{1}) \right) \hat{\rho} \Bigg].
\label{eUseful21}
\end{align}
where we defined the operator coefficient $\hat{\zeta}_{\xi'\xi''}(t'-t'')$ by
\begin{align}
\hat{\zeta}_{\xi'\xi''}(t'-t'')
\equiv \tilde{h}_{\xi'}^{\dagger}(\phi,t,t'') \tilde{h}_{\xi''}(\phi,t,t').
\label{eUseful22}
\end{align}
Assuming that incident field is quasi-monochromatic and that $t_{1}, \ldots, t_{N}$ are sufficiently larger than the optical period $2\pi/\omega_{\mathrm{o}}$, each double time integral can be evaluated as in Section \ref{sAppPhotoionization}.
Consequently, we obtain
\begin{align}
{\mathcal{Z}}_{N}
& = t_{1} \cdots t_{N}\,{\mathrm{tr}}\Bigg[\,\sum_{\xi'_{N}}^{\{x,y,z\}} \sum_{\xi''_{N}}^{\{x,y,z\}} \cdots \sum_{\xi'_{1}}^{\{x,y,z\}} \sum_{\xi''_{1}}^{\{x,y,z\}} 
\nonumber\\
& ~~~~ \hat{\kappa}_{\xi'_{N}\xi''_{N}} \cdots \hat{\kappa}_{\xi'_{1}\xi''_{1}}
\left({\tilde{A}}^{(-)}_{{\mathrm{T}}\xi''_{1}}({\bm r}_{1},t_{1}) \cdots {\tilde{A}}^{(-)}_{{\mathrm{T}}\xi''_{N}}({\bm r}_{N},t_{N}) \right)
\nonumber\\
& ~~~~ \left({\tilde{A}}^{(+)}_{{\mathrm{T}}\xi'_{N}}({\bm r}_{N},t_{N}) \cdots {\tilde{A}}^{(+)}_{{\mathrm{T}}\xi'_{1}}({\bm r}_{1},t_{1}) \right)\hat{\rho}\,\Bigg],
\label{eUseful23}
\end{align}
where the operator coefficient is defined by
\begin{align}
& \hat{\kappa}_{\xi'_{n}\xi''_{n}}(\omega_{\mathrm{o}})
= \int_{-\infty}^{\infty} \hat{\zeta}_{\xi'_{n}\xi''_{n}}(\tau) \exp\left(-{\mathrm{i}}\omega_{\mathrm{o}}\tau\right) d\tau,
\label{eUseful24}
\end{align}
and the explicit form of $\hat{\kappa}_{\xi'_{n}\xi''_{n}}(\omega_{\mathrm{o}})$ is shown in Eq.\eqref{ePhotoionization32}.
Thus, Eq.\eqref{eUseful8} is equivalent to Eq.\eqref{eUseful5} when calculating the trace of Eqs.\eqref{eUseful3}--\eqref{eUseful4}.

\section{Counting statistics}
\label{sAppCounts}
In this section, we derive the standard formulas for the counting statistics~\cite{Kelly1964}.
This calculation demonstrates how Boolean logic translates a complex calculation procedure into a compact compound proposition.
Specifically, we determine the form of the proposition `${\mathcal N},t$' that represents `${\mathcal N}$ output signals are observed during time $t$', evaluating its probability using Boolean logic.
This calculation is cumbersome for two reasons: it goes beyond the original scope of the first-order perturbation approximation, and repetitive measurements require multiple photoionization sites.
Below, we focus on a single photodetector case for simplicity.
\par
To begin with, the previous discussions through Section \ref{sAppPhotoionization} to Section \ref{sAppUseful} suggests that the first-order perturbation approximation is justified when photoionization occur only once in a given time interval.
Thus, we divide the time interval $t$ into $M$ small subintervals,
\begin{align}
s_{m} \equiv m\,\Delta{t}
,~~ \Delta{t} \equiv \frac{t}{M}
,~~ m = 1,\ldots,M,
\label{eAppPhotodetector1}
\end{align}
and reduces the problem of finding the expression of `${\mathcal N},t$' to the problem of selecting ${\mathcal N}$ subintervals from $M$ possible candidates.
This hypothesis is mathematically expressed as
\begin{align}
\mbox{`${\mathcal N},t$'}
= \bigvee_{(m_{1},m_{2},\ldots,m_{{\mathcal N}})}^{{m_{1}}<{m_{2}}<\cdots<{m_{\mathcal N}}} 
\mbox{`$(m_{1},m_{2},\ldots,m_{{\mathcal N}})$'},
\label{eAppPhotodetector2}
\end{align}
where the proposition on the right-hand side represents `output signals are observed only in the $m_{1}$-th, $m_{2}$-th, $\ldots$, and $m_{{\mathcal N}}$-th subintervals', and the tuple $(m_{1},m_{2},\ldots,m_{{\mathcal N}})$ denotes any combination of subinterval indices satisfying ${m_{1}}\,{<}\,{m_{2}}\,{<}\,\cdots\,{<}\,{m_{\mathcal N}}$.
\par
From Boolean logic and the physics suggested by traditional stationary state models, the proposition `$(m_{1},m_{2},\ldots,m_{{\mathcal N}})$' should have the following form:
\begin{align}
& \mbox{`$(m_{1},m_{2},\ldots,m_{{\mathcal N}})$'}
\equiv \!\!\!\!\!\!\!\!\! \bigwedge_{m}^{(m_{1},m_{2},\ldots,m_{{\mathcal N}})} \!\!\!\!\!\!\!\!\!
\mbox{`$\mathrm{output},[s_{m},s_{m+1})$'}
\nonumber\\
& ~~~~~~~~~~~~~
\wedge \!\!\!\!\!\!\!\!\! \bigwedge_{m}^{{\notin}\,(m_{1},m_{2},\ldots,m_{{\mathcal N}})} \!\!\!\!\!\!\!\!\!
\lnot\,\mbox{`$\mathrm{output},[s_{m},s_{m+1})$'},
\label{eAppPhotodetector3}
\end{align}
where the proposition `$\mathrm{output},[s_{m},s_{m+1})$' represents whether an output signal is observed during the $m$-th subinterval.
Since photodetectors must have multiple photoionization sites to produce multiple detection signals, the proposition `$\mathrm{output},[s_{m},s_{m+1})$' must be a disjunction over all photoionization sites:
\begin{align}
\mbox{`$\mathrm{output},[s_{m},s_{m+1})$'}
& \equiv \bigvee_{l=1}^{L} \mbox{`$\mathrm{output}_{l},[s_{m},s_{m+1})$'}.
\label{eAppPhotodetector4}\\
\mbox{`$\mathrm{output}_{l},[s_{m},s_{m+1})$'}
& \equiv \mbox{`$\mathrm{output}_{l},s_{m+1}$'}
\wedge \lnot \mbox{`$\mathrm{output}_{l},s_{m}$'}.
\label{eAppPhotodetector5}
\end{align}
Here, the proposition `$\mathrm{output}_{l},t$' in Eq.\eqref{eAppPhotodetector5} is the basic proposition derived in Eq.\eqref{5-132}.
Note that stationary state occupations are exclusive only within each photoionization site, while Eq.\eqref{eAppPhotodetector3} implies the exclusivity of `$(m_{1},m_{2},\ldots,m_{{\mathcal N}})$':
\begin{align}
& \mbox{`$(m_{1},m_{2},\ldots,m_{{\mathcal N}})$'$\wedge$`$(m'_{1},m'_{2},\ldots,m'_{{\mathcal N}})$'}
= \mathrm{False}
\nonumber\\
& ~~\mbox{unless $(m_{1},m_{2},\ldots,m_{{\mathcal N}})=(m'_{1},m'_{2},\ldots,m'_{{\mathcal N}})$}.
\label{eAppPhotodetector6}
\end{align}
\par
Our general assumptions restore the standard formulas for the counting statistics from the heuristic expression in Eq.\eqref{eAppPhotodetector2}.
Using the Born rule, Boolean logic, the definition of `${\mathcal N},t$' (Eq.\eqref{eAppPhotodetector2}), and the exclusivity of `$(m_{1},m_{2},\ldots,m_{{\mathcal N}})$' (Eq.\eqref{eAppPhotodetector6}), we obtain
\begin{align}
\left\langle \xvar{`${\mathcal N},t$'}{} \right\rangle
& = {\mathrm{tr}}\left[\,\zop{`${\mathcal N},t$'}{}\,\hat{\rho}\,\zopd{`${\mathcal N},t$'}{}\,\right]
\label{eAppPhotodetector11}\\
& = \!\!\!\!\!\! \sum_{(m_{1},m_{2},\ldots,m_{{\mathcal N}})}^{m_{1}<m_{2}<\cdots<m_{{\mathcal N}}} \!\!\!\!\!\!
{\mathrm{tr}}\Big[\,\zop{`$(m_{1},m_{2},\ldots,m_{{\mathcal N}})$'}{} 
\nonumber\\
& ~~~~~~~~ \,\hat{\rho}\,\zopd{`$(m_{1},m_{2},\ldots,m_{{\mathcal N}})$'}{}\,\Big].
\label{eAppPhotodetector11a}
\end{align}
Using de Morgan's laws and Eqs.\eqref{eAppPhotodetector3}--\eqref{eAppPhotodetector4}, the operator $\zop{`$(m_{1},m_{2},\ldots,m_{{\mathcal N}})$'}{}$ can be expressed as
\begin{align}
& \zop{`$(m_{1},m_{2},\ldots,m_{{\mathcal N}})$'}{leftright}
\nonumber\\
& = \hat{z}\left( 
\bigwedge_{m}^{(m_{1},m_{2},\ldots,m_{{\mathcal N}})}
\Bigg(\bigvee_{l=1}^{L} \mbox{`$\mathrm{output}_{l},[s_{m},s_{m+1})$'}\Bigg)\right.
\nonumber\\
& ~~~~ \wedge \!\!\!\! \bigwedge_{m}^{{\notin}\,(m_{1},m_{2},\ldots,m_{{\mathcal N}})} \!\!\!\!
\lnot \left( \bigvee_{l=1}^{L} \mbox{`$\mathrm{output}_{l},[s_{m},s_{m+1})$'}
\right)
\label{eAppPhotodetector12}\\
& = \hat{z}\left( 
\bigwedge_{m}^{(m_{1},m_{2},\ldots,m_{{\mathcal N}})}
\lnot \Bigg(\bigwedge_{l=1}^{L} \lnot \mbox{`$\mathrm{output}_{l},[s_{m},s_{m+1})$'}\Bigg)\right.
\nonumber\\
& ~~~~ \wedge \!\!\!\!\!\! \bigwedge_{m}^{{\notin}\,(m_{1},m_{2},\ldots,m_{{\mathcal N}})}
\left( \bigwedge_{l=1}^{L} \lnot\mbox{`$\mathrm{output}_{l},[s_{m},s_{m+1})$'}
\right).
\label{eAppPhotodetector13}
\end{align}
Then Boolean logic yields
\begin{align}
& \zop{`$(m_{1},m_{2},\ldots,m_{{\mathcal N}})$'}{}
\nonumber\\
& = \mathcal{T} \!\!\!\!\!\!\!\!\!\! \prod_{m}^{(m_{1},m_{2},\ldots,m_{{\mathcal N}})} \!\!
\left( \hat{1} - \prod_{l=1}^{L} (\hat{1} - \zop{`$\mathrm{output}_{l},[s_{m},s_{m+1})$'}{} ) \right)
\nonumber\\
& \prod_{m}^{\notin(m_{1},m_{2},\ldots,m_{{\mathcal N}})} \!\!
\left( \prod_{l=1}^{L} (\hat{1} - \zop{`$\mathrm{output}_{l},[s_{m},s_{m+1})$'}{} ) \right).
\label{eAppPhotodetector14}
\end{align}
If the number of subintervals $M$ is infinitely large and the number of photoionization sites $L$ is large but finite, the terms involving
\begin{align}
\mbox{`$\mathrm{output}_{l},[s_{m},s_{m+1})$'}
\wedge \mbox{`$\mathrm{output}_{l'},[s_{m},s_{m+1})$'}
,~~ l \neq l'
\label{eAppPhotodetector15}
\end{align}
are negligibly small.
We thus obtain
\begin{align}
& \zop{`$(m_{1},m_{2},\ldots,m_{{\mathcal N}})$'}{}
\nonumber\\
& \cong \mathcal{T} \!\!\!\!\!\!\!\!\!\! \prod_{m}^{(m_{1},m_{2},\ldots,m_{{\mathcal N}})} \!\!
\left( \sum_{l=1}^{L} \zop{`$\mathrm{output}_{l},[s_{m},s_{m+1})$'}{} \right)
\nonumber\\
& \prod_{m}^{\notin(m_{1},m_{2},\ldots,m_{{\mathcal N}})} \left( \hat{1} - \sum_{l=1}^{L} 
\zop{`$\mathrm{output}_{l},[s_{m},s_{m+1})$'}{} \right).
\label{eAppPhotodetector16}
\end{align}
By virtue of the time ordering symbol, we also obtain
\begin{align}
& \xop{`$(m_{1},m_{2},\ldots,m_{{\mathcal N}})$'}{}
\nonumber\\
& \equiv \zopd{`$(m_{1},m_{2},\ldots,m_{{\mathcal N}})$'}{} \zop{`$(m_{1},m_{2},\ldots,m_{{\mathcal N}})$'}{} 
\label{eAppPhotodetector21}\\
& \cong \mathcal{T} \!\!\!\!\!\!\!\!\!\! \prod_{m}^{(m_{1},m_{2},\ldots,m_{{\mathcal N}})} \!\!
\left( \sum_{l=1}^{L} \xop{`$\mathrm{output}_{l},[s_{m},s_{m+1})$'}{} \right)
\nonumber\\
& \prod_{m}^{\notin(m_{1},m_{2},\ldots,m_{{\mathcal N}})} \left( \hat{1} - \sum_{l=1}^{L} 
\xop{`$\mathrm{output}_{l},[s_{m},s_{m+1})$'}{} \right),
\label{eAppPhotodetector22}
\end{align}
where the operator on the right-hand side has the following form:
\begin{align}
& \xop{`$\mathrm{output}_{l},[s_{m},s_{m+1})$'}{}
\nonumber\\
& \equiv \zopd{`$\mathrm{output}_{l},[s_{m},s_{m+1})$'}{}\,\zop{`$\mathrm{output}_{l},[s_{m},s_{m+1})$'}{}
\label{eAppPhotodetector23}\\
& \cong \eta_{\mathrm{amp}}\,\zopd{`$\stateset{E<0}{l},s_{m}$'}{}\,\zopd{`$\stateset{E>0}{l},s_{m+1}$'}{}\,
\nonumber\\
& ~~~~ \zop{`$\stateset{E>0}{l},s_{m+1}$'}{}\,\zop{`$\stateset{E<0}{l},s_{m}$'}{}
\label{eAppPhotodetector24}\\
& = \eta_{\mathrm{amp}} \sum_{\statevectorR{\phi}{l}}^{\stateset{E>0}{l}} \hat{\pi}_{l}^{\dagger}(\phi,t) \hat{\pi}_{l}(\phi,t).
\label{eAppPhotodetector25}
\end{align}
\par
If we define the photoionization rate at time $s_{m}$ by 
\begin{align}
\hat{\mathcal{R}}(s_{m}) 
& \equiv \frac{\hat{\mathcal{N}}(s_{m+1}) - \hat{\mathcal{N}}(s_{m})}{\Delta{t}}
\label{eAppPhotodetector26a}\\
& = \frac{1}{\Delta{t}} \sum_{l=1}^{L} \hat{x}(\mbox{`$\mathrm{output}_{l},[s_{m},s_{m+1})$'})
\label{eAppPhotodetector26}
\end{align}
and set $\hat{\mathcal{N}}(0)\,{=}\,\hat{0}$, the first factor in Eq.\eqref{eAppPhotodetector22} casts into the form:
\begin{align}
\prod_{m}^{(m_{1},m_{2},\ldots,m_{{\mathcal N}})} \!\!\!\!\!\!\!\! \hat{\mathcal{R}}(s_{m})\Delta{t}
= \frac{t^{\mathcal{N}}}{M^{\mathcal{N}}} \prod_{m}^{(m_{1},m_{2},\ldots,m_{{\mathcal N}})} \!\!\!\!\!\!\!\! \hat{\mathcal{R}}(s_{m}).
\label{eAppPhotodetector27}
\end{align}
Since the detector count ${\mathcal N}$ is negligibly small compared to the number of subintervals $M$, the second factor in Eq.\eqref{eAppPhotodetector22} can be approximated as
\begin{align}
& \prod_{m}^{\notin(m_{1},m_{2},\ldots,m_{{\mathcal N}})} \left( \hat{1} - \sum_{l=1}^{L} 
\xop{`$\mathrm{output}_{l},[s_{m},s_{m+1})$'}{leftright} \right)
\nonumber\\
& \cong \prod_{m=0}^{M-1} \left( \hat{1} - \hat{\mathcal{R}}(s_{m}) \Delta{t} \right)
\cong e^{-\sum_{m=0}^{M-1} \hat{\mathcal{R}}(s_{m}) \Delta{t}}.
\label{eAppPhotodetector28}
\end{align}
As a result, Eq.\eqref{eAppPhotodetector22} can be expressed as
\begin{align}
\xop{`$(m_{1},m_{2},\ldots,m_{{\mathcal N}})$'}{leftright}
\cong \frac{t^{\mathcal{N}}}{M^{\mathcal{N}}} \,
{\mathcal{T}} \!\!\!\!\!\!\!\! \prod_{m}^{(m_{1},m_{2},\ldots,m_{{\mathcal N}})} \!\!\!\!\!\!\!\!\!\! \hat{\mathcal{R}}(s_{m})\,e^{-\hat{\mathcal{N}}(t)}.
\label{eAppPhotodetector29}
\end{align}
\par
Substituting Eq.\eqref{eAppPhotodetector21} and Eq.\eqref{eAppPhotodetector29} into Eq.\eqref{eAppPhotodetector12} and summing over different combinations, we obtain
\begin{align}
\langle \xvar{`${\mathcal N},t$'}{} \rangle
& = {\mathrm{tr}}\left[\,\xop{`${\mathcal N},t$'}{}\,\hat{\rho}\,\right]
\label{eAppPhotodetector31}\\
\xop{`${\mathcal N},t$'}{}
& \cong \frac{{}_{M}{\mathrm C}_{{\mathcal N}}}{M^{{\mathcal N}}}\,{\mathcal{T}} \left(\prod_{m=1}^{\mathcal{N}} \hat{\mathcal{R}}(s_{m}) \,t \right) e^{-\hat{\mathcal{N}}(t)}
\label{eAppPhotodetector32}\\
& \cong \frac{1}{{\mathcal N}!} \,{\mathcal{T}} \left(\prod_{m=1}^{\mathcal{N}} \hat{\mathcal{R}}(s_{m}) \,t \right) e^{-\hat{\mathcal{N}}(t)},
\label{eAppPhotodetector33}
\end{align}
where we assume that the photoionization rate does not change appreciably during $t$.
We also used Stirling's formula to approximate the binomial coefficient ${}_{M}{\mathrm C}_{{\mathcal N}}\,{\equiv}\,{M!}/((M\,{-}\,{\mathcal N})!{\mathcal N}!)$.
The operators $\hat{\mathcal{R}}(s_{m})$ mutually commute due to Eq.\eqref{eAppPhotodetector5} and Eq.\eqref{eAppPhotodetector26}.
Consequently, Eq.\eqref{eAppPhotodetector33} can be further approximated as
\begin{align}
\xop{`${\mathcal N},t$'}{}
& \cong \frac{1}{{\mathcal N}!} \hat{\mathcal{N}}(t)^{\mathcal{N}} e^{-\hat{\mathcal{N}}(t)}.
\label{eAppPhotodetector34}
\end{align}
\par
To express the operator $\hat{\mathcal N}(t)$ in terms of field operators, we first express it using de-excitation and excitation operators as
\begin{align}
\hat{\mathcal N}(t)
& \cong \eta_{\mathrm{amp}} \sum_{l=1}^{L} \zop{`$\occupation{\stateset{E>0}{l}}{},t$'}{}
\label{eAppPhotodetector41}\\
& = \eta_{\mathrm{amp}} \sum_{l=1}^{L} \sum_{\statevectorR{\phi}{l}}^{\stateset{E>0}{l}} \hat{\pi}^{\dagger}_{l}(\phi,t) \hat{\pi}_{l}(\phi,t),
\label{eAppPhotodetector42}
\end{align}
where the de-excitation and excitation operators mutually commute for $l\,{\neq}\,l'$, but not for $l\,{=}\,l'$.
Accordingly, Eq.\eqref{eAppPhotodetector34} do not simply cast into the standard form in which $\hat{\pi}_{l}(\phi,t)$ and $\hat{\pi}^{\dagger}_{l}(\phi,t)$ are arranged in the normal order (see Section \ref{sAppUseful}).
However, if we expand Eq.\eqref{eAppPhotodetector34} as
\begin{align}
\xop{`${\mathcal N},t$'}{}
& \cong \sum_{\mu=0}^{\infty} \frac{(-1)^{\mu}}{{\mathcal N}!\mu!} \hat{\mathcal N}(t)^{\mu+{\mathcal N}},
\label{eAppPhotodetector43}
\end{align}
the operator on the right-hand side takes the form:
\begin{align}
& \hat{\mathcal N}(t)^{\mu+{\mathcal N}} 
\cong \eta_{\mathrm{amp}}^{\mu} \sum_{l_{1}=1}^{L} \cdots \sum_{l_{\mu+{\mathcal N}}=1}^{L} 
\nonumber\\
& ~~  \zop{`$\occupation{\stateset{E>0}{l_{1}}}{},t$'}{} \cdots \zop{`$\occupation{\stateset{E>0}{l_{\mu+{\mathcal N}}}}{},t$'}{}.
\label{eAppPhotodetector44}
\end{align}
Then the terms that have the same $l$ only makes the contribution of $O(L^{1})$ within the sum over $L^{2}$ elements, which is negligible when $L$ is sufficiently large.
Consequently, the operator $\hat{\mathcal N}(t)$ can be expressed in the normal order of de-excitation and excitation operators.
Applying Eq.\eqref{eUseful8}, we obtain
\begin{align}
\hat{\mathcal N}(t)
& = t ~ \sum_{l=1}^{L} \sum_{\xi}^{\{x,y,z\}}\sum_{\xi'}^{\{x,y,z\}}
\hat{\mathcal{K}}_{\xi\xi',l}\,{\tilde{A}^{(-)}_{{\mathrm{T}}\xi}}({\bm r}_{l},t)\,{\tilde{A}^{(+)}_{{\mathrm{T}}\xi'}}({\bm r}_{l},t),
\label{eAppPhotodetector45}
\end{align}
where the operator coefficient is given by
\begin{align}
& \hat{\mathcal{K}}_{\xi\xi',l}(\omega_{\mathrm{o}})
= \eta_{\mathrm{amp}}
\sum_{\statevectorR{\phi_{+}}{l}}^{\stateset{E>0}{l}} \sum_{\statevectorR{\phi_{-}}{l}}^{\stateset{E<0}{l}} 
2\pi\,\delta\left(\omega_{\mathrm{o}}-\frac{E(\phi_{+})-E(\phi_{-})}{\hbar}\right) 
\nonumber\\
& \left(E(\phi_{+})- E(\phi_{-})\right)^{2} 
\statevectorL{\phi_{-}}{l}\hat{D}_{\xi}\statevectorR{\phi_{+}}{l}
\statevectorL{\phi_{+}}{l}\hat{D}_{\xi'}\statevectorR{\phi_{-}}{l}
\,\statevectorR{\phi_{-}}{l}\statevectorL{\phi_{-}}{l}.
\phantom{\Big|}
\label{eAppPhotodetector46}
\end{align}
\par
If we further assume that (i) the density operator is separable as $\hat{\rho}\,{=}\,\hat{\rho}_{\mathrm{A}} \hat{\rho}_{\mathrm{F}}$, (ii) the field amplitude does not vary significantly over the detector dimension and the measurement time $t$, and (iii) the detector response is isotropic ($\hat{\mathcal{K}}_{\xi\xi',l}\,{=}\,\hat{\mathcal{K}}_{l}$), Eq.\eqref{eAppPhotodetector45} yields
\begin{align}
\hat{\mathcal{N}}(t)
\cong \frac{\eta_{\mathrm{det}}}{\hbar\omega_{\mathrm{o}}} \hat{I}({\bm r}_{0},t)\,t,
\label{eAppPhotodetector47}
\end{align}
where ${\bm r}_{0}$ is the detector location, and $\hat{I}({\bm r}_{0},t)$ is the intensity operator defined by
\begin{align}
& \hat{I}({\bm r}_{0},t)
\equiv S \, \varepsilon_{0} c\, {\tilde{\bm E}}^{(-)}_{{\mathrm{T}}}({\bm r}_{0},t) \cdot {\tilde{\bm E}}^{(+)}_{{\mathrm{T}}}({\bm r}_{0},t).
\label{eAppPhotodetector48}\\
& {\tilde{\bm E}}^{(\pm)}_{{\mathrm{T}}}({\bm r}_{0},t)
= {\pm}\,{\mathrm{i}}\omega_{\mathrm{o}}{\tilde{\bm A}}^{(\pm)}_{{\mathrm{T}}}({\bm r}_{0},t)
\label{eAppPhotodetector48a}
\end{align}
The quantum efficiency of the photodetector is given by
\begin{align}
{\eta}_{\mathrm{det}}
& \equiv \frac{\hbar}{\varepsilon_{0} c\,\omega_{\mathrm{o}}}\,\mathrm{tr}\left[ \underline{\hat{\mathcal{K}}}\,\hat{\rho}_{\mathrm{A}}\right],
\label{eAppPhotodetector49}
\end{align}
where $\underline{\hat{\mathcal{K}}}\,{\equiv}\,(1/{S})\sum_{l=1}^{L} \hat{\mathcal{K}}_{l}$ represents an average response function per unit area.
The attenuation of optical fields in photosensitive materials is beyond the scope of this approximation; dispersive effects require higher-order perturbation approximations.
\end{document}